\documentclass[traditabstract]{aa} 
\usepackage{graphicx}
\usepackage{txfonts}
\usepackage{natbib} 
\bibpunct{(}{)}{;}{a}{}{,} 

\begin{document}
   \title{NIR imaging spectroscopy of the inner few arcseconds of NGC~4151 with OSIRIS at Keck}

   \author{Christof Iserlohe\inst{1},
           Alfred Krabbe\inst{2}, 
           James E. Larkin\inst{3},
           Matthew Barczys\inst{4}, 
           Michael W. McElwain\inst{5}, 
           Andreas Quirrenbach\inst{6},
           Jason Weiss\inst{3}, 
           Shelley A. Wright\inst{7}
          }

   \institute{I. Physikalisches Institut, Universit\"at zu K\"oln, Z\"ulpicher
     Strasse 77, 50937 Cologne, Germany\\
              \email{ciserlohe@ph1.uni-koeln.de}
       \and
              Deutsches SOFIA Institut, Universit\"at Stuttgart,
              Pfaffenwaldring 29, 70569 Stuttgart, Germany
       \and 
              Division of Astronomy, University of California, Los Angeles,
              CA, 90095-1562, USA
       \and   
              University of Rochester, Laboratory for Laser Energetics, PO Box
              278871, Rochester, NY 14627-8871, USA 
       \and
              Exoplanets and Stellar Astrophysics Laboratory, Code 667, NASA Goddard Space Flight Center, Greenbelt, MD 20771, USA
       \and
              Landessternwarte, Zentrum f\"r Astronomie der Universit\"at Heidelberg, K\"onigstuhl 12, 69117 Heidelberg, Germany
       \and
              Dunlap Institute for Astronomy \& Astrophysics,
              50 St. George Street, Toronto, ON, M5S 3H4, Canada
}

   \date{}

  \abstract
{We present H- and K-band data from the inner arcsecond of
  the Seyfert 1.5 galaxy NGC~4151 obtained with 
the adaptive optics assisted near-infrared imaging field spectrograph OSIRIS
at the Keck
Observatory. The angular resolution is about a few parsecs on-site and thus
competes easily with optical images taken previously with the Hubble Space
Telescope. We present the morphology and dynamics of most
species detected but focus on the morphology and dynamics of the narrow line
region (as traced by emission of [FeII]$\lambda$1.644~$\mu$m), the interplay between
plasma ejected from the 
nucleus (as traced by 21 cm continuum radio data) and hot H$_2$ gas and characterize the
detected nuclear HeI$\lambda$2.058~$\mu$m absorption feature as a narrow absorption
line (NAL) phenomenon.\\
\noindent Emission from the narrow line region (NLR) as traced by
[FeII] reveals a biconical morphology and we compare
the measured dynamics in the [FeII] emission line 
with models proposing acceleration of gas in
the NLR and simple ejection of gas into the NLR. 
In the inner 2.5 arcseconds the acceleration model reveals a better fit to our
data than the ejection model. 
We also see evidence that the jet very locally
enhances emission in [FeII] at certain positions in our field-of-view such
that we were able to distinct the kinematics of these clouds from clouds 
generally accelerated in the NLR. Further, the radio jet is aligned with the
bicone surface rather than the 
bicone axis such that we assume that the jet is not the dominant mechanism
responsible for driving the kinematics of clouds in the NLR.
The hot H$_2$ gas is thermal with a temperature of about 1700 K.
We observe a remarkable correlation between individual H$_2$ clouds at
systemic velocity with the 21 cm continuum radio jet. 
We propose that the radio jet is at least partially embedded
in the galactic disk of NGC~4151 such that deviations from a linear radio
structure are invoked by interactions
of jet plasma with H$_2$ clouds that are moving into the path of the jet
because of rotation of the galactic disk of NGC~4151. Additionally, we observe a
correlation of the jet as traced by the radio data, with gas as traced in
Br$\gamma$ and H$_2$, at velocities between systemic and $\pm$ 200 km/s at
several locations along the path of the jet. 
The HeI$\lambda$2.058~$\mu$m line in
NGC~4151 appears in emission with a blueshifted absorption component from an outflow.
The emission (absorption) component has a velocity offset of 10 km/s (-280
km/s) with a
Gaussian (Lorentzian) full-width (half-width) at half maximum of 160 km/s
(440 km/s).  
The absorption component
remains spatially unresolved and its kinematic measures differ from that of UV
resonance absorption lines. From the amount of absorption we derive a lower
limit of the HeI 2$^1$S column density of 1 $\times$ 10$^{14}$ cm$^{-2}$ with
a covering factor along the line-of-sight of C$_{los}\simeq 0.1$.
}


   \authorrunning{C. Iserlohe et al.}
   \maketitle
%

\section{Introduction}

\noindent NGC~4151 is a Seyfert 1.5 galaxy \citep{2006A&A...455..773V}
originally cataloged by 
\cite{1943ApJ....97...28S}. Due to its proximity of about 13 Mpc it is an ideal
testbed 
for the physics of active galactic nuclei (AGN)  
and therefore one of the most intensively studied Seyfert galaxies.
In the AGN  model proposed
by \cite{1993ARA&A..31..473A} a supermassive 
black hole (SMBH) is surrounded by a dusty molecular torus. 
This torus is assumed to collimate the ionizing radiation field from the
accretion disk that surrounds the SMBH. Seyfert galaxies as a class of AGNs
are distinguished according to the viewing angle of the observer onto the obscuring torus. 
One of the most important topics in AGN research has been the search for the
obscuring torus \citep[for NGC~4151 see e.g.][]{2010ApJ...715..736P}, although the
torus model has  
evolved significantly in the last decade \citep[e.g.][]{2000ApJ...545...63E}. 
This torus is usually held responsible for the sometimes spectacular biconical
emission morphologies of the narrow line region (NLR) \citep[see e.g.][]{1989Natur.341..422T}.
In NGC~4151 emission of highly ionized species such as [OIII]$\lambda$501nm  
reveal an outflow into a bicone with a projected opening angle of 75$\degr$
along a position angle (PA) of 60$\degr$ \citep[e.g.][]{2000ApJ...528..260K}. 
\citet{1995ApJ...445..700T} showed, that most of the iron in the NLR is in the
gaseous phase, and because the kinematics observed in [OIII] is similar to that
observed in [FeII]$\lambda$1.257~$\mu$m and Pa$\beta$$\lambda$1.282~$\mu$m \citep{1996AJ....112...81K}, emission of
[OIII] and [FeII] is most likely due to photoionization by the AGN. 
However, [FeII] emission line profiles appear broader than e.g. Pa$\beta$ line
profiles along the NLR such that additional 
mechanisms may be necessary to locally enhance emission of [FeII], like
shocks in a wind
from the AGN or a jet. A faint spatially resolved jet emerging from the
nucleus has been observed by \citet{2003ApJ...583..192M} and others. 
The jet extends several arcseconds at a PA of about $77\degr$ 
and is therefore misaligned with the NLR bicone axis by 
about $15\degr$.
However, the inclination of the jet with respect to the galactic plane of
NGC~4151 is assumed to be low such that interactions of the
jet with the interstellar medium in the galactic disk of NGC~4151 are likely to
occur and, indeed, some correlation between the jet and 
[FeII]$\lambda$1.644~$\mu$m emission knots has been observed \citep{2009MNRAS.394.1148S}.
The dynamics within the NLR has been modeled by several authors
\citep{2000ApJ...545L..27C, 2005AJ....130..945D, 2010MNRAS.402..819S} 
assuming acceleration in an outflow and simple ejection with no 
acceleration. But the dynamics of gas within the NLR maybe more complex. 
The mass outflow rate into each cone is about 100 times higher than the
accretion rate \citep{2010MNRAS.402..819S}, which indicates that the origin of
the outflowing gas is not the AGN but surrounding gas of the galaxy
interstellar medium that is accelerated in a nuclear outflow.
Additionally, the activity of the nucleus of NGC~4151 is also known to vary on
timescales of months, and the velocities of the nuclear outflow (as observed in P-Cygni
HeI$\lambda$388.8 nm and Balmer absorption) change on the same
timescale \citep{2002AJ....124.2543H}. However, the dynamics and morphology of
the NLR is still puzzling and the role of 
the radio jet is not fully understood (e.g. why does the radio jet
show an S-like structure as reported by \citet{2003ApJ...583..192M}?).
In the inner 250 parsec of the galactic disk of NGC~4151 a large
reservoir of hot H$_2$ gas is seen
\citep[e.g.][]{1999MNRAS.305..319F}. This gas is in 
thermal equilibrium and is most likely heated by X-rays
\citep{2009MNRAS.394.1148S} or by shocks from 
gas inflowing from larger distances \citep{1999MNRAS.304..475M, 1999MNRAS.304..481M}. 
This reservoir is usually considered responsible for feeding the AGN while
the outflow into the NLR is considered as the feedback
\citep{2010MNRAS.402..819S}, although the final feeding stage (at the sphere
of influence) has not been observed yet.\\

\noindent In this paper we present data obtained during one of the first
commissioning runs with the near-infrared
imaging field spectrograph OSIRIS at the Keck observatory.
Our observations have angular resolutions corresponding to a few parsecs
on-site, such that we can address the questions mentioned above.
Our integral field data are perfectly
suited to derive the dynamics from within a truly two-dimensional
field-of-view (FoV). Additionally, our near-infrared observations are less hampered by dust
extinction and allow angular resolutions at the diffraction limit of the
Keck telescope when using adaptive optics. \\
\noindent The paper is organized as follows:
In section 2 we summarize the instrumental setup and our
observations. Section 3 briefly summarizes the data reduction methods. 
Section In section 4 we discuss continuum
emission. In section 5 we present emission line morphologies of all prominent
species, and we discuss their dynamics in section 6. In section 7 we describe
the excitation mechanisms of H$_2$ and calculate column densities of the ro-vib
transitions of H$_2$ and HeI. We conclude in section 8 with a summary of our
findings.\\ 

\noindent In this paper we use $h_0$=75 km/s/Mpc, implying linear distances of 64
pc/arcsecond at a distance of 13.25 Mpc.

\section{Observations}
NGC~4151 was observed in 2005 February and May as one of the first
commissioning targets for the OSIRIS spectrograph (\underline{O}H 
\underline{S}uppressing \underline{I}nfra-\underline{R}ed \underline{I}maging
\underline{S}pectrograph) at the W. M. 
Keck Observatory. OSIRIS is an integral field 
spectrograph \citep{2006SPIE.6269E..42L} for 
the near-infrared (z, J, H and K band) with a nominal spectral 
resolution of $\lambda / \Delta\lambda=3700$ (corresponding 
to ~5\AA$_{ }$ or ~60 km/s) mounted on the Nasmyth platform of
the Keck II telescope, which works with the
Keck Adaptive Optics System \citep{2000SPIE.4007....2W}. The design of the 
instrument is based on concepts developed for the TIGER spectrograph 
\citep{1995A&AS..113..347B} and uses an
infrared transmissive microlens array that samples a rectangular 
FoV of the adaptive optics focal plane. The selectable plate scales are 
20, 35, 50, and 100 milli-arcseconds (mas) per angular resolution element. 
Each microlens focuses the light
into a pupil image that serves as input to the actual spectrograph which
consists of a collimator (a three-mirror-anastigmat), a diffraction grating, a 
three-mirror camera, and a Hawaii II HgCdTe detector (with 2048 $\times$ 2048 pixel
and 32 output channels). The readout noise of the detector is 13 $e^-$ per
single read, and the detector is read in sampling-up-the-ramp mode. 
OSIRIS offers several near-infrared broad- and narrowband 
filters. The number of angular resolution elements, hence the size of the FoV,
is filter dependent (see Appendix A). 
Additionally, OSIRIS is equipped with an imaging camera
hosting a HAWAII I detector (1024 $\times$ 1024 pixel). The
FoV of the camera is 20~arcseconds and is offset by 20~arcseconds from 
the center of the FoV of the spectrograph. \\

\noindent We performed spectroscopic H narrowband and K broadband adaptive
optics (AO) assisted observations of the inner few arcseconds of the Seyfert
1.5 galaxy NGC~4151 (see table \ref{observations} for details about the
observations). The plate scales were 
35 mas (in H band) and 50 mas (in K band) per angular resolution element
corresponding to linear distances of 2.2 pc and 3.2 pc at the
location of NGC~4151. The total on-source integration times were 8.3
minutes in H band and 50 minutes in K band.
The AO system\footnote{The Keck Adaptive Optics system uses a Shack-Hartmann
wavefront sensor in the visible spectrum.} 
was operated in natural guide star (NGS) mode with the bright central AGN of
NGC~4151 as point spread function (PSF) reference source. The AO correction rate during both nights was
higher than 100 Hz and the 
measured full-width at half maximum (FWHM) of the AGN in K broadband after data reduction varied between 80
mas (@2.3 micron) and 140 mas (@2 micron)
(the theoretical FWHM of the diffraction spike of the Keck 
telescope is approximately 60 milli-arcseconds in the K band). 
In K-broadband mode the FoV format is 3.20 $\times$ 0.95~arcseconds and
covers 64 $\times$ 19 field points. We observed using the classical AB pattern
with a sky offset of 20~arcseconds. 
Because the FoV in the K-broadband mode is extremely narrow, we observed NGC~4151 at
different position angles (PA) keeping the  
AGN always centered to have a well-defined reference position for 
mosaicking. 
In H narrowband the FWHM is about 80 milli-arcseconds. The FoV format
is 2.31 $\times$ 1.79~arcseconds and covers 66 $\times$
51 field points, and we used the same sky offset as in the K broadband.
We observed the A0V stars HD~140729 and HD~105601 as tellurics.


\begin{table*}
\caption{Observation summary: Date of observations, wavelength coverage of our
     data cubes, wavelength band, and name of the filter used, plate scale,
     on-source integration time and PA (positive from north to east). 
Observations from 2/23 and 2/24 have been used to create the
     KBB mosaic and the corresponding integration mask is shown in figure 
     \ref{imask}.}
\label{observations}      
\centering          
\begin{tabular}{c c c c c c  }
\hline\hline       
Date & Wavelength & Wavelength & Plate  & Integration time   & PA\\
     & coverage   &  band      & scale  & (on-source)        & \\
     & [$\mu$m]&               & [mas]  & [sec]              & [$\degr$]\\
\hline
2/23/2005  & 1.965-2.381 &KBB& 50 & 2 $\times$ 600 & -90\\
2/24/2005  & 1.965-2.381 &KBB& 50 & 3 $\times$ 600 & -45\\
5/27/2005  & 1.594-1.676 &HN3& 35 & 1 $\times$ 500 & 90\\
           & 1.965-2.381 &KBB& 20 & 1 $\times$ 600 & 0\\
\hline                  
\end{tabular}
\end{table*}

\begin{figure}
\centering
\includegraphics[width=8cm]{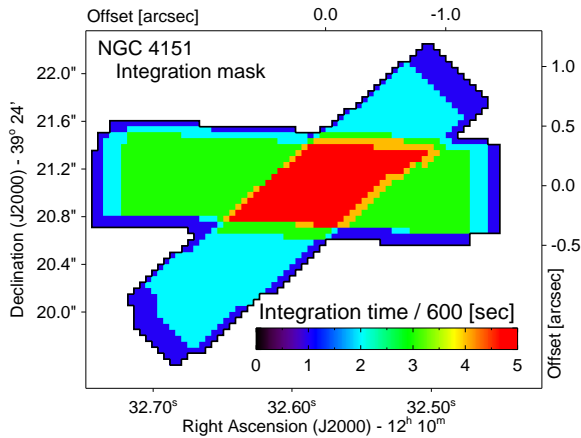}
\caption{Size of our FoV after mosaicking and total on-source
   integration time in the K band.}
\label{imask}
\end{figure}

\section{Data reduction and integrated flux values}

In this section we briefly summarize the various data reduction steps.  
We refer to appendix A for a more detailed discussion of the data
reduction concept for OSIRIS commissioning data, about the OSIRIS data reduction
pipeline (DRP) in general, and about details of the various data reduction steps
involved. \\

\begin{figure}
\centering
\includegraphics[width=8cm]{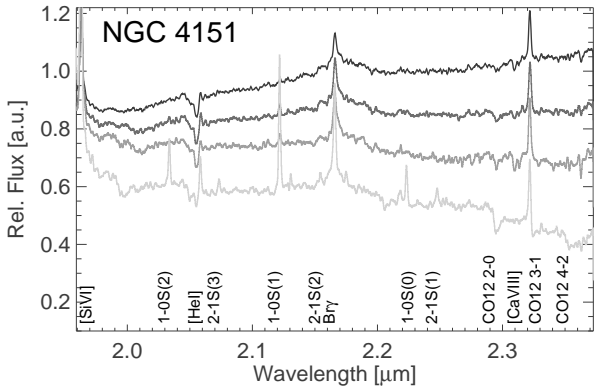}
\caption{K-band spectra of NGC~4151 extracted from circular apertures with a
radius of 125~milli-arcseconds. From top to bottom: Nucleus, region
160~milli-arcseconds southwest of the nucleus, region 160~milli-arcseconds
north of the nucleus, and region 600~milli-arcseconds northwest of the nucleus.  
The spectra are normalized to unity at 2.245~$\mu$m and shifted by multiples
of 0.15. Detected species are indicated.}
\label{spectra_K}
\end{figure}

\noindent The first data reduction step is the subtraction of the sky followed
by a 
bad-pixel search and a one-dimensional interpolation of the bad pixels along the
dispersion axis. Supplementary reduction steps to cope with detector
artifacts such as time variable direct current (DC) biases in the 32
multiplexer readout channels or electronic ghosts followed. The next
step was reconstructing the distribution of incident light on the
microlens array from the recorded detector image in a process called
rectification. In this process we used PSFs of each microlens
on the detector, recorded during daytime. After rectification, the
light in each reconstructed spectrum was assigned to a specific
microlens. The spectra were then wavelength 
calibrated and interpolated onto a regular wavelength grid with a
sampling close to the intrinsic sampling (2.5~\AA$_{ }$ in K broadband and
2.0~\AA$_{ }$ in H narrowband, corresponding roughly to 30~km/s and 35~km/s per
spectral channel respectively). 
The spectra were rearranged according to their proper positions on the sky
to a data cube with an $\alpha$, a $\delta$ and a $\lambda$ axis. 
Once the data cube was constructed, additional
reduction steps, e.g., three-dimensional bad-pixel identification and
interpolation, or correction for atmospheric differential refraction were
applied.\\ 
From here, our K- and H-band observations were treated differently.
For our K-broadband observations, the shift of the FoV with increasing
wavelength in the data cube due to atmospheric differential refraction was
determined to be less than 1/4 of an angular resolution element between 
2.0 and 2.3~$\mu$m, and were not corrected. The  
spectra of the telluric standard stars were extracted from a
400~milli-arcsecond-wide circular aperture. The intrinsic Br$\gamma$ 
absorption line at 2.166~$\mu$m in the extracted telluric spectrum (A0V star)
was fitted with a Lorentzian, and the fit was subtracted from the spectrum. Then we 
divided the extracted spectrum by a blackbody (T = 9600~K) to correct for the
intrinsic continuum emission of the telluric stars. 
In the NGC~4151 datacubes some spectra show residuals after division with the
telluric spectrum due to an imbalanced sky subtraction, especially in wavelength
regimes of low atmospheric transmission. We added appropriate 
biases to these spectra prior to division by the telluric
spectrum to correct for this imbalance.
The resulting five data cubes of NGC~4151 were finally rotated to a common PA,
shifted to a common reference position, in our case the point-like AGN, 
and were averaged.
We applied no telluric correction to our H-narrowband observations,
since in this wavelength band A0V stars exhibit quite a few strong absorption 
lines from hydrogen and helium and the transmission of the atmosphere 
in this wavelength range is nearly constant and higher than 95~\%.
However, we compared our telluric spectrum with an A0V spectrum from 
\citet{1998PASP..110..863P} to check for instrument-specific transmission 
effects. Within the signal-to-noise ratio achieved in every spectrum, neither 
atmospheric nor instrumental transmission affect the result.\\
\noindent Finally, the telluric standards were used to flux-calibrate 
our data by applying the Vega flux densities listed in Allen (2000).\\
\noindent We follow convention and refer to a spectrum corresponding to one
spatial position in the data cube as a \textit{spaxel}. The individual data
 elements of a spaxel are refered to as \textit{spexels}. The integration
 mask of the mosaicked K-broadband data cube is presented in
 Figure \ref{imask}. In that figure the effective integration time for each
 spaxel is coded in color, which also provides an assessment about the
 relative signal-to-noise ratio (S/N) throughout the FoV. From the distribution
 of the colors one can also trace the contribution of individual data cubes to
 the x-shaped final mosaic. Maximum overlap occurs within the central 
500~milli-arcseconds. The H-band integration map is not shown, since only one data cube was
obtained.
The designated coordinates in right ascension and declination are derived from the radio map
presented in \citet{2003ApJ...583..192M}, where the authors identified the
position of the bright AGN with component D of the radio jet.
On-nucleus and off-nucleus K-band spectra are displayed in
Figure \ref{spectra_K}. Table \ref{whole_flux}
lists fluxes of all species detected in our H- and K-band data.
Channel maps of the most prominent line emission/absorption features
are presented in the appropriate sections below.\\

\begin{table*}
\caption{Flux values of detected species derived from integrating the whole
        field-of-view and from a circular aperture with a radius of
        300~milli-arcseconds centered on the bright H$_2$ region in the
        southeast. HeI absorption component with redshift closer to systemic
        velocity.} 
\label{whole_flux}      
\centering          
\begin{tabular}{ l l l l }
\hline\hline       
Species & $\lambda$   & Total flux                   & Flux on eastern H$_2$ knot\\
        & [$\mu$m] & [10$^{-17}$ W/m$^2$]           & [10$^{-17}$ W/m$^2$]\\
\hline
[FeII]                & 1.644   & 10.5 $\pm$ 0.1      & - \\
$[$SiVI$]$            & 1.9634  & 6.8  $\pm$ 0.2      & 0.39 $\pm$ 0.04\\ 
1-0S(2)               & 2.0338  & 0.98 $\pm$ 0.05     & 0.13 $\pm$ 0.03\\ 
HeI (emission)        & 2.0581  & 1.5  $\pm$ 0.2      &  0.07 $\pm$ 0.02\\
HeI (absorption)      & 2.0581  & 3.9  $\pm$ 0.3      & - \\
2-1S(3)               & 2.0735  & $<$ 0.15 $\pm$ 0.05 & $<$ 0.03\\ 
1-0S(1)               & 2.1218  & 2.9  $\pm$ 0.08     & 0.38 $\pm$ 0.02\\ 
2-1S(2)               & 2.1542  & $<$ 0.1 $\pm$ 0.05  & $<$ 0.01 \\ 
Br$\gamma$ total      & 2.166   & 24.7 $\pm$ 1.3      & - \\ 
Br$\gamma$ narrow     & 2.166   & 5.0  $\pm$ 0.25     & 0.19 $<$ 0.03\\ 
1-0S(0)               & 2.2235  & 0.68 $\pm$ 0.04     & 0.10 $\pm$ 0.002\\ 
2-1S(1)               & 2.2477  & 0.15 $\pm$ 0.06     & $<$ 0.02\\ 
$[$CaVIII$]$          & 2.3213  & 2.6  $\pm$ 0.2      & $<$ 0.09\\ 
\hline                  
\end{tabular}
\end{table*}





\section{Continuum emission}

\subsection{Shape and spectral profile} 
\begin{figure}
\centering
\includegraphics[width=8.5cm]{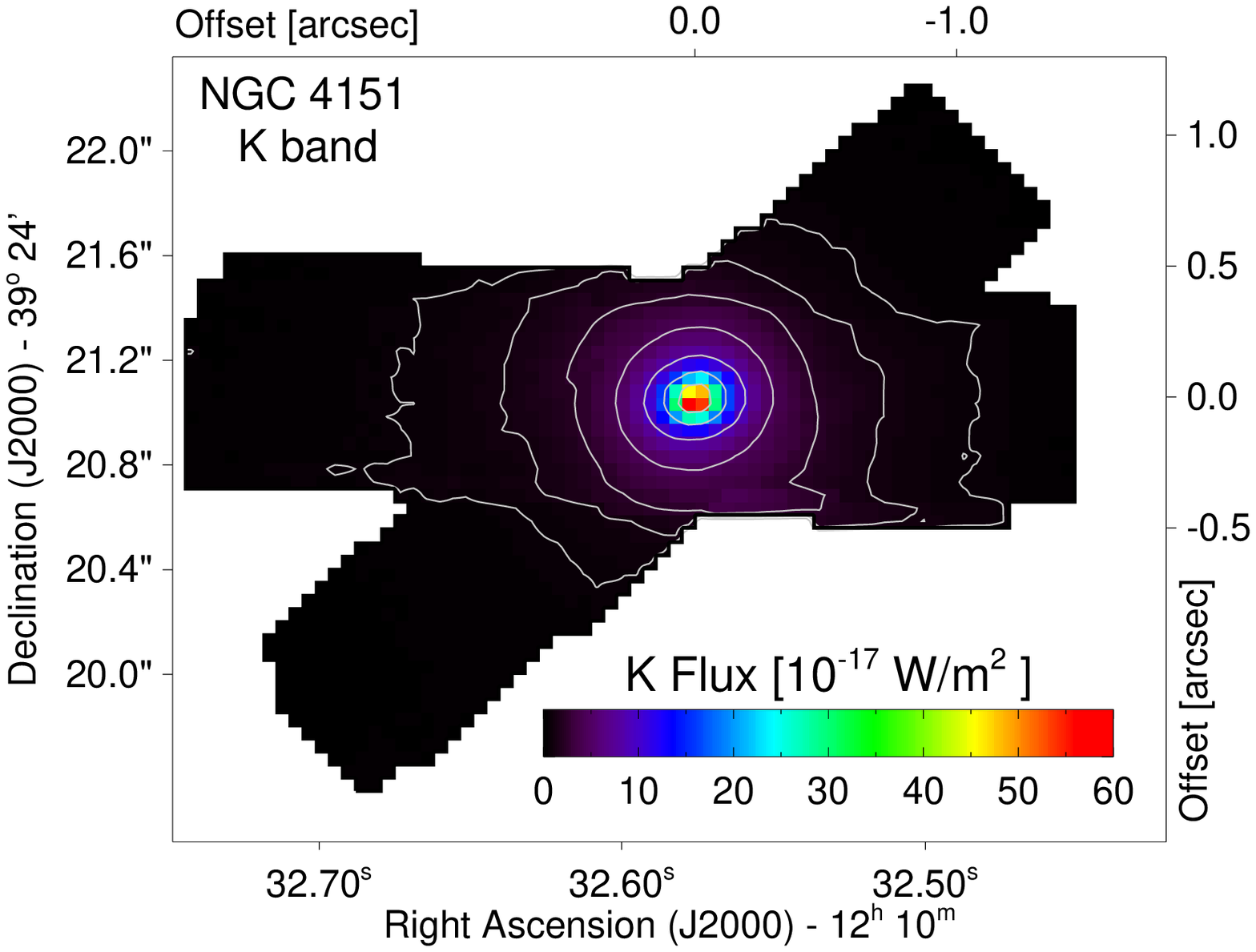}
\includegraphics[width=8.5cm]{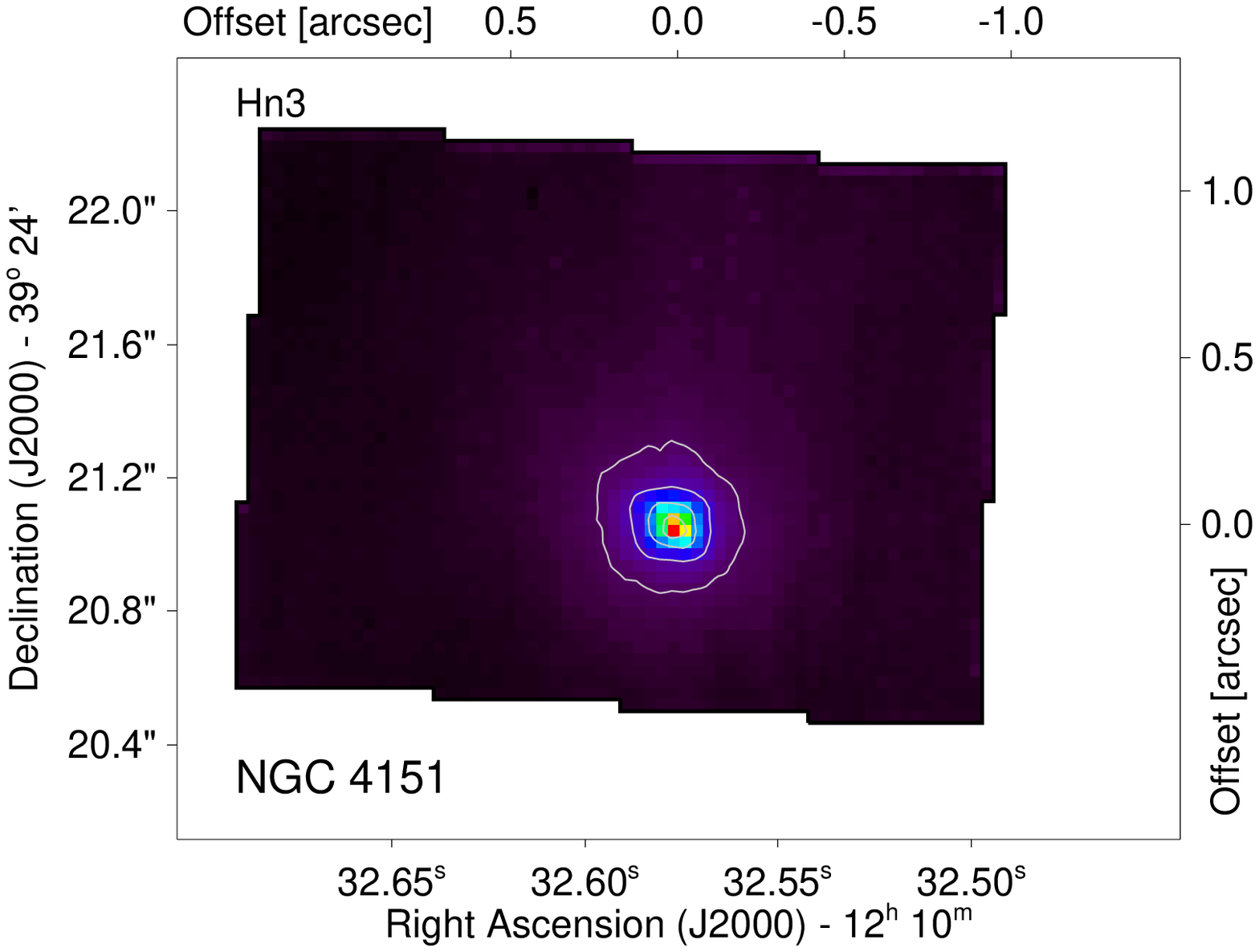}
\caption{ 
K-band image (1.965 - 2.381 $\mu$m) (top) 
and H-band image (1.594 - 1.676 $\mu$m) (bottom) of NGC~4151
with corresponding continuum contours at (1\%, 2\%, 4\%, K only), 8\% ,16\%,
32\% and 64\% of the
maximum level. North is up and east to the left.}
\label{bands}
\end{figure}

Pseudo H- and K-band images of the NGC~4151 nuclear region were created by collapsing the 
respective data cubes along their wavelength axes. They are displayed
in figure \ref{bands}. The K-band flux is 
$F_K = 2.6 \times 10^{-14}$ W/m$^2$ from our full
aperture and $F_K =3.95 \times 10^{-15}$ W/m$^2$ from a circular
 aperture with a radius of 100 mas centered on the fitted continuum peak 
 between 1.965 - 2.381~$\mu$m. 

\noindent In both wavelength bands the continuum contours are almost
circular, in particular within the region of maximum overlap where the
S/N peaks. The contours are extending the circular contours observed
by \citet{pel99} on larger scales. Small deviations
from circularity are, however, notable in both bands. We also
note discontinuities in the K-band continuum image along the lower  
rim of the horizontal stretch
and close to the southeastern end of the tilted stretch.
These features can be attributed to different seeing conditions
prevailing during the observations of the five individual data cubes 
(see also Appendix B).\\ 

\begin{figure}
\centering
\includegraphics[width=8cm]{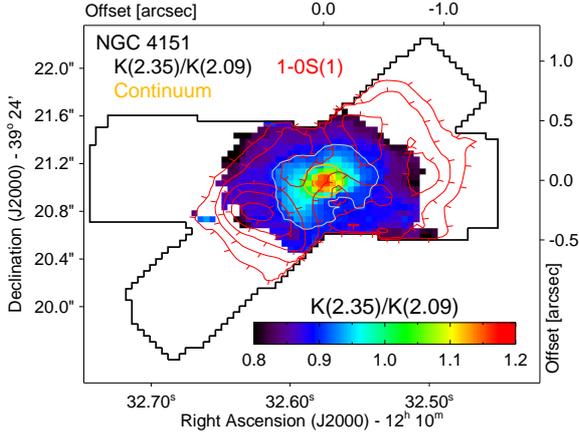}
\caption{ Flux density ratio of emission- and absorption-line-free
  continuum emission around 2.35 and 2.09~$\mu$m. H$_2$ (1-0S(1)) contours in
  red, flux ratio contour of 0.9 in white. Continuum contours (20\% and 50\%
  of the peak intensity) in yellow. North is up and east to the left.
}
\label{klks}
\end{figure}

\noindent Figure \ref{klks} shows the flux density ratio 
around 2.35 and 2.09~$\mu$m extracted from continuum emission free of
absorption and emission lines. The ratio is
about 1.3 on the nucleus and drops to 0.8 at larger distances to the
nucleus. 
Figure \ref{klks} also reveals that the flux density ratio contours (the white
contours in figure \ref{klks} correspond to a flux density ratio of 0.9)
deviate significantly from the almost circular shape
of the K-band continuum contours in figure \ref{bands} (top) and are elliptical (even
boxy).  The effect of variable seeing conditions on the mosaicked
  datacube is also discussed in Appendix B and cannot be accounted for by the
  observed asymmetry.
The spectra are redder along a long axis PA 
of -60$\degr$ with an extent of approximately 15 pc
and a long/short axis ratio of 1.5/1. 
Interestingly, this direction roughly complies with the direction
of the H$_2$-emitting bicone (as shown in figure \ref{klks} and
described below). \\


\subsection{Spectral decomposition and stellar content}
\begin{figure}
\centering
\includegraphics[width=8cm]{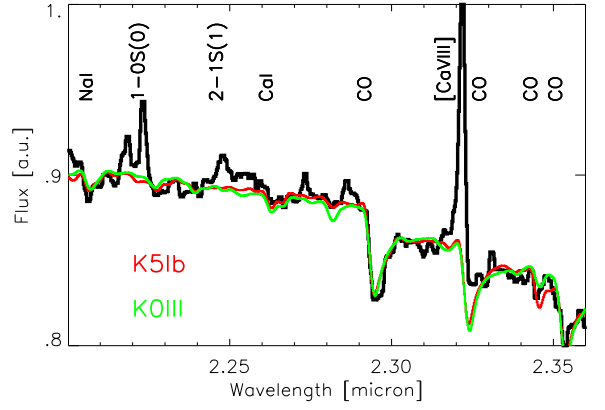}
\caption{Spectrum (black) extracted from an annuli with an inner radius of 300
mas and an outer radius of 600 mas. The artificial spectrum using the
template stars HR8726/HR8694 is overplotted in red/green.
}
\label{decomp}
\end{figure}

The continuum emission is composed of stellar and non-stellar
contributions. The stellar contribution can be quantified using pronounced
stellar absorption features like the CO absorption bandheads around 2.3~$\mu$m, 
while contributions from
non-stellar continuum emission sources can only be quantified by evaluating
the overall continuum slope over a wide wavelength range.
Following \citet{kra00}, we decomposed the continuum into stellar components and
non-stellar components, e.g., emission of hot dust, AGN power-law emission, and
free-free emission (see Appendix C for a detailed description of the
decomposition 
algorithm). For every measured spectrum we subsequently generated a synthetic 
continuum spectrum, reddened it, and compared it with the measured spectrum in
terms of minimizing the $\chi^2$ difference between the two.\\
Three effects complicate a proper decomposition of the continuum
emission for every pixel in the FoV, the broadening of the PSF with
wavelength, broad emission lines such as Br$\gamma$, 
and the limited wavelength coverage of our data. 
For instance, \citet{2009ApJ...698.1767R} decomposed the nuclear continuum spectrum 
of NGC~4151 using a wavelength coverage from 0.4~$\mu$m to 2.4~$\mu$m.
Nevertheless, it
is well possible to decompose the continuum emission into stellar and
non-stellar contributions 
(on the basis of the depth of the CO absorption bandheads), and we constrain
ourselves here to find the position of the nuclear star cluster. We did not 
attempt to decompose the non-stellar emission into its individual
components.\\
Before decomposing we needed to estimate the dominating
spectral type of the stellar continuum. Prominent features in the stellar
continuum in the near-infrared are the CO absorption bandheads around 2.3~$\mu$m, the NaI absorption 
doublet around 2.208~$\mu$m, and the CaI triplet around 2.264~$\mu$m . 
The last two are either too weak to be identified in each individual spectrum
or are strongly diluted by the non-stellar continuum. We summed all spectra
from an annulus with an inner radius of 300
mas and an outer radius of 600 mas and applied our decomposition
algorithm using all spectra from the Wallace and Hinkle stellar library
\citep{1997ApJS..111..445W}. K and M supergiants match the total
off-nuclear spectrum best, and we used the spectrum of HR8726, a K5Ib star,
as stellar template (see figure \ref{decomp}).\\
The stellar and non-stellar continuum emission peaks coincide.
At this position the stellar continuum emission drops to a very few percent 
(of the total continuum), and both peaks appear to be unresolved, although the 
stellar continuum emission can be traced farther out, where it reveals the stellar
disk of NGC~4151. 




\begin{figure}
\centering
\includegraphics[width=8cm]{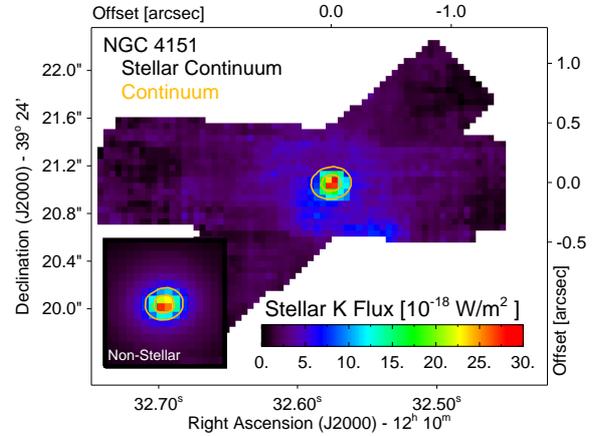}
\caption{Stellar flux derived from the spectral decomposition
  described in the text. The non-stellar continuum is not on the same flux
  scale. Corresponding continuum contours (20\% and 50\%
  of the peak intensity) are denoted in yellow. North is up and east to the left.}
\label{stellar_cont}
\end{figure}

\section{Morphology of emission and absorption lines}
\begin{figure}
\centering
\includegraphics[width=6.75cm]{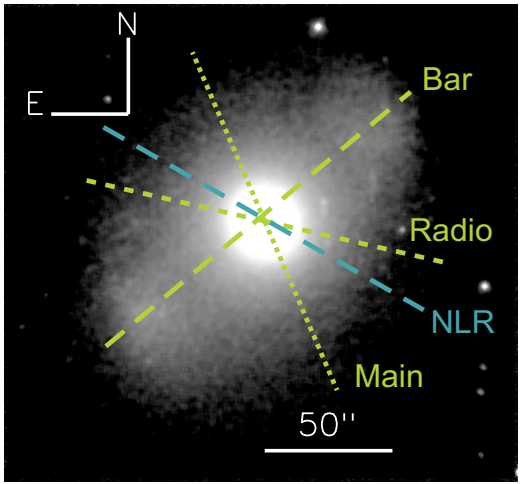}
\caption{K-band image of NGC~4151 taken from \citet{2003MNRAS.344..527K}. The
axes indicate the main kinematical axis (Main), the radio axis (Radio),
the orientation of the bar (Bar) and the direction of the NLR bicone axis 
(NLR).}
\label{4151}
\end{figure}

\noindent NGC~4151 as a whole has been classified as a (R')SAB(rs)ab barred
spiral \citep{1965PASP...77..287S}.
The inclination and PA of the galactic disk have been determined from HI
observations and optical observations to be $i \simeq 23\degr$ with a PA of
22$\degr$ \citep[e.g.][]{1973MNRAS.161P..25D, 1992MNRAS.259..369P, 1975ApJ...200..567S}.  
The observed HI dynamics and the orientation of the spiral arms indicate that
the galactic plane is closer to the observer at the southeast.
There is an inner bar with a PA $\simeq 130\degr$ deduced from
brighter isophotes \citep{1973MNRAS.161P..25D, 1975ApJ...200..567S, 1992MNRAS.259..369P}.
\citet{1999MNRAS.304..475M} and \citet{1999MNRAS.304..481M} conducted HI
observations that revealed an elliptical HI structure with a diameter of a very
 few arcminutes around the nucleus along a PA of about $-60\degr$, referred to
 as the oval. They concluded that the oval is a kinematically weak bar along
 which the authors detected inflow of gas to the nucleus, which may represent an early
 stage of the fueling process. Farther inside at distances of a few arcseconds
 around the nucleus, \citet{1999MNRAS.305..319F} observed emission of ro-vib
 transitions of H$_2$. Observations made with the Hubble Space Telescope in the
 optical ([OIII]$\lambda$501~nm) revealed a biconical emission morphology of the NLR with an opening angle
 of approximately 30$\degr$ at a PA of $60\degr$,
extending several arcseconds to either side of the nucleus 
\citep{1993ApJ...417...82E,1998ApJ...492L.115H,2000ApJ...528..260K}.  
The orientation of the galactic disk and the bicone as they appear on the sky
is shown in the next section in figure \ref{cones}, where we discuss in detail 
 the dynamics seen in individual emission lines. 
Finally, \citet{2003ApJ...583..192M} observed the 21 cm continuum radio
jet that emerges from the nucleus and extends several
arcseconds from the nucleus with a PA of about $77\degr$. 
The orientation of the line-of-nodes of the galactic disk,
 the bar, the radio jet, and the NLR bicone axis are shown in
 figure \ref{4151}. 

\subsection{Extraction}

Flux maps were extracted with FLUXER, an analyzation and visualization tool
for astronomical data cubes written in IDL. 
Since most line profiles deviate strongly from a Gaussian, we used the following
method to extract flux maps: The continuum
around an emission line was fitted with a parabola and subtracted from 
the spectrum. From the center of intensity position within the line profile
all spexels with increasing/decreasing wavelength were summed as long as the
spexel value in the continuum subtracted spectrum was positive. 
Prior to extracting the fluxes, we smooth each slice of the datacube
with constant wavelength with a boxcar to increase the S/N ratio, if
noted. Since the angular resolution therefore
depends on the above smoothing process, we usually added continuum isophotes of 50\%
and 20\% of the continuum peak flux at the
wavelength of the emission line to our images. The wavelength dependence of
our angular resolution can also 
be inferred from the continuum isophotes of figure \ref{coronal}, where no
smoothing was applied.\\ 

\subsection{Morphology}

\noindent In the following we present and compare near-infrared emission
line maps of e.g. Br$\gamma$$\lambda$2.16~$\mu$m, ro-vib transitions of molecular
hydrogen, [FeII]$\lambda$1.644~$\mu$m, [CaVIII]$\lambda$2.321~$\mu$m, 
[SiVI]$\lambda$1.963~$\mu$m, and HeI$\lambda$2.058~$\mu$m (the latter also appears
in absorption in nuclear spectra).  


\begin{figure}
\centering
\includegraphics[width=8.5cm]{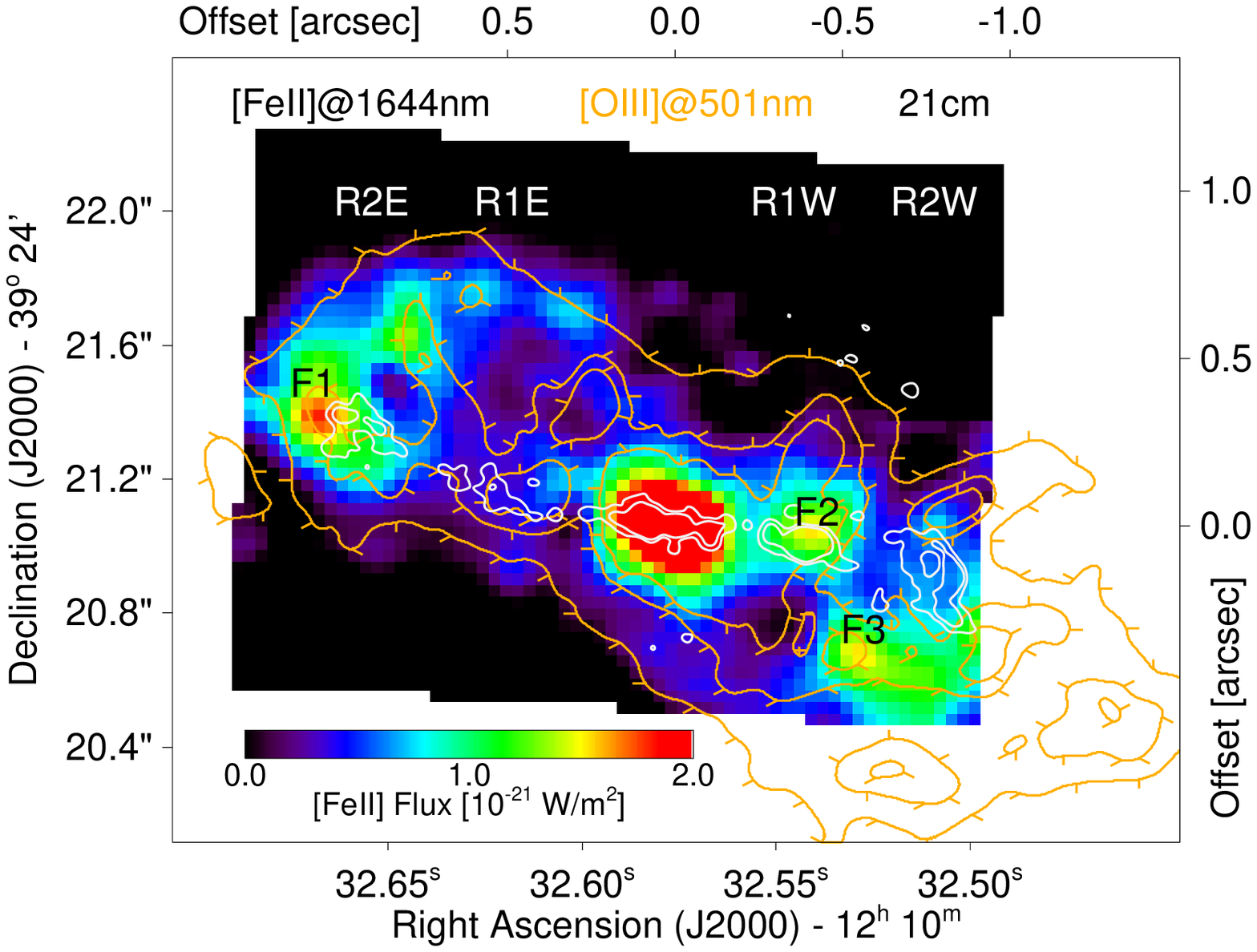}
\includegraphics[width=8.5cm]{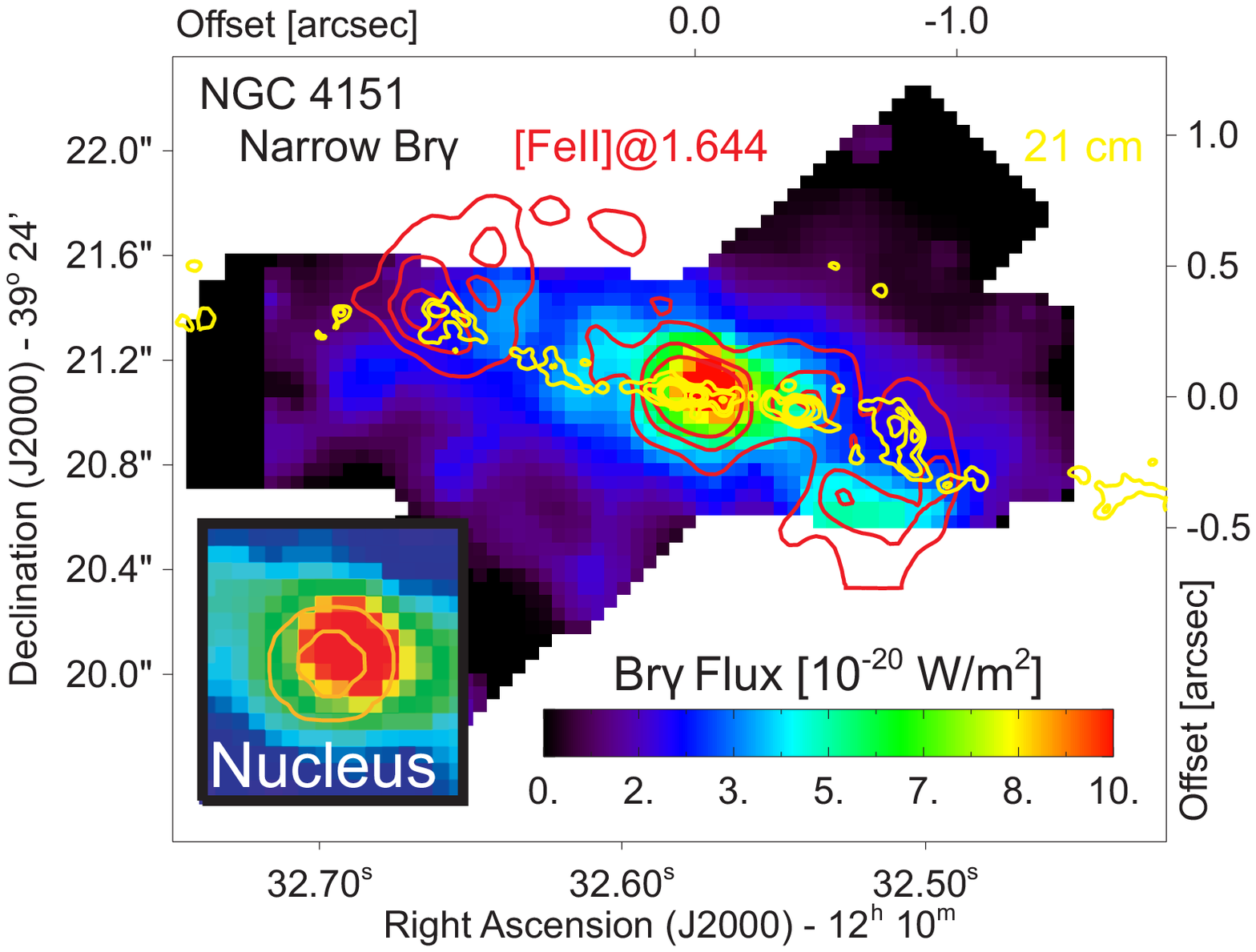}
\includegraphics[width=8.5cm]{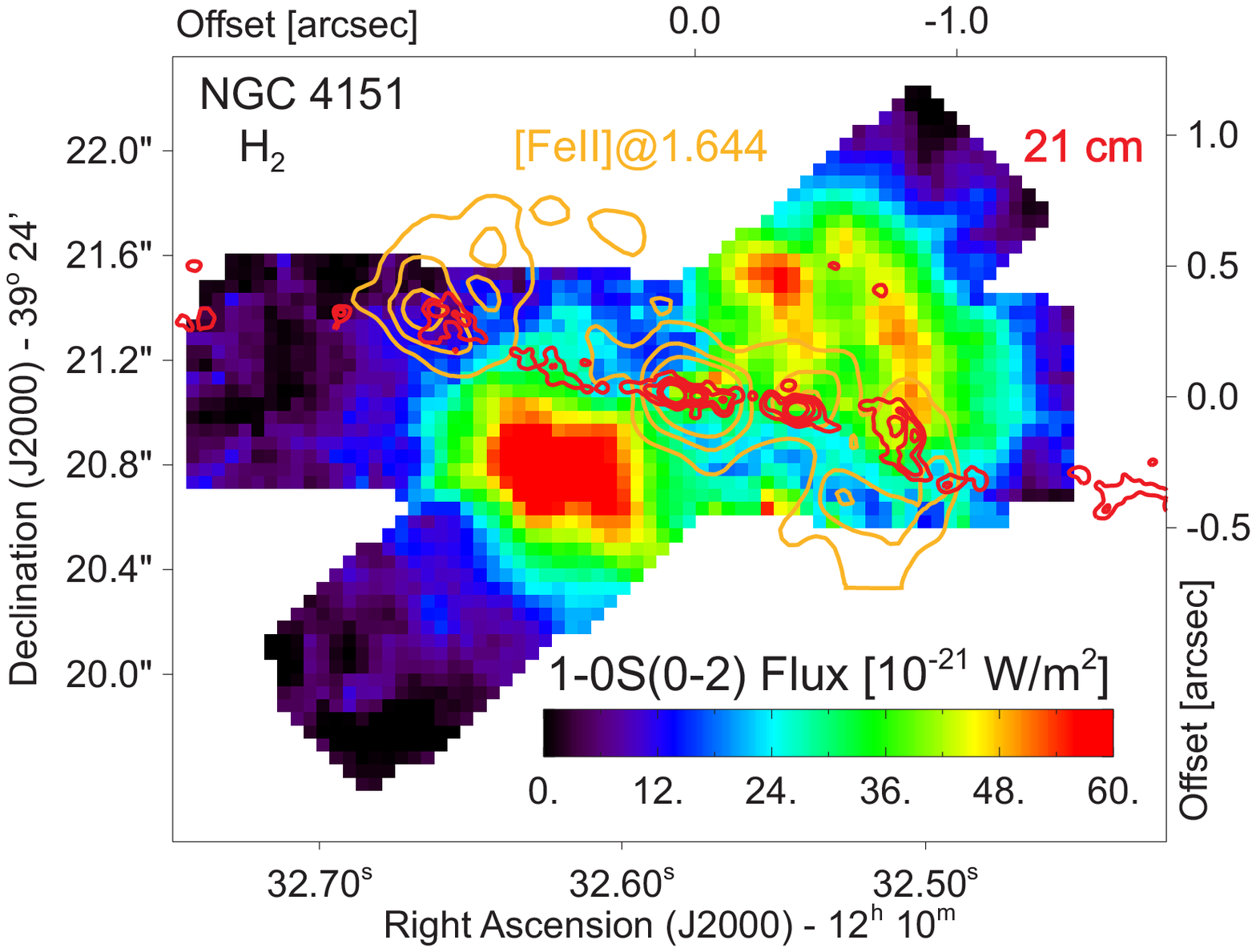}
\caption{
\textbf{All figures: }North is up and east to the left. 
Radio contours in levels of .5, 1, 2, and 4 mJy beam$^{-1}$ with a
  beam size app. four times smaller than our angular sampling.
\textbf{Top: }[FeII]$\lambda$1.644 $\mu$m emission, [OIII]$\lambda$501~nm
  emission contours taken from 
  \citet{2000ApJ...528..260K} in yellow and 21 cm radio contours taken from 
  \citet{2003ApJ...583..192M} in white. Our naming scheme for the prominent
  [FeII] emitting regions (F1-3) and the 21 cm radio continuum knots
  (R2E, R1E, R1W, R2W) is indicated. 
  Radio contours in levels of .5 and 1 mJy beam$^{-1}$ only.
\textbf{Middle: }Br$\gamma$ emission map extracted from a running
  aperture of 3 $\times$ 3. 
  [FeII] emission contours in red, 21 cm
  emission contours in yellow. Inset: The inset shows Br$\gamma$ emission from
  the nuclear region on the same flux scale. The
  orange contours represent continuum isophotes (50\% and 20\% of the
  peak value) at the wavelength of Br$\gamma$.
\textbf{Bottom: }Integrated 1-0S(0-2) flux, [FeII]$\lambda$1.644~$\mu$m, and
  21 cm radio data taken from \citet{2003ApJ...583..192M}. The H$_2$ data were
  extracted in the same way as the Br$\gamma$ emission map, but using a running
  aperture of 3 $\times$ 3.}
\label{emissionall}
\end{figure}

\clearpage

\subsubsection{Emission from within the NLR}
We used emission of [FeII]$\lambda$1.644~$\mu$m
(see figure \ref{emissionall}, top) to trace the NLR. 
The emission arises from a biconical
structure with a position angle of approximately 60$\degr$ with a projected
opening angle of 75$\degr$. 
The emission morphology appears to be clumpy and most of the emission is
detected from the position of the continuum peak. There is a clumpy arc of
[FeII] located approximately 1 arcsecond to the northeast extending north
(region F1 hereafter), a knot approximately 0.5
arcseconds to the west (region F2), and 
another extended structure 0.75 arcseconds to the southeast (region F3).
Since [FeII] and [OIII] emission are both assumed to be due to photoionization
by the AGN \citep{1996AJ....112...81K}, we overplotted the contours of
[OIII]$\lambda$501~nm taken from \citet{2000ApJ...528..260K} on our [FeII] flux
map.  The [OIII] emission appears to be clumpy as well, with emission from the same
[FeII] regions as mentioned above, but also from beyond.
The 21 cm continuum radio data taken from \citet{2003ApJ...583..192M} are also
overplotted in figure \ref{emissionall} (top). Their radio component D correlates with the position of the AGN,
and our data are aligned such that the 
position of this radio component is identical with the position of the bright
continuum peak that we assume to be the AGN. The radio emission extends
linearly from the nucleus at a PA of 77$\degr$.
In the case of NGC~4151 the radio jet
is misaligned by 15$\degr$ with the NLR's major axis if deduced
solely from [FeII] or [OIII] morphologies. 
Within our FoV we detect four bright radio knots (plus the bright radio 
knot at the position of the continuum peak), approximately 0.5
and 1.0 arcseconds to the east (regions R1E and R2E) and 0.4 and 0.75
arcseconds to the west (regions R1W and R2W) from
the continuum peak. The knots closer
to the nucleus (the inner knots, with index 1) appear to extend roughly
linearly from the nucleus at a PA of app. 80$\degr$,
while the knots at larger distances (outer knots with index 2) seem to deviate
from this straight line to a PA of about 72$\degr$, causing 
the jet to appear S-shaped. The outer eastern radio knot (R2E) correlates well
with the south end of the [FeII] arc (F1) and the inner western radio knot (R1W)
correlates well with F2. Together with the above mentioned misalignment of the
jet with the main axis of the NLR (based on 
the [FeII] and [OIII] morphologies), we conclude from these velocity-integrated
flux maps that the excitation in the NLR
maybe partially enhanced by the jet. We discuss the correlation of the radio
jet with [FeII] emission at individual velocities in the next section. We did
not observe enhanced [FeII] emission from the locations where deviations from
the straight jet axis occur.\\
\noindent At the nucleus the Br$\gamma$ emission line shows a prominent broad
component (with a Gaussian FWHM of 
several thousand km/s) with a narrow component at the top. In the following we
discuss the narrow emission of Br$\gamma$ only. 
The morphology of Br$\gamma$ (figure \ref{emissionall}, middle) appears to be less
clumpy than [FeII] with smoothly distributed emission along the
NLR. The emission peak in Br$\gamma$ is
located approximately 10 pc to the northwest at a PA of -50$\degr$ 
of the continuum peak (see the inset in figure \ref{emissionall}, middle).
We detect Br$\gamma$ emission knots at the locations of the [FeII] knots
F2 and F3 (all to the west of the nucleus), but not at locations where the 21
cm radio continuum is prominent. 
However, Br$\gamma$ isophotes seem to extend along a
PA of about $70\degr$ and to locations where the inner radio knots are
observed.\\
\noindent The HeI $n=2^1P-n=2^1S$ at 2.0581~$\mu$m emission line in nuclear 
spectra is accompanied 
by a blueshifted optically thin absorption complex that
is marginally resolved in the spectra presented by \citet{2006A&A...457...61R}.  

\begin{figure}
\centering
\includegraphics[width=8cm]{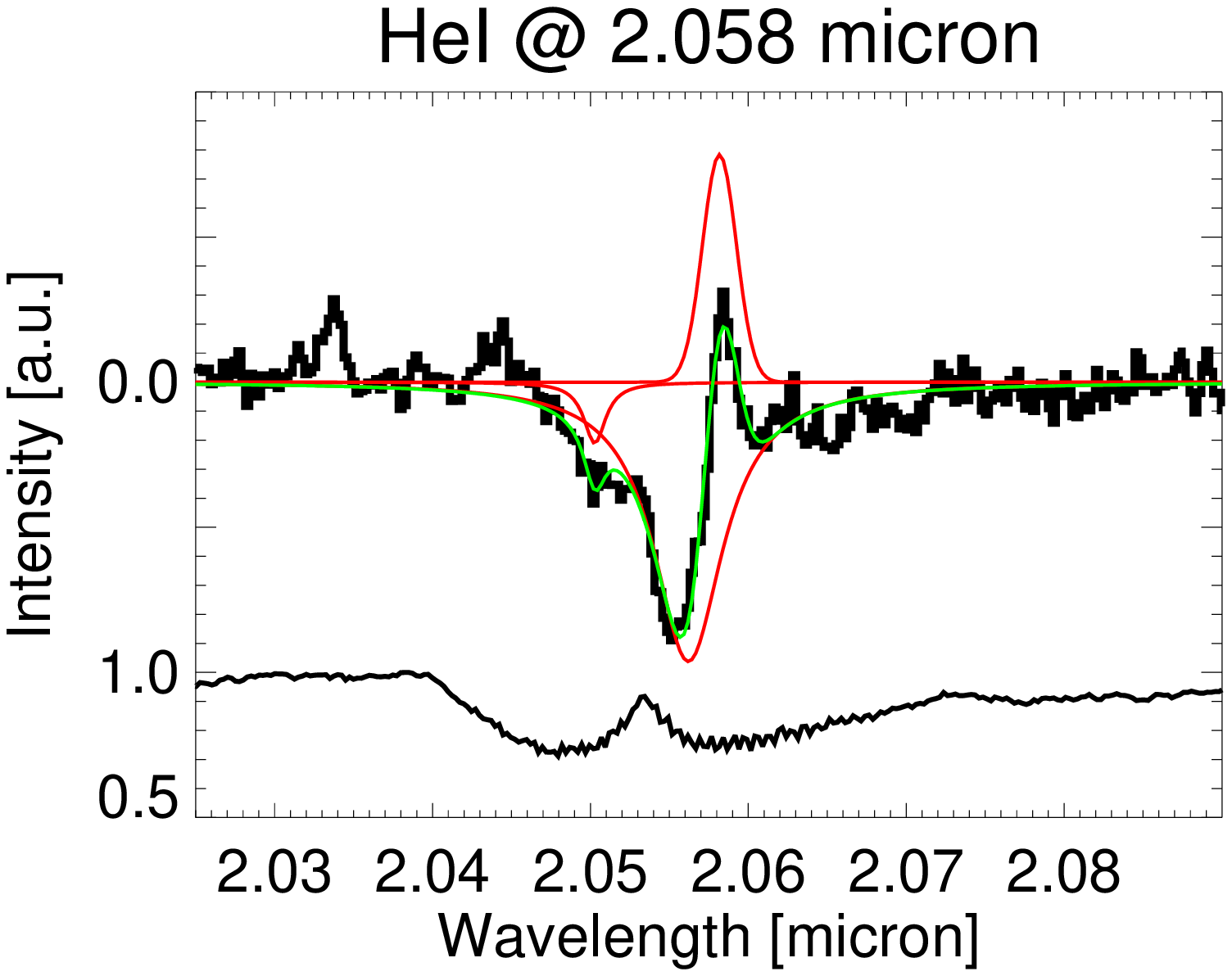}
\includegraphics[width=8.5cm]{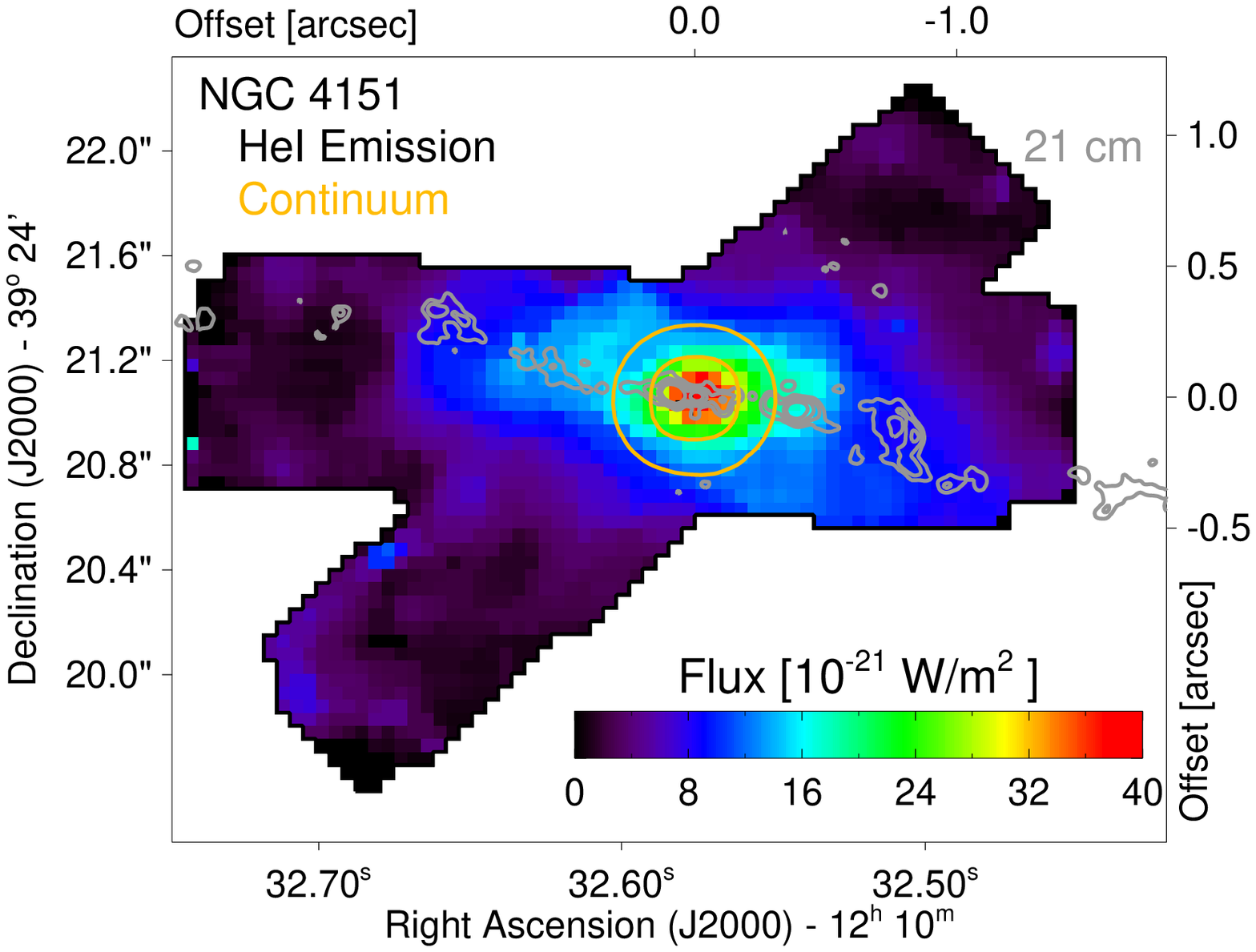}
\includegraphics[width=8.5cm]{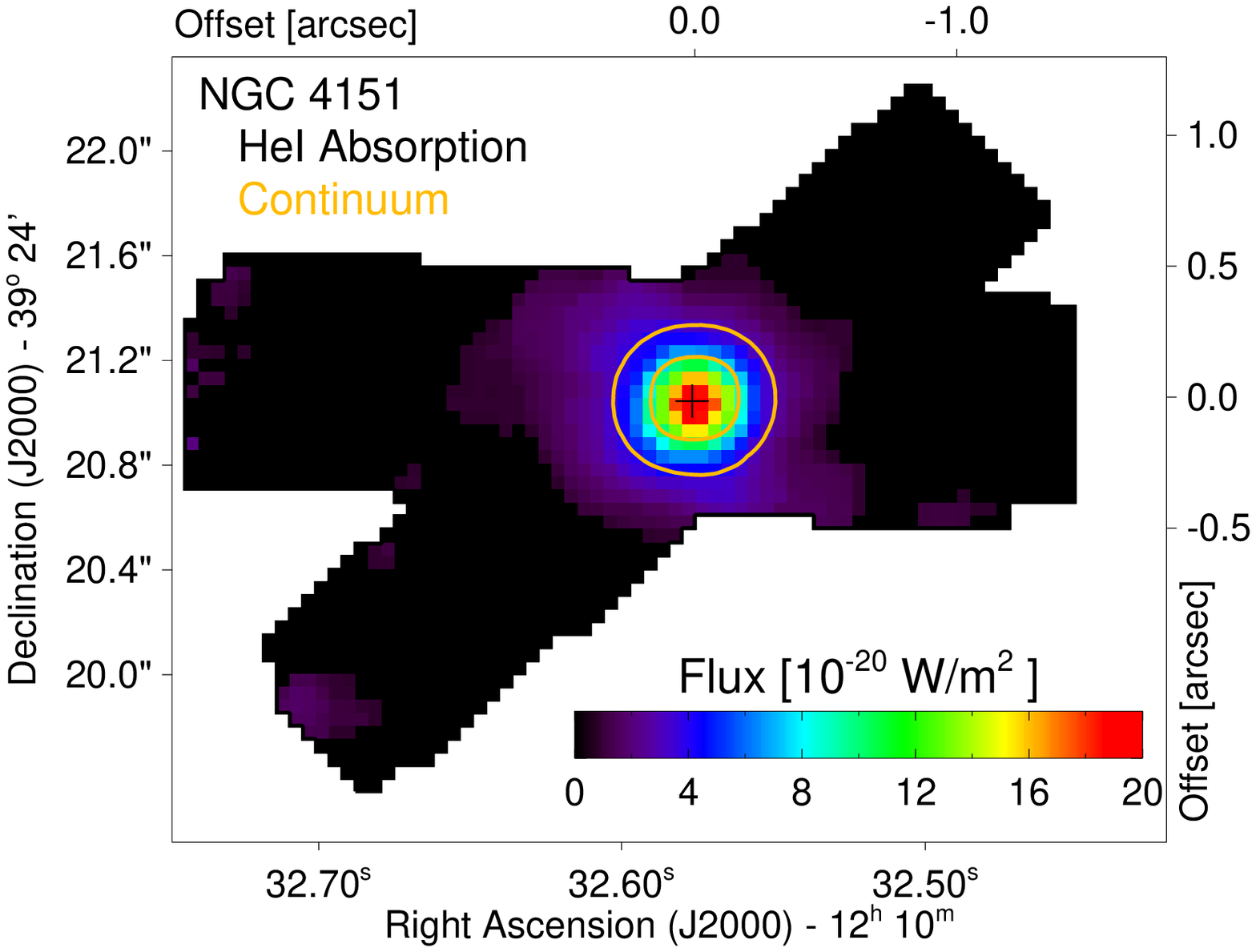}
\caption{
HeI$\lambda$2.0581~$\mu$m line complex.\newline
\textbf{Top: }Continuum-subtracted line complex of HeI
  extracted from a circular aperture with a radius of 400 milli-arcseconds 
centered on
  the nucleus; data are plotted in black, the fitted Lorentzian absorption and Gaussian
  emission components are plotted in red, total fit plotted in green. The continuum was
  fitted with a parabola using emission/absorption-free wavelength channels to
  the left and to the right of the line complex. 
  Below the extracted line complex we plot the telluric spectrum
  around the HeI line complex. In this spectral regime
  the atmospheric transmission drops to about 75\%, as indicated.\newline
\textbf{Middle and bottom: }HeI emission (middle) and absorption flux (bottom) 
derived from the three-component fit, as explained in the text. To increase the
S/N we used a running aperture of 5 $\times$ 5 prior to extraction. Continuum
isophotes are marked in yellow at the wavelength of HeI for 20\% and 50\% of the continuum peak
intensity. North is up and east to the left. Radio contours in
emission map in levels of .5, 1, 2, and 4 mJy beam$^{-1}$ with a
  beam size about four times smaller than our angular sampling.}
\label{emission2}
\end{figure}
\clearpage
\noindent The atmospheric transmission around that emission line highly depends on
wavelength (see the telluric spectrum in figure \ref{emission2}, top).
Hence, a proper telluric correction is critical and was performed as described
in the data reduction section to ensure a smooth continuum around the line
complex. Figure \ref{emission2} (top) shows the continuum-subtracted spectrum
around the HeI complex extracted from a circular aperture with a radius of
400 milli-arcseconds centered on the nucleus. To extract the flux and 
determine the amount of absorption, we fit a Gaussian emission
component and two Lorentzian absorption components to the line complex.
The results of our fits (flux, Gaussian and Lorentzian widths, equivalent widths,
and column densities derived from the absorption component) are summarized in
table \ref{he_fit}. The 
deeper absorption component is blueshifted by -280 km/s and the weaker by -1150
km/s. However, we detect the absorption component with the larger velocity
offset everywhere in our FoV, which indicates that this component is a residual
in our data reduction.

\begin{table*}
\caption{Results of the fit to the nuclear HeI$\lambda$ 2.0581~$\mu$m complex. The corresponding
spectrum is extracted from a circular aperture with a radius of
400~milli-arcseconds centered on the nucleus. The fit is explained in the text. The
calculation of the column density of the lower state of this transition is
described in section \ref{subsec:helium}.}   
\label{he_fit}      
\centering          
\begin{tabular}{l l l }
\hline\hline       
   Component      &  Emission & Absorption \\
    v [km/s]  & 10      & -280  \\
    Gaussian $\sigma$ or Lorentzian HWHM [km/s] & 160   & 404     \\
    Flux [10$^{-18}$ W/m$^2$]                     & 7.4      & 27.1      \\
    EW [\AA]                              & 2.1      & 7.7       \\
    Column density for C$_{los}$=1 [cm$^{-2}$]     &          & 5.4 $\times$ 10$^{12}$\\
    Minimum C$_{los}$                             &         & 0.09\\ 
    Column density for minimum C$_{los}$ [cm$^{-2}$] &       & 1.1 $\times$ 10$^{14}$\\
\hline                  
\end{tabular}
\end{table*}

\begin{figure}
\centering
\includegraphics[width=8.75cm]{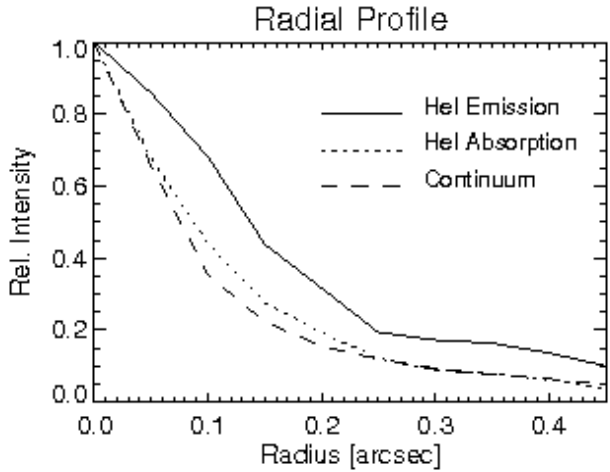}
\caption{Radial profiles of the HeI$\lambda$2.0581~$\mu$m emission and
  absorption and continuum flux at the position of the HeI line.}
\label{he_radial}
\end{figure}

\noindent Figure \ref{emission2} (middle and bottom) shows the derived HeI
morphology (integrated emission and absorption) deduced from our fits. 
The emission component extends into the direction of the NLR and is well
comparable with the Br$\gamma$ emission morphology with 
two exceptions: the HeI emission peaks at the position of the continuum peak,
and we observe no significant HeI emission at F3, where [FeII]
and Br$\gamma$ are bright.
The integrated absorption component peaks at the
nucleus but remains unresolved (as indicated by a radial 
plot of the continuum and the integrated absorption flux in
figure \ref{he_radial}). 
The HeI absorption seems to be a local feature of the nucleus, occurring closer
to it than the resolution limit of our observations (a few parsecs).

\subsubsection{Emission from beyond the NLR}

In contrast to the morphologies of Br$\gamma$ and
HeI, which extend into the NLR, the H$_2$ morphology is completely different and
extends nearly 
perpendicular to the NLR, thus avoiding the NLR (see Fig. \ref{emissionall}, bottom).
The H$_2$ emission is most pronounced along a PA of about 
-50$\degr$ and is roughly aligned with the bar and the galaxy minor axis.
In the northwestern region the H$_2$ emission appears to be clumpy on scales
of a few pc and reveals a shell-like 
structure. In the southeast the emission appears to be less clumpy. 
Toward the south of the inner shell-like structure in the northwest, at the edge of our field-of-view, the
H$_2$ emission increases such that the H$_2$ emitting region is not fully
sampled by our FoV.
There is a remarkable correlation between parts of the H$_2$ emission
morphology and the 21 cm continuum radio jet. Deviations from the straight
radio jet axis (drawn from the inner jet knots) occur exactly at the edges of 
bright H$_2$ clouds (see especially the radio knot associated with the southern
extension of the outer shell in the northwest). 
Following \citet{1982ApJ...263..576W}, we propose that the 21 cm radio is indeed
arising from regions in space where H$_2$ gas in the galactic disk is rotating
clockwise \citep{1995MNRAS.272..355M}
into the path of the jet, giving rise to the remarkable interaction zones. 
The [FeII] emission does not originate from the same regions on the sky as the H$_2$
emission, which is compatible with
the assumption that emission from the NLR originates from regions
exposed to the radiation field of the AGN where H$_2$ molecules dissociate.\\
Although our FoV does not sample the full inner two arcseconds due to its
X-shape, the H$_2$ emission does not extend significantly 
beyond our FoV \citep[see][]{2009MNRAS.394.1148S}. It is possible that the
pronounced H$_2$ emission  
morphology represents another feeding stage of the HI bar observed
by \citet{1999MNRAS.304..481M} but closer to the nucleus.
These authors detected an oval distortion in the inner two
arcminutes resembling a rotating, inclined HI ring or disk (see their figure
5), referred to as the oval. However, the velocity field of the oval differs
significantly from a rotating, 
inclined disk because the isovelocity contours are elongated along the oval. 
Additionally, \citet{1999MNRAS.304..481M} detected bright HI regions close to the leading
edges of the oval whose kinematics shows signature of a bar shock. 
From this they concluded that the oval is a kinematically weak bar. 
Subtracting a radial velocity field of purely circular rotation reveals that gas may
stream along that bar at a PA of -60$\degr$ toward the nucleus
\citep{1999MNRAS.304..475M}, which
they interpreted as an early stage of the fueling process.
The position angle of this bar is -60$\degr$ and thus roughly coincides with the
position angle drawn from the most prominent H$_2$ regions. It is possible
that the observed emission of H$_2$ is at least enhanced by 
this inflow of gas to the central 200 parsecs.
There is also a weak bridge of hot H$_2$ gas between the two large gas reservoirs
to the southeast and the northwest that runs through the nucleus.

\subsubsection{Coronal emission lines}

\begin{figure}
\centering
\includegraphics[width=9cm]{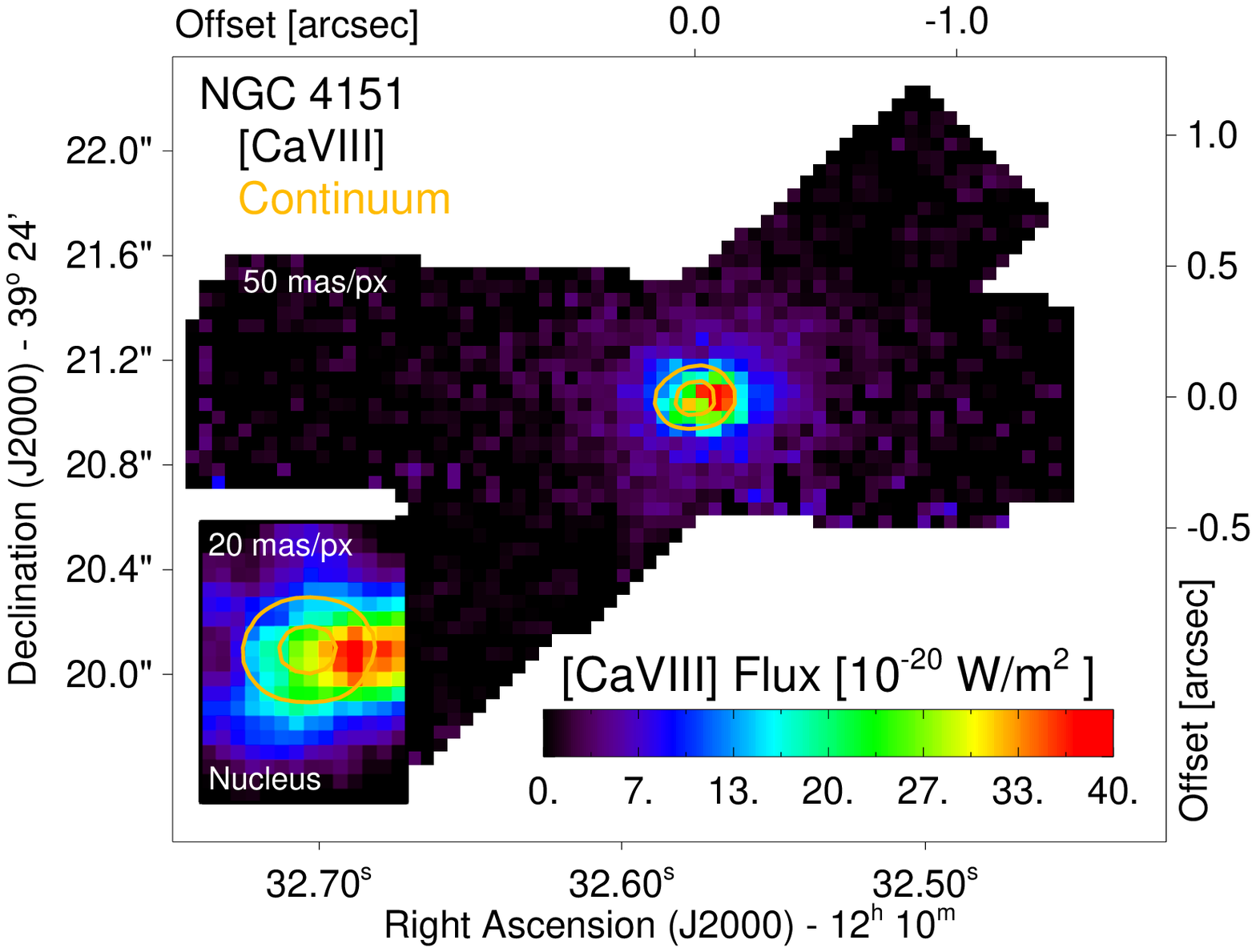}
\includegraphics[width=9cm]{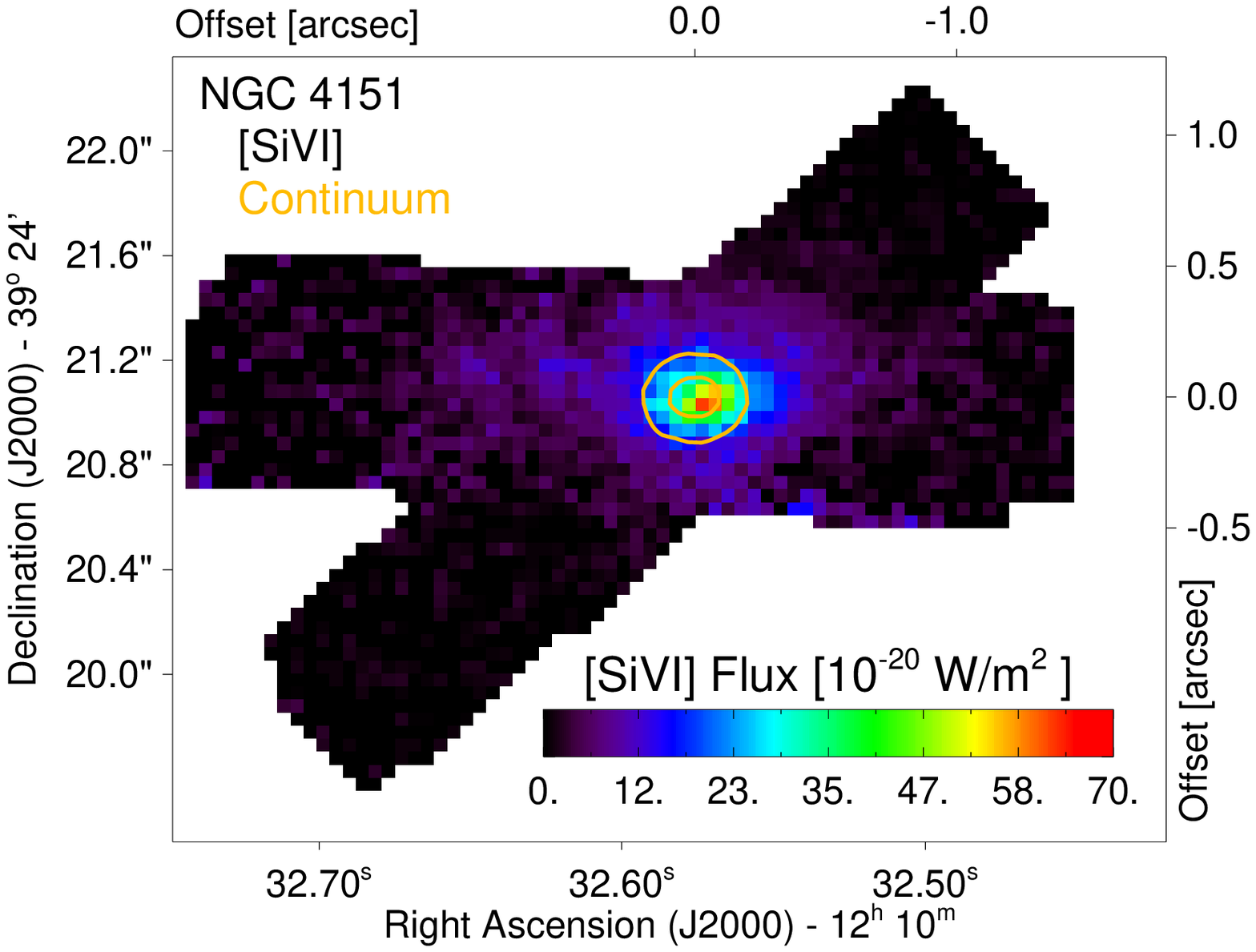}
\caption{[CaVIII] (top) and [SiVI] (bottom) emission morphology extracted from
every spaxel. The 
  yellow contours represent continuum isophotes at the wavelength of
  [CaVIII] for 20\% and 50\% of the continuum peak intensity. The
  inset (top figure) shows a [CaVIII] emission map derived from the data set with a
  sampling of 20~milli-arcsecond per pixel. North is up and east to
  the left.}
\label{coronal}
\end{figure}

\begin{figure}
\centering
\includegraphics[width=9.25cm]{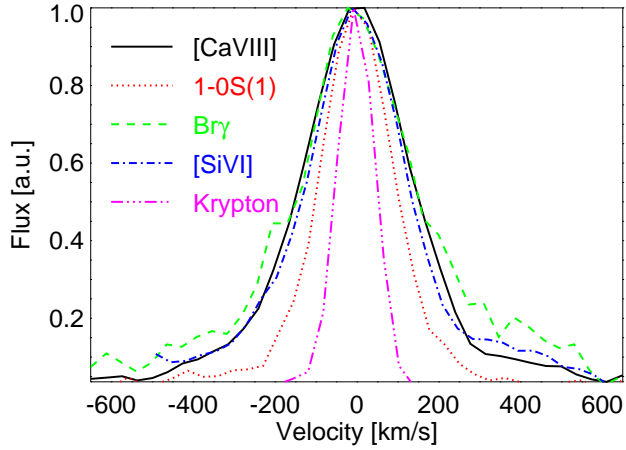}
\caption{Emission line profiles of [CaVIII], [SiVI], Br$\gamma$,
  (extracted from the nucleus with a circular aperture with a radius
  of 400 milli-arcseconds), 1-0S(1) (extracted from the eastern
  H$_2$ region) and a krypton calibration lamp line at 2.148~$\mu$m. 
  The spectra were continuum-subtracted by fitting a
  parabola to wavelength channels to the left and right of the
  emission lines. In the case of [CaVIII] the
  continuum was modeled as described in Appendix C to account
  for absorption of stellar CO bandheads. The emission lines were
  shifted in velocity to the center of intensity.}
\label{ca8_profiles}
\end{figure}

\noindent NGC~4151 is an archetypical Seyfert galaxy, and strong
emission of coronal lines in [CaVIII] and [SiVI] is detected. 
The [CaVIII] emission profile is strongly affected by the
CO3-1 absorption bandhead. 
The continuum emission was modeled with the spectral decomposition
algorithm prior to extracting the [CaVIII] channel map. 
The emission line profiles of [CaVIII] and [SiVI] (see
figure \ref{ca8_profiles}) are very similar, but show a 
very weak blueshifted hump compared to the much narrower 1-0S(1) line.
Fitting two Gaussians to the
coronal emission line reveals a broader component that is marginally
blueshifted by one spectral channel (app. 35 km/s). The FWHMs are 950 km/s of
the broader and 230 km/s of the narrower component (not corrected for
instrumental resolution). The considerable
broadening of the blueshifted component may indicate that the [CaVIII]
emission at least partially arises from a shock-excited
region close to the nucleus \citep{2005MNRAS.364L..28P}.\\
\noindent The morphology of the [CaVIII] and [SiVI] emission is shown in 
figure \ref{coronal}. Extended emission from both species is detected (at
systemic velocity) from the
NLR at distances of up to 50 pc from the nucleus. This may imply an 
energetic radiation field in the galactic plane of NGC~4151 that is possibly due 
to a clumpy obscurer of the nuclear radiation field.
Emission of both species peaks close to the nucleus. However, a closer
inspection of the dataset with 20~milli-arcsecond sampling (see inset in figure \ref{coronal}, top)
reveals that e.g. the [CaVIII] peak is located approximately 6 parsecs
to the west of the continuum peak and that both peaks are resolved. 
The peak in the [SiVI] emission map seems to be less pronounced than the [CaVIII]
peak, but the performance of the AO is wavelength dependent, as indicated by the
more extended continuum contours in the [SiVI] image that pretend different morphologies.

\noindent To test whether the two species are differently distributed, 
the [CaVIII] emission map was convolved with a Gaussian seeing disk such that the continuum maps at
the wavelengths of [SiVI] and [CaVIII] match (not shown). The convolution of
the maps reveals that both emission peaks
are almost equally extended and are
located to the west of the continuum peak, which very roughly
coincides with the nuclear Br$\gamma$ peak. 
These emission peaks may represent interaction zones of outflowing material
with gas a few parsecs away from the nucleus.\\


\noindent We summarize that we did see [FeII], Br$\gamma$ and HeI most pronounced
along the NLR, with the H$_2$ gas strictly avoiding the NLR. Coronal emission
lines indicate a wind and peak a few parsecs to the west of the nucleus.
A comparison with the results presented in \citet{2009MNRAS.394.1148S} and \citet{2010MNRAS.402..819S}
indicates that our observations have a slightly higher angular resolution such
that elongated isophotes in the coronal line emission appear to be resolved in our observations
(see e.g. the [CaVIII] peak in figure \ref{coronal}, top).

\clearpage
\section{Dynamics of gas in the NLR}

The dynamics of gas in the NLR of NGC~4151 has been subject of various
studies, e.g. \citet{2000AJ....120.1731C}, \citet{2005AJ....130..945D}
and \citet{2010MNRAS.402..819S}.  
Here, the centroid velocities along the line-of-sight of individual clouds were
determined by fitting the emission line profiles with, e.g., multi-Gaussian
profiles. The geometry of the volume into which the clouds expand was 
modeled as a radial outflow where the central obstructing torus confines the
flow into a bicone. Assumptions about the mechanism that drives the movement of
clouds in the NLR (acceleration or simple ejection) are formulated as the
velocity-distance law that relates the radial velocity of clouds with the 
distance from the nucleus. From the geometry and the velocity distance law one
can set up a model to predict possible velocities of clouds in the NLR that
can be compared with measured centroid velocities. \\

\noindent The aim of this section is to determine the dynamics of the NLR as
traced by [FeII]$\lambda$1.6440~$\mu$m and to compare our measurements with the
results presented in \citet{2005AJ....130..945D}
and \citet{2010MNRAS.402..819S}. We adopted the geometries and velocity-distance
laws presented in these 
papers and fit two Gaussians to the emission line profiles in our FoV to
determine the centroid velocities. We compared the measured centroid
velocities and model predictions in terms of position velocity diagrams.\\

\noindent We begin discussing the dynamics of [FeII] with presenting emission
line profiles and quantities derived from single-Gaussian fits. Then we
discuss channel maps, position velocity diagrams, and results
from our fits to emission line profiles in our FoV and compare the
found centroid velocities with the model predictions proposed
by \citet{2005AJ....130..945D} and \citet{2010MNRAS.402..819S}. 
Finally, channel maps and position velocity diagrams of narrow 
Br$\gamma$$\lambda$2.166~$\mu$m and H$_2$ 1-0S(1)$\lambda$2.122~$\mu$m
are presented and discussed.

\subsection{Emission line profiles, channel maps, and position velocity diagrams}

\begin{figure}
\centering
\includegraphics[width=8.25cm]{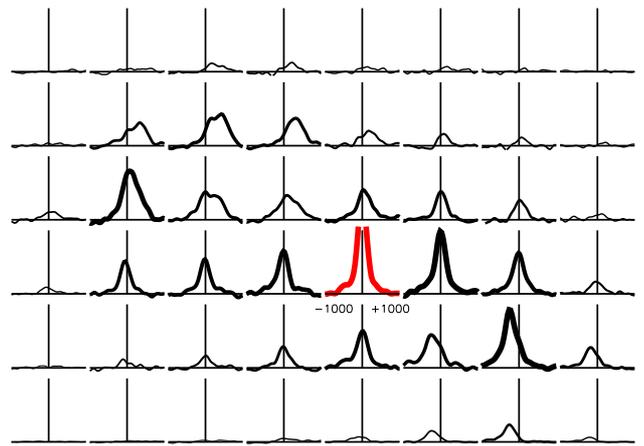}
\caption{[FeII] emission line profiles extracted from apertures of 385 $\times$
385~milli-arcseconds. [FeII] emission from the nucleus in red. 
North is up, east to the left. The x-axis represents -1000~km/s to +1000~km/s.
The thickness of the emission line profiles scales with flux to guide the eye.}
\label{nlr}
\end{figure}

The emission line profiles of [FeII] in our FoV generally deviate from a single-Gaussian profile (figure \ref{nlr}).
At distances larger than approximately 0.5~arcseconds from the nucleus
and along the NLR at a position angle of 60$\degr$,
line profiles with a component at the systemic velocity of the
galaxy and another component that shows up as a blueshifted or redshifted hump to
the line profile can be observed. Within our FoV we detect emission 
at velocities of up to +500~km/s in the northeast and up to -250~km/s in the
southwest. At and around the nucleus the line profile appears to be symmetrical, 
while at other positions the line profile is composed of two (or more) 
components. 

\begin{figure}
\centering
\includegraphics[width=8.5cm]{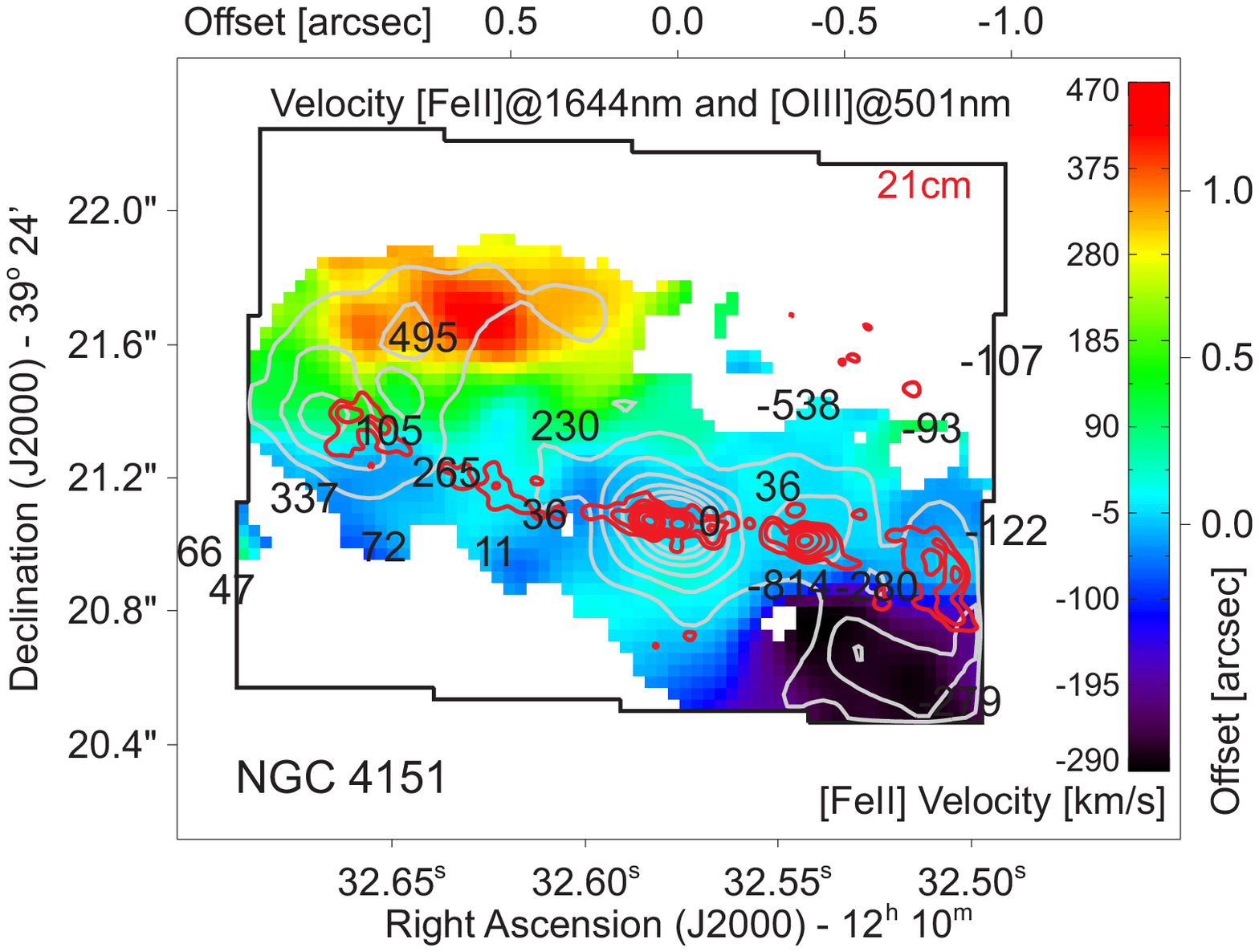}
\includegraphics[width=8.5cm]{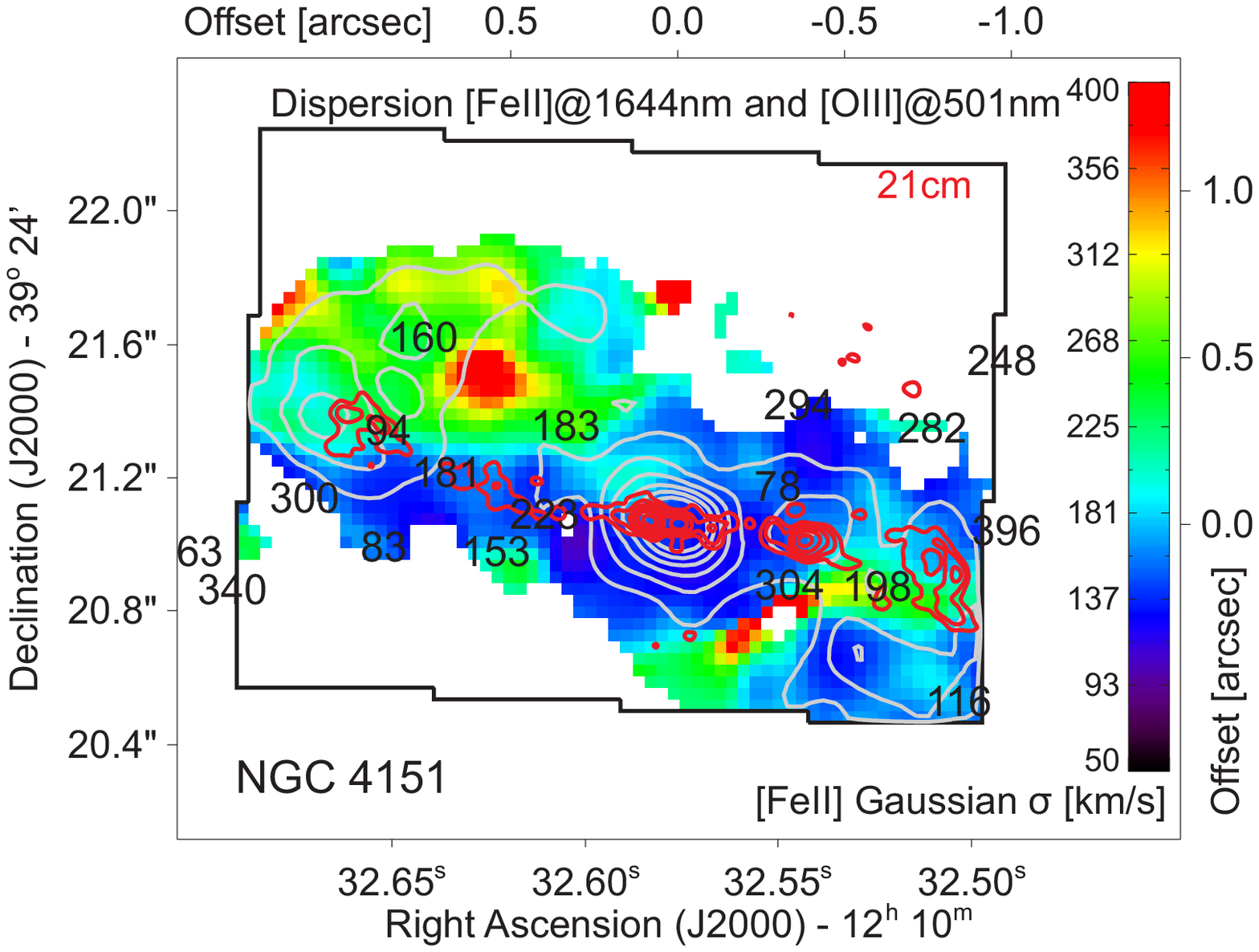}
\caption{[FeII]$\lambda$1.644~$\mu$m velocity and dispersion field derived from
  single-Gaussian fits. The black box indicates our FoV. North is up and east
  to the left. Gray contours represent [FeII] flux isophotes to guide the
  eye.\newline 
\textbf{Top: }[FeII]$\lambda$1.644~$\mu$m velocity field, 21 cm radio 
  contours in red (.5, 1, 2 and 4 mJy beam$^{-1}$ with a
  beam size about four times smaller than our angular sampling) taken from
  \citet{2003ApJ...583..192M}, and [FeII]$\lambda$1.644~$\mu$m contours in
  gray. Numbers in the plot indicate velocities of prominent [OIII] clouds
  \citep[taken from][]{2000ApJ...528..260K}.\newline  
\textbf{Bottom: }[FeII]$\lambda$1.644~$\mu$m dispersion field, 21 cm radio 
  contours in red (.5, 1, 2 and 4 mJy beam$^{-1}$ with a
  beam size app. four times smaller than our angular sampling) taken from
  \citet{2003ApJ...583..192M}, and [FeII]$\lambda$1.644~$\mu$m contours in
  gray. Numbers in the plot indicate dispersions of prominent [OIII] clouds
  \citep[taken from][]{2000ApJ...528..260K}.}
\label{fevels}
\end{figure}


\noindent The velocity and Gaussian dispersion of [FeII] derived from single-Gaussian fits are shown in figure \ref{fevels}.  
The velocities are generally redshifted to the northeast and 
blueshifted to the southwest with respect to the systemic velocity. 
At the nucleus the emission generally emerges with systemic velocity.
At a distance of 1~arcsecond to the nucleus velocity offsets of about 400~km/s (see
figure \ref{fevels}, middle) are observed in the direction of the NLR at a
position angle of 60$\degr$.
Because the most extreme velocities observed in [FeII] exceed those
predicted by simple rotation by a factor of 10
\footnote{assuming that the dynamics is dominated by a massive point-like 
object at the position of the nucleus with a mass 
of $5 \times 10^7$ M$_{\odot}$ , see e.g. \citet{2007ApJ...670..105O}, the
rotational velocity for an edge-on disk at a projected radial distance of 70
pc is only 56~km/s instead of the observed 500~km/s}, this kinematical
signature is attributed to an outflow. 
The derived [FeII] velocities and dispersions together with [OIII]$\lambda$501nm velocities
and dispersion from selected clouds taken from \citet{2000ApJ...528..260K} are
shown in figure \ref{fevels}.
Velocities and dispersions agree well except that the [OIII]
velocities exceed the [FeII] velocities at some very few positions and at the edge
of the NLR by several hundred km/s. These high-velocity [OIII]
clouds have been subject to speculations, but because generally the morphology
and dynamics of [FeII] and [OIII] are comparable, the excitation mechanism for
the two species is the same (photoionization by the central AGN).
Generally, the jet may drive this outflow, and 
the 21 cm radio continuum jet \citep{2003ApJ...583..192M} is additionally
shown in figure \ref{fevels}. The jet is clearly misaligned with the
direction of the most extreme 
velocities in our velocity fields such that the jet may not be the dominant
source for the kinematical features observed. Moreover, no increase in
dispersion caused by higher turbulance where the jet collides with the ISM is
observed (see figure \ref{fevels}, bottom). Instead, the
dispersion derived from single-Gaussian fits is increased at locations where
the [FeII] velocity field shows strong gradients, indicating that the line
profile becomes asymmetrical.\\

\begin{figure*}
\centering
\includegraphics[bb=170 54 558 738,angle=90,width=\textwidth]{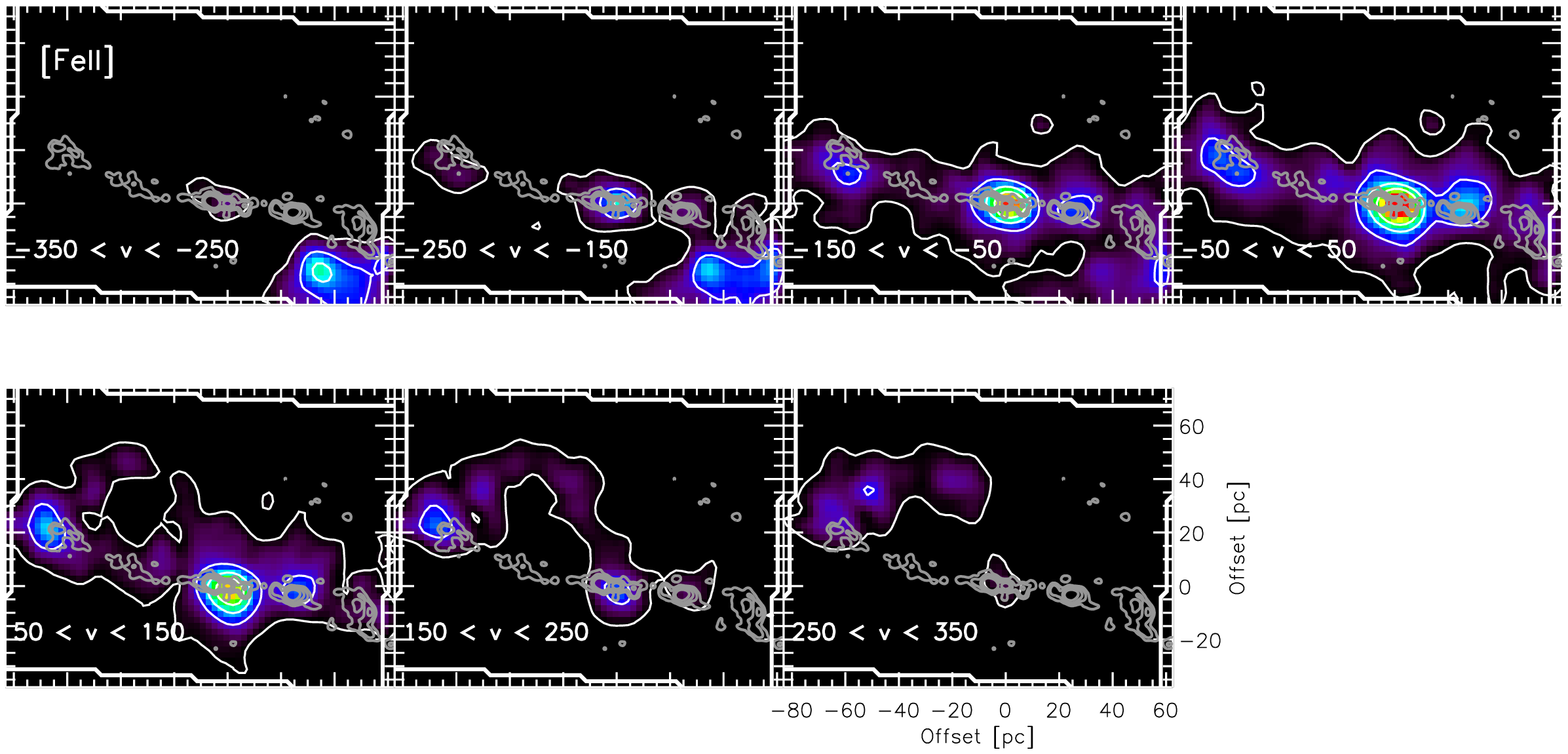}
\includegraphics[bb=125 54 558 738,angle=90,width=\textwidth]{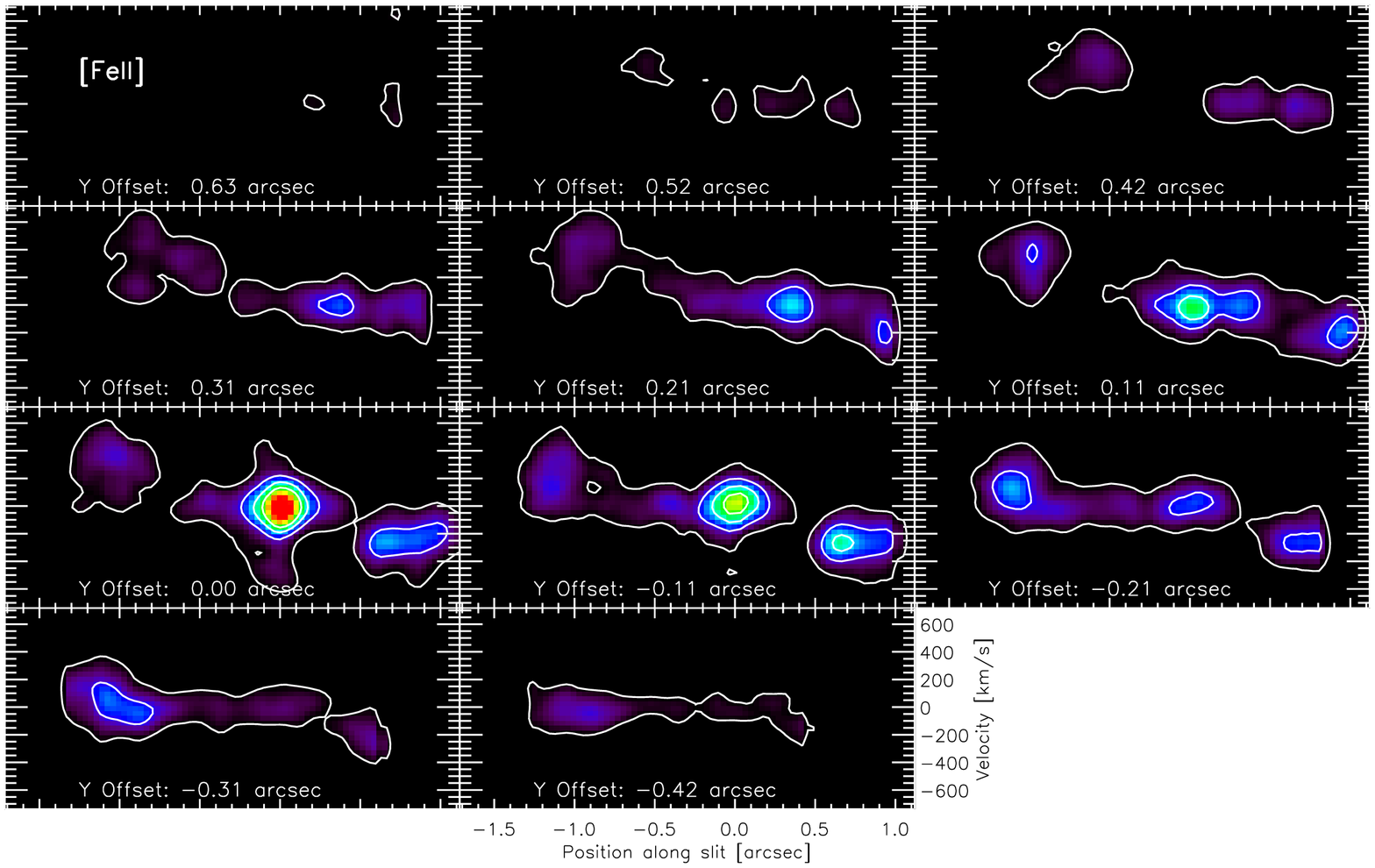}
\caption{
\textbf{Top: }[FeII]$\lambda$1.644~$\mu$m channel maps. Radio contours
overplotted in gray 
  (.5, 1, 2 and 4 mJy beam$^{-1}$ with a
  beamsize about four times smaller than our angular sampling). The white contours are 10\%, 25\%,
  40\%, and 55\% of the [FeII] peak intensity. Velocity ranges are given in
km/s. North is up and east to the left.\newline
\textbf{Bottom: }Position velocity diagrams of [FeII]$\lambda$1.644~$\mu$m
  extracted from pseudoslits oriented along the NLR with a 
PA of 60$\degr$. The Y Offset is the angular offset
of the pseudoslit with respect to the nucleus. Along the pseudoslit, 
position 0 indicates
the position of the continuum peak. Contours represent 10\%, 25\%, 
40\%, and 55\% of the [FeII] peak intensity.}  
\label{femaps}
\end{figure*}

\noindent To better illustrate the dynamics in [FeII] and its relation to the
jet, channel maps and position velocity diagrams of [FeII] are shown in figure \ref{femaps}. 
The channel maps are separated by 100~km/s, the white contours
represent 10\%, 25\%, 40\%, and 55\% of the [FeII] peak intensity to guide the
eye, and the nucleus is located at position 0,0. 
The channel maps centered on -300~km/s and +300~km/s show an outflow from the
nucleus to the southeast directed toward the observer and to the northwest directed
away from the observer with prominent emission at systemic velocity. It is possible that
the jet locally enhances emission in [FeII], and the 21 cm
continuum radio contours are also shown in our channel maps. At about systemic velocity, [FeII] and radio 
emission correlate at positions F1 and F2. In channel maps centered on high-velocity offsets (e.g. -300~km/s), the correlation is rather weak such that the
high-velocity gas seems to be unaffected by the jet.
If emission at systemic velocity is enhanced by the 
jet, the inclination of the jet must be comparable to the inclination
of the galactic disk.\\
To determine how the velocity of redshifted and blueshifted emission
components varies with distance from the nucleus, we assumed that the outflow is
directed toward the brightest [FeII] emitting knots at a PA of 
60$\degr$ and constructed the position velocity diagrams shown in
figure \ref{femaps}. These are extracted from pseudoslits with a width of
0.105~arcseconds aligned at a PA of 60$\degr$ and are
offset by multiples of the pseudoslitwidth. For all pseudoslit offsets 
prominent emission at systemic velocity, i.e., emerging from the galactic disk,
is observed. Emission at other velocities along the pseudoslit seems to
increase with distance to the nucleus along the pseudoslit. For the
pseudoslit centered on the nucleus (Y offset of 0~arcseconds), for instance,
we observe
emission at +400~km/s at approximately -1.25~arcseconds from the nucleus along
the pseudoslit and at -250~km/s at a distance of +0.7~arcseconds. 
These numbers indicate that the velocities observed may scale linearly
with distance to the nucleus, but we investigate this below in more
detail.\\

\subsection{Multi-Gaussian fits to the [FeII] emission line profile}

\noindent 
To separate the dynamics of the disk from that of the NLR, we followed
\citet{2005AJ....130..945D} and \citet{2010MNRAS.402..819S} and decomposed  
the emission line profiles in the NLR into multiple Gaussians.

\begin{figure*}
\centering
\includegraphics[angle=90,width=18cm]{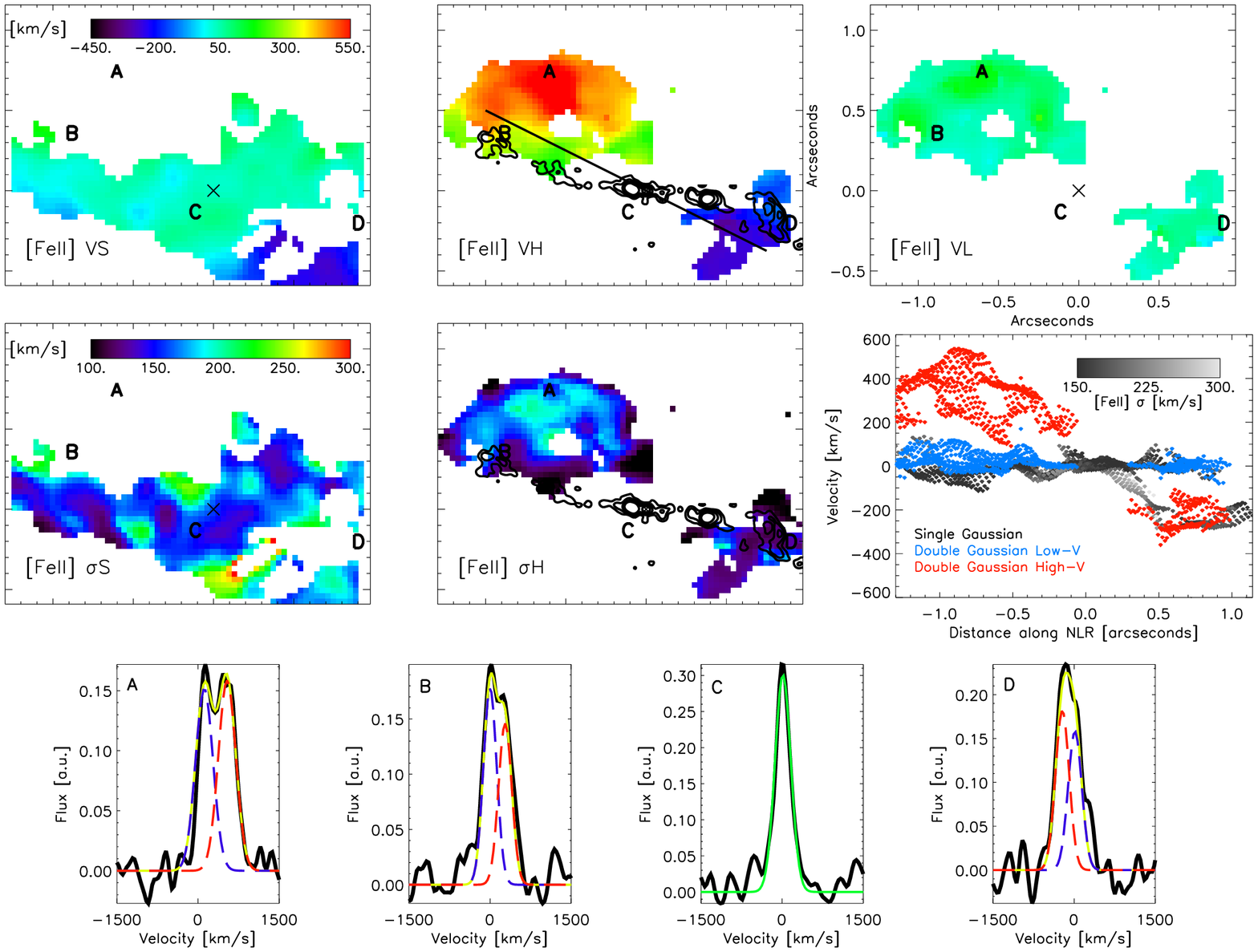}
\caption{
[FeII] dynamics as derived from single- and double-Gaussian fits. 
All slices with constant
wavelength of the datacube were convolved with a Gaussian seeing disk
with a FWHM of 0.07~arcseconds. The cross denotes the position
of the continuum peak. Mayor tick marks in the velocity and dispersion fields
correspond to 0.50~arcseconds. Letters A to D mark positions from which the
spectra shown in the last row were extracted.
The black line indicates the position angle of the bicone of 60$\degr$.
\newline
\textbf{Top: }Centroid velocities from single-Gaussian fits, VS (left), high/low
velocity component from double-Gaussian fits, VH/VL (middle/right). The
contours in the middle image in this row represent the 21 cm radio continuum
contours \citep{2003ApJ...583..192M} at
levels of .5, 1, 2, and 4 mJy beam$^{-1}$ with a
  beam size about four times smaller than our angular sampling.\newline
\textbf{Middle left and middle: }Gausian dispersions of the single- and double-Gaussian fits ($\sigma$S and $\sigma$H).\newline
\textbf{Middle right: }Centroid velocity of our fits as a function of
distance from the continuum peak along the NLR at a position angle of
60$\degr$. Black/red/blue points represent
the centroid velocities derived from single-Gaussian fits and high- and
low-velocity components in case of double-Gaussian fits.\newline
\textbf{Bottom: }[FeII] line profiles extracted from various regions with
Gaussian fits overplotted. Measured line profile in black, single-Gaussian
fits in green, combined double-Gaussian fits in yellow, the low/high velocity
component of the double-Gaussian fit in red/blue. }
\label{fedissect}
\end{figure*}

\noindent The spectral resolution of our data allows a decomposition into two
Gaussian components, although \citet{2005AJ....130..945D} reported the need for
more than three components on selected locations but using spectra with
R$\simeq$9000. To additionally increase the S/N in our data, each
slice of the datacube with constant wavelength was convolved with a
Gaussian seeing disk with a FWHM of 0.14~arcseconds.
The underlying continuum was fitted locally with a parabola in each
spectrum and was subtracted before fitting the emission line profile. 
To facilitate the comparison between this and previous studies, the fit was
constrained to a double-Gaussian fit where both Gaussians 
have an identical FWHM where applicable. \\
\noindent The results of the fits to the [FeII] line profiles are shown in
figure \ref{fedissect}. To illustrate the quality of the fits, emission
line profiles and the corresponding fits from selected positions (A-D) in our
FoV are also shown. 
In the case of single-Gaussian fits the centroid velocities VS are usually
close to systemic with the exception of the very 
southwest, close to region F3 (figure \ref{fedissect}, top left).
The single-Gaussian dispersion $\sigma$S varies between 150 and 200~km/s, but
is increased in regions where the line profile becomes asymmetrical
(figure \ref{fedissect}, middle row, left). 
In the case of a double-Gaussian fit, the centroid velocities 
closer to systemic VL (figure \ref{fedissect}, top row, right) 
are usually close to systemic too, indicating that emission from the
outflowing [FeII] is usually blended with emission from the galactic
disk. However, at position A, which corresponds to the western end of region F1,
VL deviates from systemic (see the corresponding 
line profile in figure \ref{fedissect}, where the
component closer to systemic is clearly redshifted). 
This component cannot be attributed to the jet or the disk but may originate 
from the front side of the eastern bicone.
The centroid velocities farther from
systemic VH (figure \ref{fedissect}, top middle) range from nearly
600~km/s at position A to -300~km/s at position D (which corresponds to region
F3). The derived velocities are typical for emission
from the back-side of the eastern bicone especially at position A.
It is notheworthy that at position A the dispersion derived from double-Gaussian fits is increased (figure \ref{fedissect}, middle middle), 
either due to intrinsically broadened emission or additional 
unresolved components from the disk. 
At the brightest [FeII] emission knot in region F1 the radio jet 
is prominent. Although VH is close to systemic there, 
it might represent gas that collides with the jet.  
To investigate whether the gas is accelerated or simply ejected from the
nucleus, we show in figure \ref{fedissect} (middle right) the derived centroid velocities as a function of distance from
the nucleus along the direction of the NLR at a PA of 60$\degr$. From this figure
acceleration in the NLR cannot be ruled out because the centroid velocities seem
to increase with distance at least in the inner 0.5~arcsecond. Beyond that
radius the velocities seem to increase no more at velocities around 400~km/s
in the northeast and -250~km/s in the southwest. However, to the southwest
the dispersion of [FeII] line profiles fitted with a single Gaussian is
increased, which implies that the two profile components remain unresolved
such that we actually do not see acceleration to the southwest.\\

\subsection{Comparison with existing NLR models}
\begin{figure}
\centering
\includegraphics[width=8.5cm]{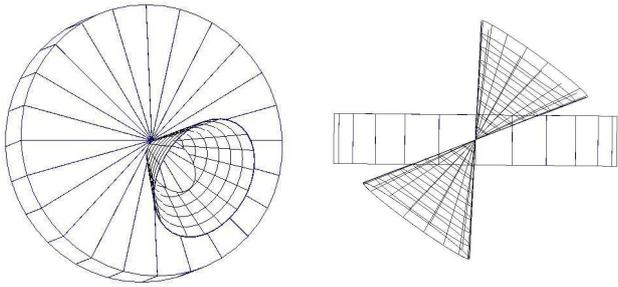}
\caption{
Orientation of the galactic plane and the NLR bicone as they
appear on the sky and viewed along the galactic plane of NGC~4151 (according to
  the model by \citet{2005AJ....130..945D}). Only the outer bicone with an
  opening angle of 33$\degr$ is shown.\newline}
\label{cones}
\end{figure}

\begin{figure*}
\centering
\includegraphics[width=15cm]{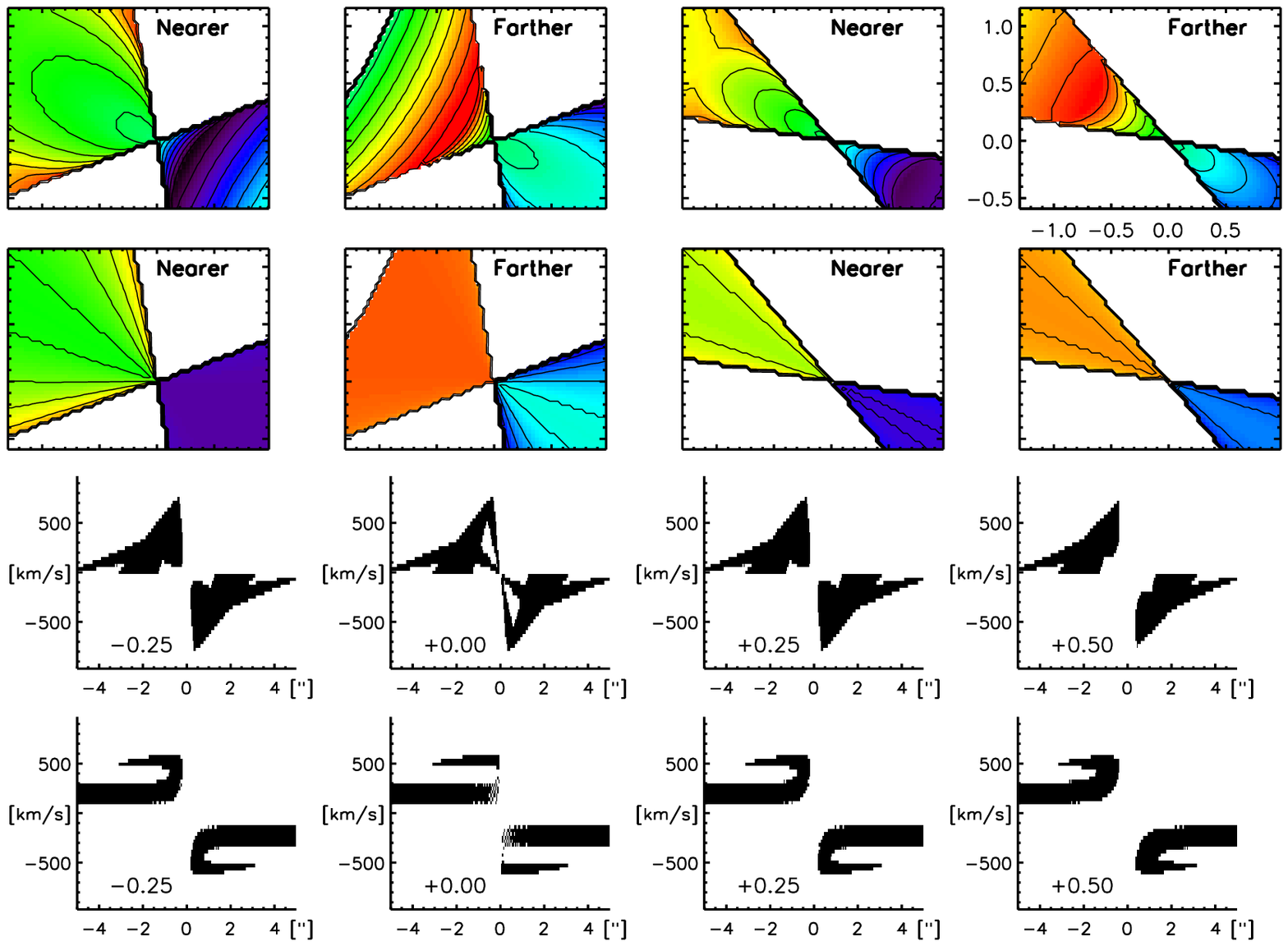}
\caption{Extreme velocities from the bicone surfaces and position velocity
diagrams extracted from the models by \citet{2005AJ....130..945D}
and \citet{2010MNRAS.402..819S}.\newline 
\textbf{Row 1: }Velocities on the bicone surfaces that are nearer and
farther to the observer according to the \citet{2005AJ....130..945D} model. 
Color scales from -700~km/s (blue) to 700~km/s (red), contours are in steps of 100~km/s.
Major tick marks correspond to 0.5~arcseconds. The surfaces shown cover our
FoV. North is up, east to the left.
\newline
\textbf{Row 2: }Same as row 1 for the constant-velocity model
by \citet{2010MNRAS.402..819S}.\newline
\textbf{Row 3: }Position velocity diagrams extracted from pseudoslits aligned
along the NLR at a position angle of 60~$\degr$ according to
the \citet{2005AJ....130..945D} model.
The slitwidth/slitlength is 0.15/5~arcseconds and the pseudoslits are moved
-0.2, 0, 0.2, and 0.4~arcseconds away from the nucleus perpendicular to the
direction of the NLR. The black area indicates the region in the pv-diagram 
from which emission is expected according to the model.
Note that our observations only cover the inner arcsecond.\newline
\textbf{Row 4: }Same as row 3, but for the constant-velocity model
by \citet{2010MNRAS.402..819S}.}
\label{models}
\end{figure*}

The dynamics of the NLR of NGC~4151 has been the subject of  
various studies, e.g. \citet{2000AJ....120.1731C}
and \citet{2005AJ....130..945D} who used the STIS 
slit-spectrograph on-bord the Hubble Space Telescope, and 
\citet{2010MNRAS.402..819S} who used near-infrared imaging spectroscopy at the
GEMINI telescope.
The models by \citet{2000AJ....120.1731C} and more recently
by \citet{2005AJ....130..945D} state that [OIII] emission primarily arises from
a hollow bicone that is inclined at 
45$\degr$ to the plane of the sky at a PA of 60$\degr$. The
inner/outer opening angles are 15$\degr$ and 33$\degr$ and
emission is primarily seen to emerge from between the two bicones. The
southwestern part of the bicone points toward the observer, the northeastern
part points away from the observer. Because the inclination of the galactic disk is about
22$\degr$ (with a PA of the line-of-nodes of about
25$\degr$, \citet{1992MNRAS.259..369P}), one 
side of the bicone is more deeply embedded in the galactic disk than the
other (see figure \ref{cones}, top, for the orientation of the galactic disk 
and the bicone). 
The [OIII] data presented in \citet{2000AJ....120.1731C} and \citet{2005AJ....130..945D} indicates acceleration in a nuclear outflow within these two 
bicones although the origin of this mechanism remains unclear. These authors
proposed a 
velocity-distance law  in which the velocity v and the radial distance r to
the nucleus are proportional. Acceleration takes place up to a maximum
radial velocity of 800~km/s at a maximum radial distance of 96 pc. At this
distance the ejected material begins to interact with the surrounding
interstellar medium such that it decelerates to systemic velocity at a
distance of 400~pc. The predicted velocities on the surfaces of the
nearer and farther surface of the inner and outer bicone are presented in
figure \ref{models}. In this model, typical
velocities observed at that side of the bicone that is closer to the plane of
the sky are about 200-300~km/s, while velocities on the side pointing
away from the galactic disk are higher than 500~km/s.\\

\noindent The models presented by \citet{2010MNRAS.402..819S} are based on the
geometry of the model by \citet{2005AJ....130..945D}. The authors present
position velocity diagrams of [SIII]$\lambda$0.9533~$\mu$m extracted from 
pseudoslits extending four arcseconds to either side of the nucleus (see their
figure 10). These indicate that deviations from the linear
velocity-distance law do occur because the gas seems to move with nearly
constant velcocity 
at distances larger than approximately 0.75~arcsecond from the nucleus in all
pv-diagrams. This motivated the authors to modify the velocity-distance law to
a constant with a velocity of 600~km/s, such that clouds are simply ejected
from the nucleus and are not accelerated/decelerated in the NLR. 
Predicted velocities on the surfaces of the nearer and farther surface of the
inner and outer bicone are shown in figure \ref{models} (row 2) and
the straight, radial isovelocity contours clearly distinguish this model from
the acceleration model. But the main difference between the two models becomes
more obvious in pv-diagrams extracted from pseudoslits aligned
with the mayor axis of the modeled bicone (see figure \ref{models}, row 3 and
4). The acceleration model allows higher velocities within 0.5~arcseconds to
the nucleus because the maximum radial velocity in this model is higher than in
the constant-velocity model and occurs at the surfaces of the farther/nearer
side of the bicone that are closer to our line-of-sight. At larger 
distances to the nucleus the velocities from the farther and nearer side of
the bicone recede to systemic. In contrast, the constant-velocity model 
predicts constant velocities for all distances to the nucleus only depending on
the inclination of the cloud path outward relative to the line-of-sight.
\citet{2010MNRAS.402..819S} also presented pv-diagrams of [FeII] in which this
constant velocity law is less 
obvious. However, the morphology and dynamics based on position velocity
diagrams of our and their [FeII] observations agree very well, 
although our FoV only samples the inner 2.5~arcseconds. Since the
[SIII]$\lambda$0.9533~$\mu$m emission line is not within our wavelength range,
we used the [FeII] emission line instead to trace the dynamics. Furthermore,
to facilitate a comparison between our results and the
results obtained by \citet{2005AJ....130..945D} and \citet{2010MNRAS.402..819S},
we used exactly the same geometry and the same velocity-distance laws as
summarized in table \ref{laws}. 

\begin{table}
\caption{Geometry and velocity-distance laws. \citet{2005AJ....130..945D}:
Acceleration to V$_{max @ R1}$, the maximum radial velocity,
occurs at radius R1 and deceleration back to systemic velocity occurs at radius R2.
\citet{2010MNRAS.402..819S}: Constant ejection velocity VC without emission
beyond a radial distance of R.}
\label{laws}      
\centering          
\begin{tabular}{ c c }
\hline\hline       
\multicolumn{2}{c}{Geometry}\\
\hline
Position angle & 60 $\degr$\\
Inclination    & 45 $\degr$ (southwestern is closer)\\
Inner/outer opening angle & 15$\degr$ / 33$\degr$\\
\hline       
\multicolumn{2}{c}{Velocity law according to \citet{2005AJ....130..945D} }\\
\hline
V$_{max @ R1}$ & 800 km/s\\ 
R1/R2 & 96 pc / 400 pc\\
\hline       
\multicolumn{2}{c}{Constant velocity law according to \citet{2010MNRAS.402..819S} }\\
\hline
VC & 600 km/s\\ 
R & 400 pc\\
\hline
\hline                  
\end{tabular}
\end{table}

\begin{figure*}
\centering
\includegraphics[bb=135 54 558 738,angle=90,width=\textwidth]{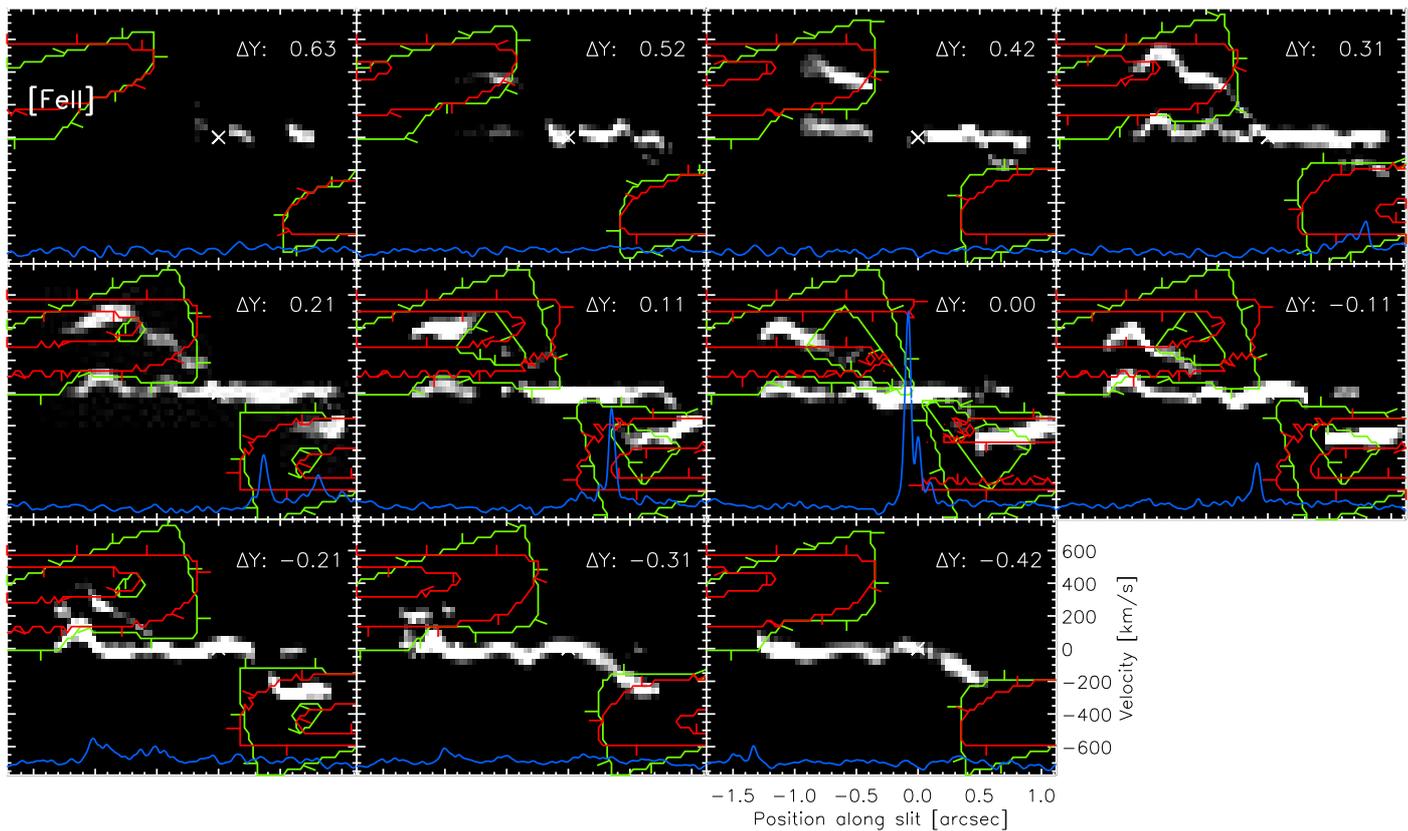}
\caption{
Position velocity diagrams of 0.105 arsecond wide pseudoslits aligned with the
NLR at a PA of 60$\degr$ extracted from the modeled datacube that 
contains emission from the single- and double-Gaussian fits to the emission line. 
$\Delta$Y denotes the separation of the pseudoslits
in arcseconds.  For $\Delta$Y=0 the pseudoslit covers the nucleus.
The green/red contours indicate the allowed regions according to the
velocity laws by \citet{2005AJ....130..945D}
and \citet{2010MNRAS.402..819S}, where tick marks point in the direction
of forbidden velocity ranges. The slightly staggered
arrangement of the model calculations is not real but is rather due to
calculations performed on a grid. The cross marks the center position of the
pseudoslit and coincides with the position of the nucleus. 21 cm radio
continuum emission along the pseudoslits is plotted in blue.}
\label{femodels}
\end{figure*}

\noindent Figure \ref{femodels} shows all derived centroid velocities in terms
of position velocity diagrams. The pseudoslits are aligned with the 
modeled bicone and have a width of 0.105 arseconds. 
The green and red contours in this figure are the allowed velocity ranges 
according to the models by \citet{2005AJ....130..945D}
and \citet{2010MNRAS.402..819S}. 
On average, both models cover the
same 80\% of the allowed velocity ranges for all pseudoslit offsets. The main
difference between the models can be observed for pseudoslit offsets around
0.2~arcseconds around the nucleus. Here, emission at $\pm$400~km/s for
distances larger than approximately 0.5~arcseconds (depending on the
pseudoslit offset) is not expected according to the constant-velocity model,
while it is allowed in the acceleration scenario. 
For positive pseudoslit offsets we actually do observe emission around
400~km/s at distances along the pseudoslit of -1.0~arcsecond (corresponding to 
position A and B in figure \ref{fedissect}), while on the opposite side of the
pseudoslit and for comparable negative pseudoslit offsets we only observe
velocities around -250~km/s (corresponding to position D).\\

\noindent Taking only centroid velocities from the high-velocity component of
the double-Gaussian fit and from single-Gaussian fits with velocity offsets greater than
150 km/s into account, the acceleration model by \citet{2005AJ....130..945D} 
complies with 79\% of all centroid velocities measured in [FeII], while the 
constant-velocity model by \citet{2010MNRAS.402..819S} only complies with 
66\% of all measured centroid velocities. Applying flux values from 
the individual profile components derived from our fits as weights results in
a match of 80\% for the \citet{2005AJ....130..945D} model and 74\% for 
the \citet{2010MNRAS.402..819S} model. The velocity signatures of the high-velocity component for all pseudoslit offsets at negative
positions along the pseudoslit rather indicate acceleration along the slit up
to a position of approximately -1 arcsecond. Beyond that point, the gas
clearly seems to decelerate. Considering the latter and the numbers presented
above, we assume that the gas is accelerated within the inner 1 arcsecond.\\

\noindent We also checked if the constant-velocity
model can be modified that it fits our result better. Changing the
ejection velocity only stretches the regions of allowed
velocity ranges in the position velocity diagrams such that emission from 
either positions A and B or D do not match the model. Changing the systemic
velocity (redshift) would shift the allowed velocity ranges in the constant-velocity model symmetrically up and down, but this would leave us with
the same problems.\\

\noindent The jet may also contribute to the observed low-velocity 
kinematics of the gas through interactions of the jet with gas in the galactic
disk. The blue profiles in figure \ref{femodels}
indicate the 21 cm radio intensity within the extracted pseudoslits.
The high-velocity gas does not correlate with the radio intensity 
in the northeast (at negative distances along the pseudoslits) for all
pseudoslit offsets but in the southwest (at positive distances along the
pseudoslits) for pseudoslit offsets of +0.31 to +0.11 arcseconds at distances
around +0.75 arcseconds. The low-velocity gas seems unaffected except for a
pseudoslit offset of -0.21 arcseconds at a position of around -1.0 arcseconds.
This position coincides with the radio knot R2E in figure \ref{emissionfe}.
Thus, the jet at least partially enhances emission in [FeII].  
The high-velocity components for pseudoslit offsets of +0.21 to 0.00 
arcseconds at negative pseudoslit positions do not correlate with radio
emission. These components are not predicted by the constant-velocity model
and cannot be explained in terms of jet enhancements. However, in the
acceleration model these components can be explained in terms of gas
decelerating because of collisions with the ISM.\\

\noindent In total, the acceleration scenario seems to fit the measured [FeII]
kinematics in the inner 1~arcsecond better than the constant-velocity scenario.


\begin{figure*}
\centering
\includegraphics[bb=170 54 558 738,angle=90,width=\textwidth]{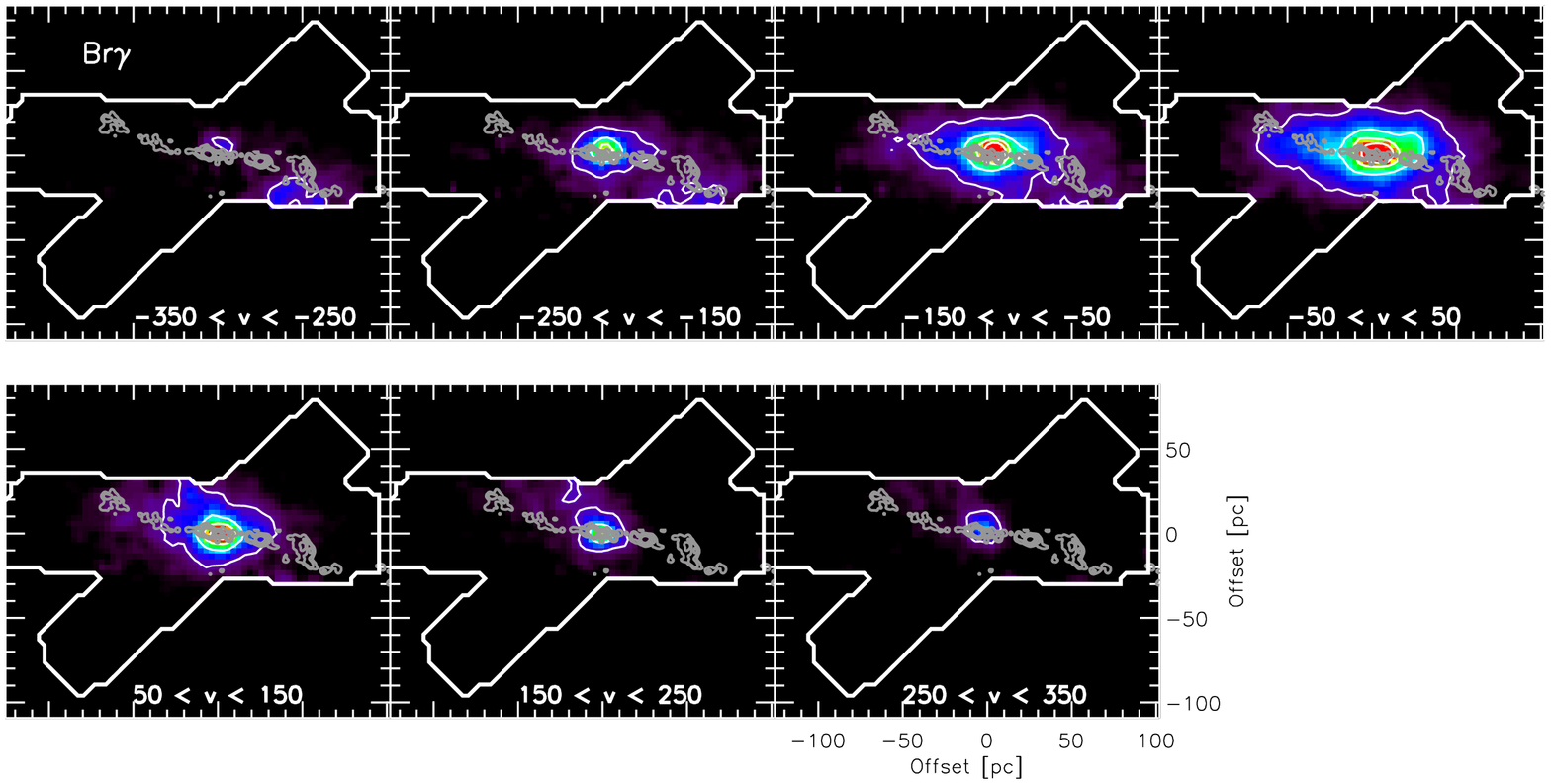}
\includegraphics[bb=125 54 558 738,angle=90,width=\textwidth]{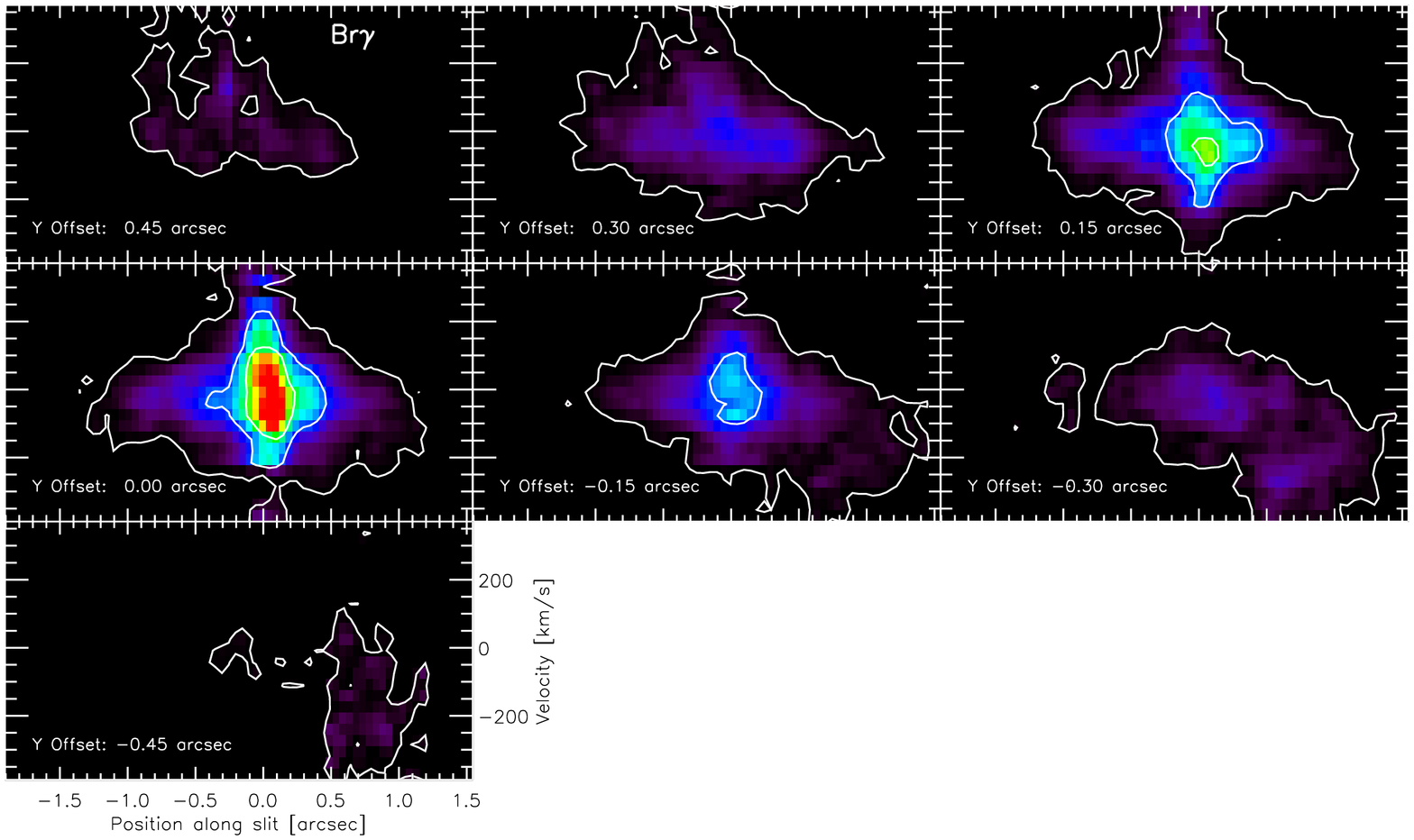}
\caption{Br$\gamma$ channel maps and position velocity diagrams. Prior to
extraction all slices with constant wavelength of the datacube were
convolved with a Gaussian seeing disk with a FWHM of 0.10~arcseconds.\newline
\textbf{Top: }Channel maps. 21 cm radio contours in gray 
  (.5, 1, 2, and 4 mJy beam$^{-1}$ with a beam size about four times smaller
  than our angular sampling) and the white contours are 10\%, 25\%,
  40\%, and 55\% of the Br$\gamma$ peak intensity. Velocity ranges are given in
km/s. North is up and east to the left.\newline
\textbf{Bottom: }Position velocity diagrams extracted from pseudoslits
  oriented from east to west. The Y offset is the angular offset
of the pseudoslit with respect to the nucleus. Along the pseudoslit, 
position 0 indicates the position of the continuum peak. Contours represent
10\%, 25\%, 40\%, and 55\% of the Br$\gamma$ peak intensity.}
\label{brgmaps}
\end{figure*}

\begin{figure*}
\centering
\includegraphics[bb=170 54 558 738,angle=90, width=\textwidth]{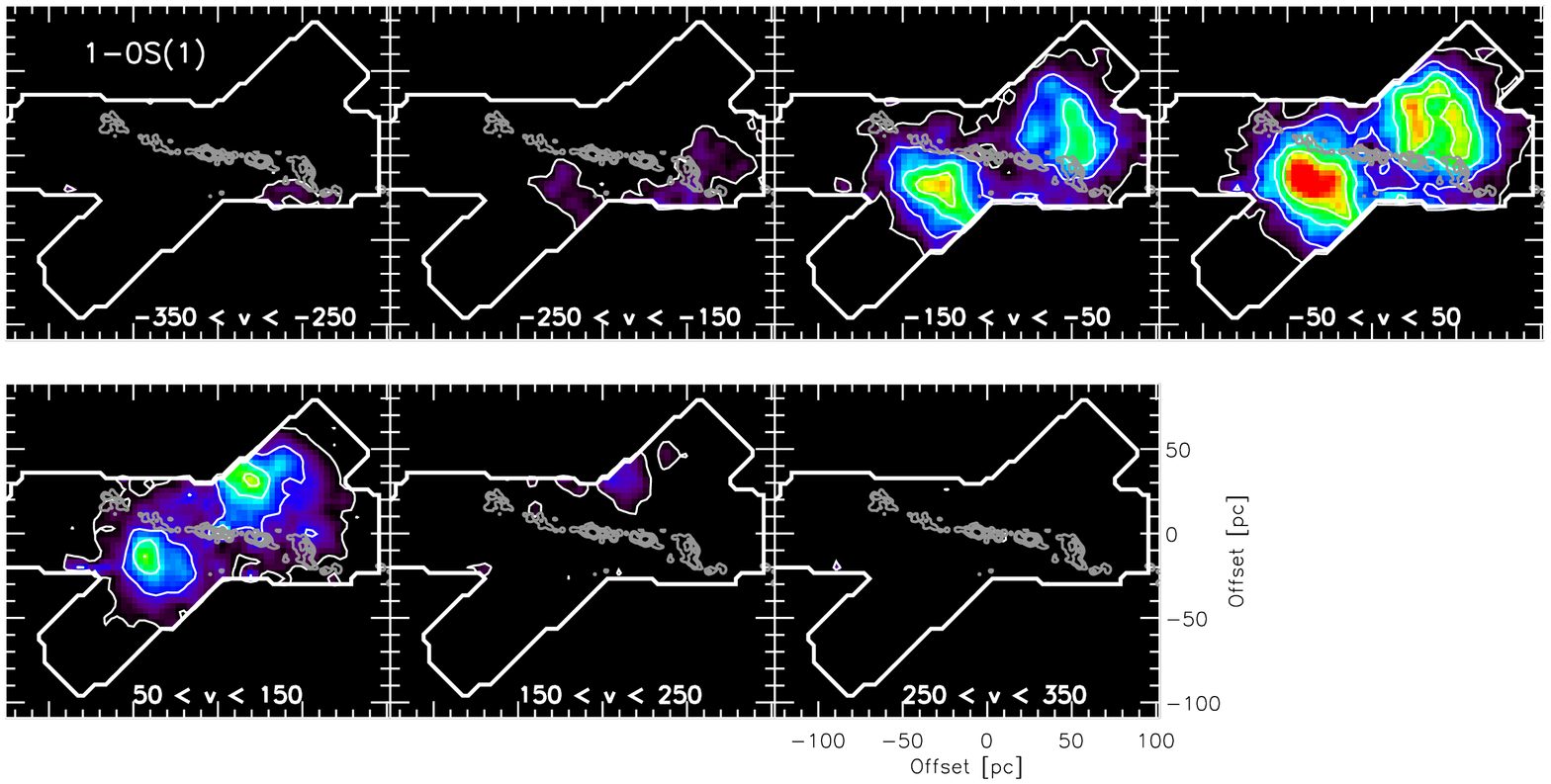}
\includegraphics[bb=125 54 558 738,angle=90,width=\textwidth]{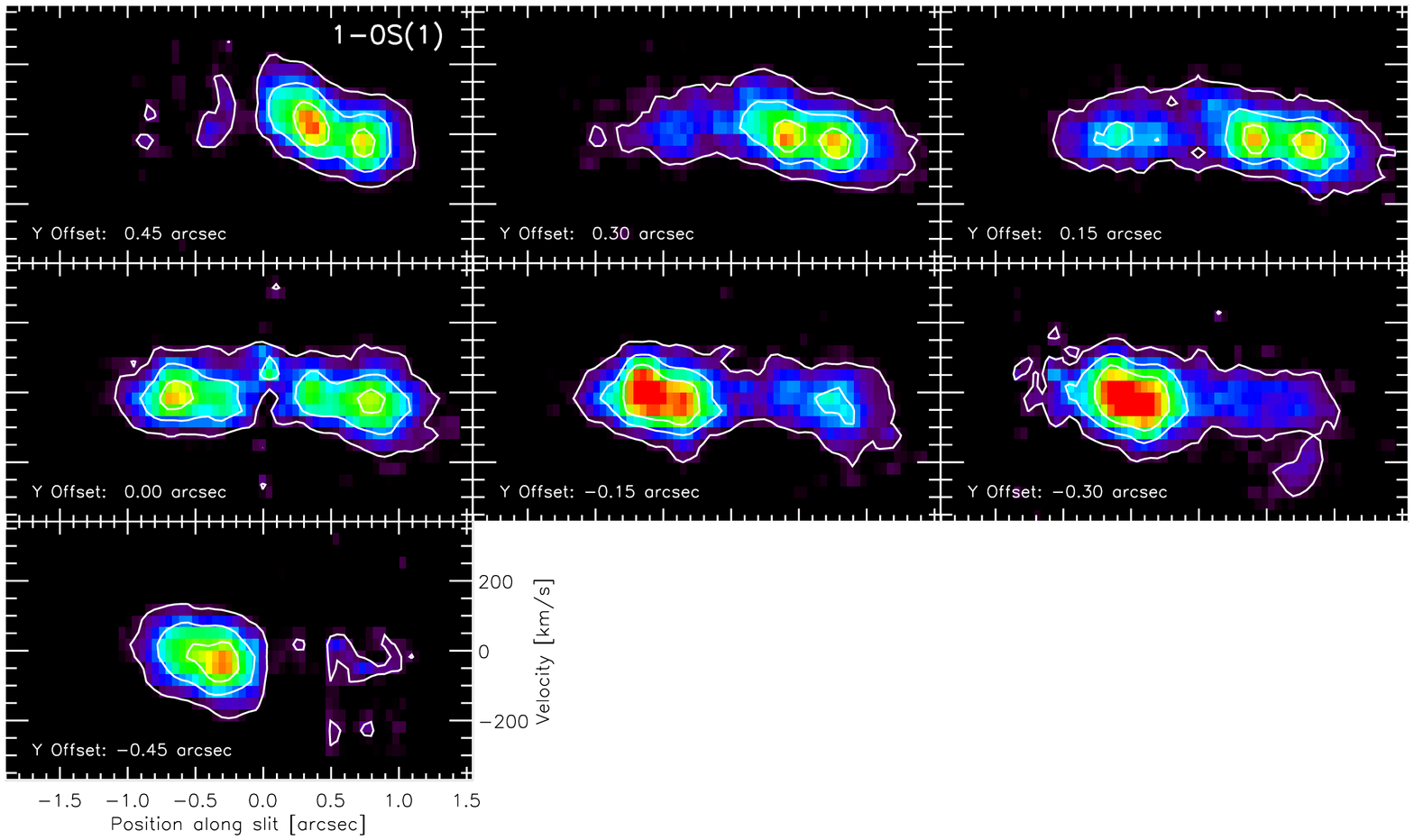}
\caption{Same as figure \ref{brgmaps}, but for 1-0S(1).}
\label{h2maps}
\end{figure*}

\subsection{Dynamics in Br$\gamma$ and H$_2$}
\begin{figure}
\centering
\includegraphics[width=7cm]{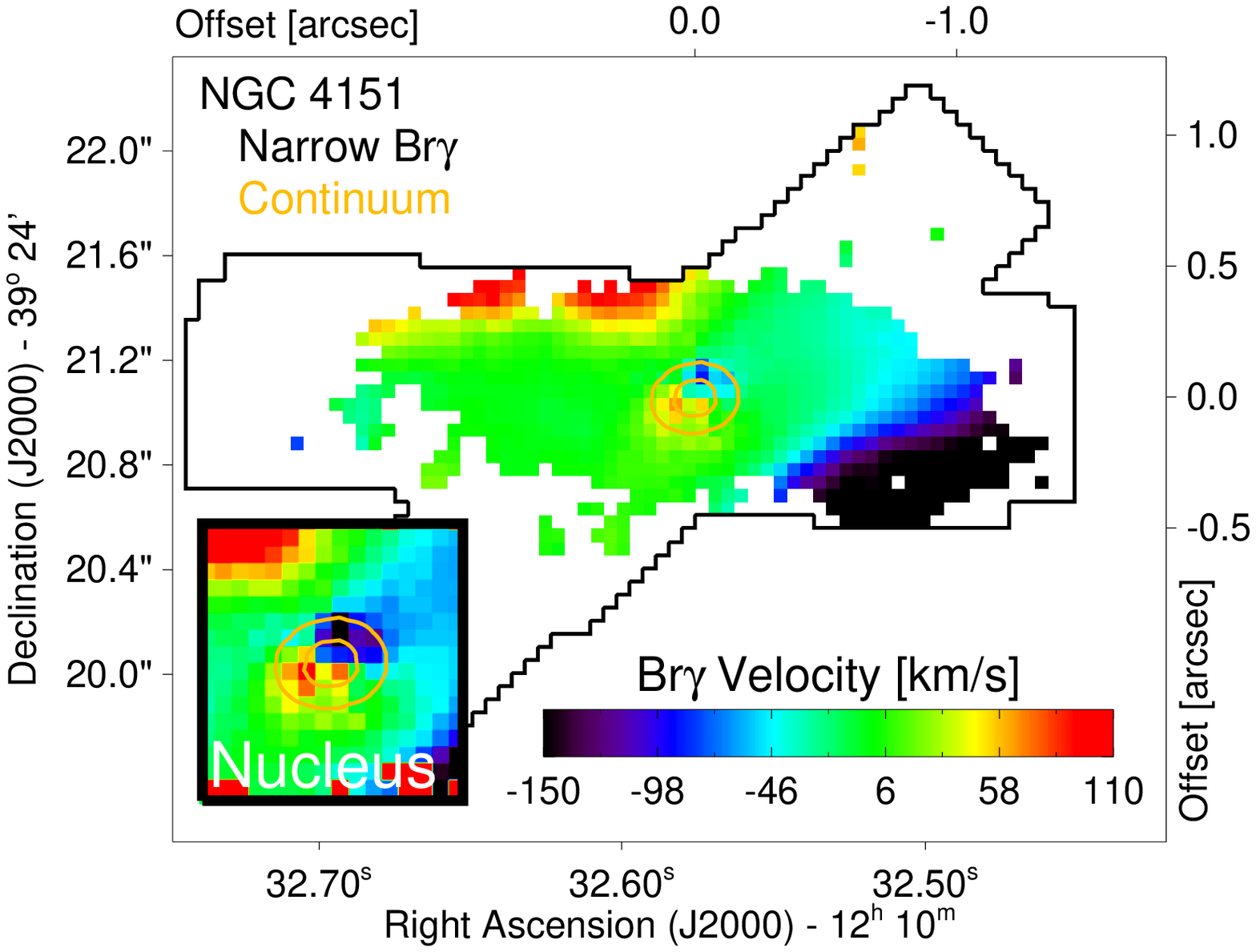}
\caption{Br$\gamma$ velocity field derived from single-Gaussian fits to
the line profile extracted from every spaxel with
continuum contours (20\% and 50\% of the peak value) in yellow.   
The inset shows the Br$\gamma$ velocity field on the nucleus scaled
in velocity from -75~km/s (blue) to 75~km/s (red).}
\label{brgvels}
\end{figure}

\begin{figure*}
\centering
\includegraphics[angle=90,width=15cm]{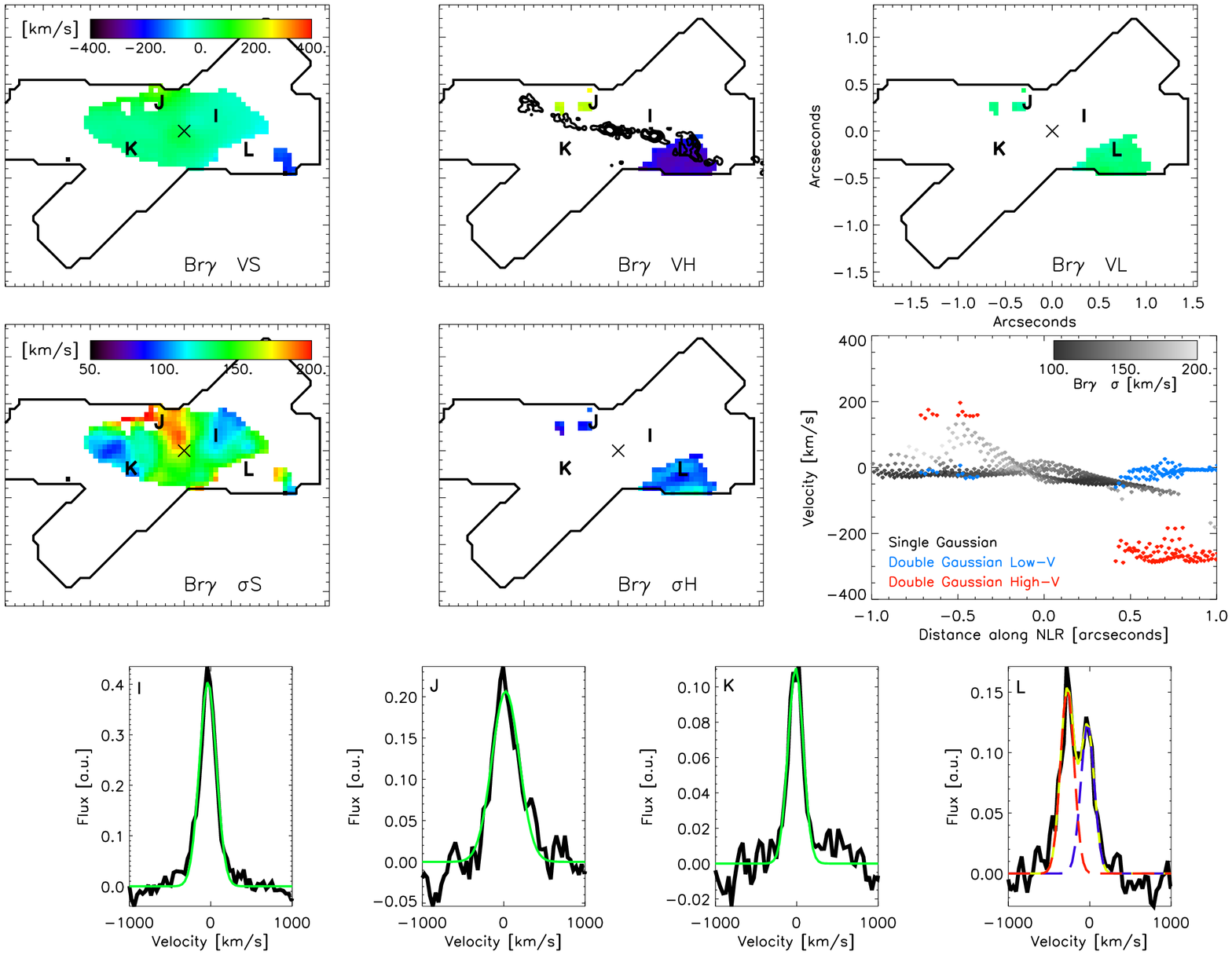}
\caption{Same as figure \ref{fedissect}, but for Br$\gamma$. All slices with
constant 
wavelength of the datacube were convolved with a Gaussian seeing disk
with a FWHM of 0.2~arcseconds.}
\label{brgdissect}
\end{figure*}

\begin{figure*}
\centering
\includegraphics[angle=90,width=15cm]{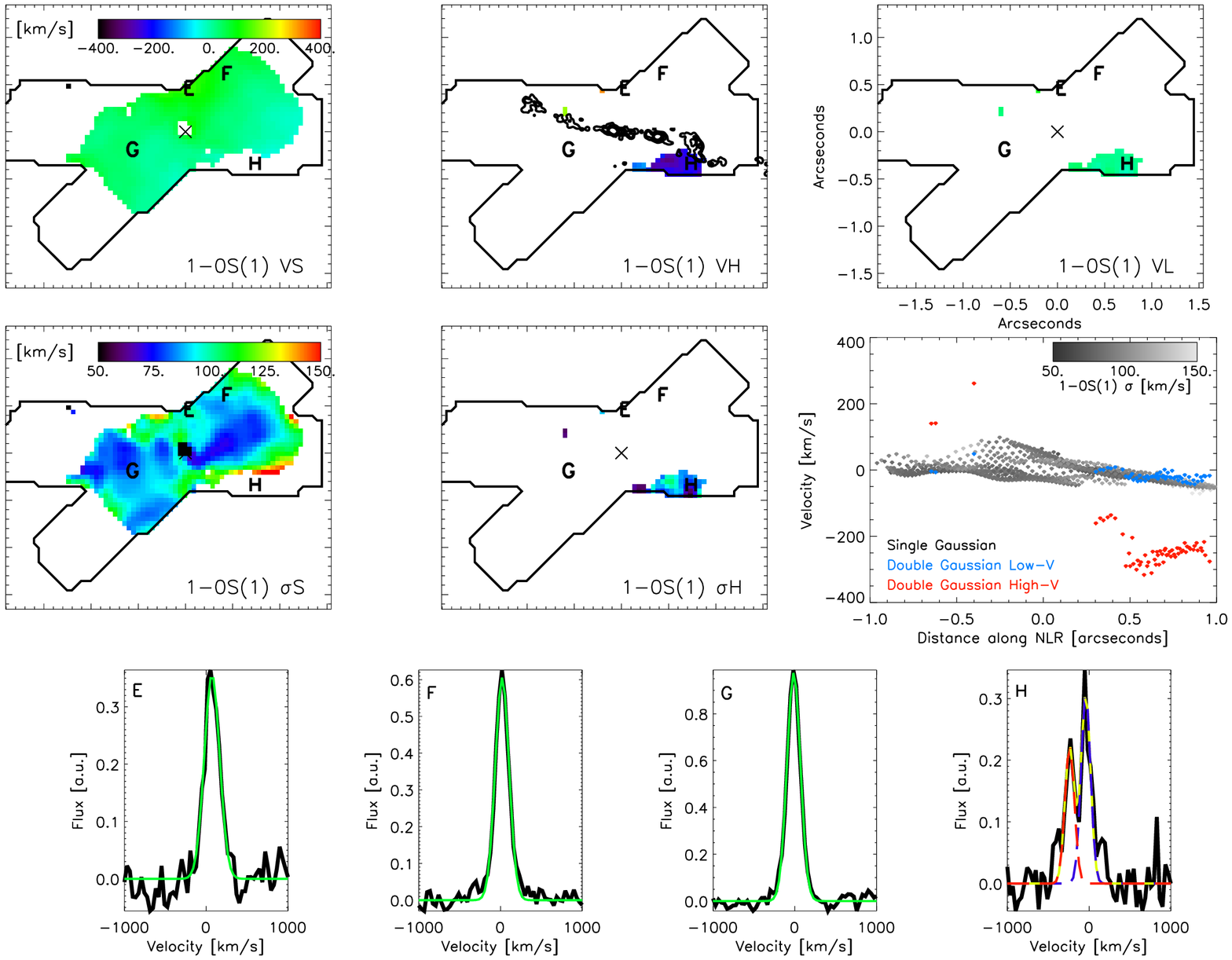}
\caption{Same as figure \ref{fedissect}, but for 1-0S(1). All slices with
constant wavelength of the datacube were convolved with a Gaussian
seeing disk with a FWHM of 0.15~arcseconds.}
\label{h2dissect}
\end{figure*}

\noindent Channel maps and pv-diagrams of Br$\gamma$ and H$_2$ are presented in
figures \ref{brgmaps} and \ref{h2maps}.
In contrast to [FeII], the Br$\gamma$ channel maps appear to be less clumpy
and emission primarily emerges with systemic velcocity extending
toward a PA of approximately 80$\degr$. A low-velocity component at
about -250~km/s can be observed  
that spatially and dynamically coincides with the southwestern [FeII] knot 60
pc to the west, but the correlation between the [FeII] and Br$\gamma$ channel
maps, especially to the northeast of the nucleus, is low. 
Instead, emission emerging with about +200~km/s nearly north to the nucleus at a PA of
20$\degr$ can be observed. This region is still located within the NLR and may be attributed
to the outflow. It is noteworthy that the Br$\gamma$ channel maps presented
by \citet{2010MNRAS.402..819S} show the very same behavior.
Farther in, at the position of nucleus, the slightly curved isophotes in the
pv-diagram (for a Y offset of 0~arcseconds) may indicate an inflow.
This inflow can also be observed in the inset of figure \ref{brgvels},
where the Br$\gamma$ velocity field as derived from single-Gaussian
fits is shown. The 
inflow can be traced to either side of the nucleus up to distances of about 20
parsecs at a PA of -45$\degr$, which is nearly perpendicular to the outflow
observed in [FeII] and deviates only about $\sim$20$\degr$ from the PA of the
mayor H$_2$ emitting regions. The orientation of the galactic disk (see
figure \ref{cones}) is well compatible with an inflow within the galactic disk
maybe from the larger $H_2$ reservoirs seen in figure \ref{h2maps}.
However, the symmetry of the velocity signature may also imply
a rotating but dynamically decoupled disk around the black hole. Considering
the angular resolutions provided by our observations, it is hardly possible
to distinguish the two scenarios.\\
The H$_2$ gas morphology is completely different from Br$\gamma$ or
[FeII]. 
H$_2$ seems to emerge primarily at systemic velocity, thus from within the
disk, but avoiding the NLR. At the position of the low-velocity component at
-250~km/s seen in Br$\gamma$ and [FeII], H$_2$ emission at about
the same velocity is observed, but not the increase in velocity to the north
of the nucleus. Investigating the H$_2$ channel maps presented
by \citet{2010MNRAS.402..819S} we see the same behaviour (see their channel map
centered on -270~km/s). 
To investigate the dynamics in Br$\gamma$ and H$_2$ in more detail, we fitted the
emission line profiles with a double Gaussian. The results are shown in figures \ref{brgdissect}
and \ref{h2dissect}. The emission line profiles at the position of the bright
western [FeII] knot are clearly double-peaked with one profile component at systemic and
another at about -250~km/s, i.e., similar to what is observed in [FeII]. Again,
emission at systemic and/or low velocities at this
position may be locally enhanced by the jet. To the north the Gaussian
dispersion in Br$\gamma$ is increased (see figure \ref{brgdissect}, Br$\gamma$
line profile at position J), indicating that the individual profile components
remain spectrally unresolved.


\clearpage
\clearpage
\clearpage
\section{Column densities and excitation mechanisms}

In this section we investigate the excitation mechanism of hot H$_2$ and derive
column densities and the total mass of hot H$_2$. 
We furthermore determine the mass of ionized hydrogen and estimate a lower limit
of the column density of the HeI $2^1$S state. 

\subsection{H$_2$}

\subsubsection{Column densities and population density diagram}

\noindent Because there is little H$_2$ emission from inside the NLR bicone, it
is reasonable to assume that the H$_2$ 
molecules dissociate in the radiation field of the AGN while H$_2$
molecules, for which the AGN is hidden behind the putative molecular torus,
survive and are excited. The emission of H$_2$ is most
pronounced along the direction of
the resolved flux density ratio map (figure \ref{klks}), which roughly
coincides with the bar. 

\noindent \citet{2009MNRAS.394.1148S} detected more than 10~H$_2$ emission 
lines from J to K~band and concluded from a population-level diagram that a
thermal excitation mechanism dominates. 
In our spectra only the 1-0 series is detectable with high
confidence, although emission from the 2-1~series is detectable with lower
confidence.\\

\noindent The column density can be derived from
$$N_{col}=\frac{4\pi I}{A h \nu}\textrm{     },$$ \citep{1978ApJ...223..464B} where I is
the measured surface brightness in the line and A is the Einstein A
coefficient for that transition. 
Table \ref{whole_flux} lists the measured fluxes for our whole FoV. The
derived column densities (see table \ref{column_dens}) are all lower than
$10^{16}$~cm$^{-2}$, characteristic for optically thin H$_2$ regions.

\begin{table*}
\caption{Detected species, transition probabilities
  \citep{1977ApJS...35..281T}, degeneracy, upper
level energy, and derived column densities for our whole FoV.}             
\label{column_dens}      
\centering          
\begin{tabular}{ l l l l l l  }
\hline\hline       
Species & $\lambda$ & A & g & T$_u$ &N$_{col}$ \\
        & [$\mu$m]   & [10$^7$ s]& &[K]&[10$^{14}$~cm$^{-2}$]\\
\hline
1-0S(0) & 2.2235  &  2.53 & 5  &  6471   & 5.2 \\
1-0S(1) & 2.1218  &  3.47 & 21 &  6956   & 15.5\\ 
1-0S(2) & 2.0338  &  3.98 & 9  &  7584   & 4.4 \\
2-1S(1) & 2.2477  &  4.98 & 21 &  12550  & 0.6 \\
2-1S(2) & 2.1542  &  5.60 & 9  &  13150  & 0.3 \\ 
2-1S(3) & 2.0735  &  5.77 & 33 &  13890  & 0.5 \\
\hline                  
\end{tabular}
\end{table*}

\begin{figure}
\centering
\includegraphics[width=8cm]{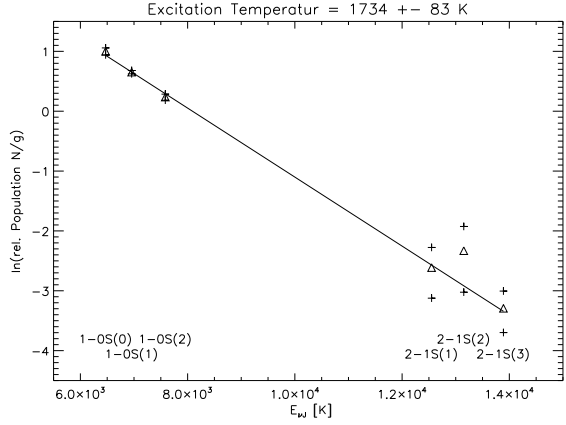}
\caption{Population density as function of upper level energy. Triangles
  denote data points, error limits are indicated by crosses. }
\label{h2_temp}
\end{figure}

\noindent The population-level diagram reveals a thermal excitation mechanism
with a temperature of about 1700 K using all species (see figure
\ref{h2_temp}) and 1500~K using 1-0~species only. Since our H$_2$ fluxes were
not corrected for reddening, the derived excitation temperature may be a
lower limit.

\subsubsection{Diagnostics of excitation mechanisms}

\begin{figure}
\centering
\includegraphics[width=8cm]{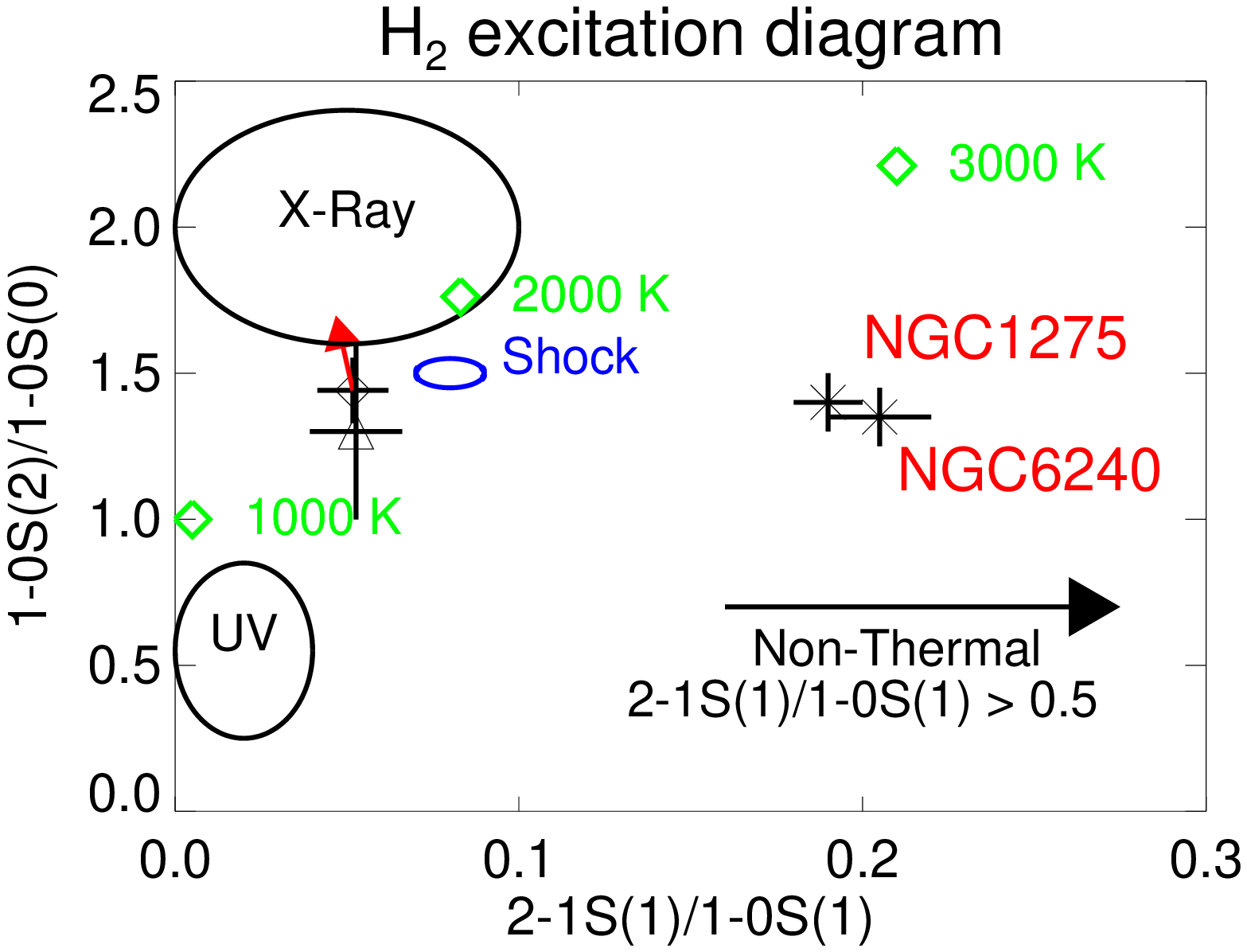}
\caption{"Mouri diagram". 
The data regions correspond to shocks \citep{1989MNRAS.236..929B}, non-thermal
(\cite{1987ApJ...322..412B}) and thermal UV \citep{1989ApJ...338..197S}, and
X-ray \citep{1983ApJ...269..560L, 1990ApJ...363..464D}.
The diamond corresponds to data extracted from our whole FoV, the triangle to
data extracted from the bright H$_2$ knot in the east, and stars to data from
other galaxies (NGC~1275 and NGC~6240 taken from \citet{kra00}).
The red arrow shows the displacement for a reddening correction of $A_K=1$.}
\label{h2_Mouri}
\end{figure}

\noindent \citet{1994ApJ...427..777M} proposed to use ratios involving ro-vib
transitions of H$_2$ to distinguish the dominant excitation mechanisms.
The measured ratios of 2-1S(1)/1-0S(1) and 1-0S(2)/1-0S(0) (see figure
\ref{h2_Mouri}) imply that shock 
(J-shock model from \citet{1989MNRAS.236..929B})
and X-ray excitation are favored, which we investigate in more detail in the following.

\noindent We used the models of \citet{1996ApJ...466..561M} to 
compare observed and predicted 1-0S(1), 2-1S(1) and [FeII]$\lambda$
1.644~$\mu$m fluxes in XDRs. 
Their figure 6 summarizes derived intensities for the above
species as a function of
the effective (attenuated) ionization parameter $\zeta_{eff}$, which is defined
as  
$$\zeta_{eff}=1.26 \times 10^{-4} \frac{f_X}{n_5 N_{22}^{0.9}}\textrm{     }.$$
Here $n_5$ [10$^5$~cm$^{-3}$] is the density of hydrogen nuclei in the
emitting cloud, $f_X$ [erg~cm$^{-2}$~s$^{-1}$] the flux of X-rays penetrating
the cloud, and $N_{22}$ [$10^{22}$~cm$^{-2}$] the total 
attenuating hydrogen column density. 
\citet{2006A&A...446..459C} found an X-ray flux of F$_X$(2-10 keV) = $4.5 \times
10^{-14}$ W m$^{-2}$ and an attenuating column density of $N=7.5 \times 10^{22}$~cm$^{-2}$.
Assuming isotropic X-ray emission, we find $\zeta_{eff}=0.162 *
n_5^{-1} * d^{-2}$, where d [pc] is the distance of a H$_2$ cloud from the nucleus. 
Table \ref{maloney} lists predicted 1-0S(1) fluxes per spaxel
as a function of $\zeta_{eff}$ and
hydrogen densities of n$_5$=0.01 and n$_5$=1 . 
For typical distances d of a few dozen parsecs, $\zeta_{eff}$ is
generally lower than 0.01. Comparing $\zeta_{eff}$ with figure 6 of
\citet{1996ApJ...466..561M} reveals that the predicted 1-0S(1) fluxes are 
generally lower than $\thicksim$ 10$^{-21}$ W m$^{-2}$, which is at least two orders of magnitude
lower than the highest observed 1-0S(1) flux.\\

\noindent Using [FeII] flux values from \citet{2009MNRAS.394.1148S} extracted from a 
$0.3 \times 0.3 $ arcsecond aperture
at the position of the bright eastern H$_2$ knot (see their table 1) reveals
a measured ratio of 1-0S(1)/[FeII] of about 0.9. In the models of \citet{1996ApJ...466..561M}
this requires $\zeta_{eff} \thicksim 0.1$, for which the predicted 1-0S(1) flux
would also be close to the observed flux values.
Therefore X-ray excitation might be the dominant excitation mechanism if
the emission of X-rays is not isotropic, implying higher $\zeta_{eff}$ at the
position of the H$_2$ clouds.
We might now speculate that the central absorber is perforated. In this case
some H$_2$ regions could be more directly exposed to radiation from the central
engine (which remains unresolved by our observations) with the nucleus hidden by
the central absorber along our line-of-sight. 
However, the X-ray emission as traced by high-ionization lines appears rather confined
to the narrow line region \citep{2000ApJ...545L..81O}, which is partially
embedded in the galactic disk. If we assume that the radio jet is roughly perpendicular to
the putative molecular torus, the torus is highly inclined with respect to the
galactic plane. In this case the central
engine is nearly hidden by the torus along our line-of-sight, giving rise to
the high attenuating column density observed by
\citet{2006A&A...446..459C}, but H$_2$ regions in the galactic disk are much
more exposed to the nuclear radiation field. We therefore conclude that X-rays
might contribute significantly to the H$_2$ and [FeII] emission observed.\\

\begin{table*}
\caption{Theoretical flux values for 1-0S(1) in W/m$^2$
  emerging from a spaxel of 50 $\times$ 50
  milli-arcseconds as a function of the hydrogen density and distance from the
  nucleus and hence $\zeta_{eff}$ from figure 6 of
  \citet{1996ApJ...466..561M}.}             
\label{maloney}      
\centering          
\begin{tabular}{ c c c c c }
\hline\hline       
\multicolumn{5}{c}{1-0S(1)}\\
\hline
       & \multicolumn{2}{c}{n=10$^5$~cm$^{-3}$} &\multicolumn{2}{c}{n = 10$^3$~cm$^{-3}$}\\
d [pc] & log($\zeta_{eff}$) & log(F$_{H_2}$) & log($\zeta_{eff}$) & log(F$_{H_2}$) \\
20   & -3.5 & -21.2 & -1.5 & -23.7 \\
30   & -3.8 & -21.4 & -1.8 & -23.5 \\
50   & -4.3 & -21.7 & -2.3 & -23.1 \\
\hline
\end{tabular}
\end{table*}





\noindent On the other hand, 1-0S(1) and [FeII] emission can be strong in shocks from supernova explosions into the interstellar medium.
Following \citet{1989ApJ...342..306H}
(their figure 5 and figure 8) the line ratio 1-0S(1)/[FeII] is about unity
for shock velocities above 40~km/s. 
These shocks, if present, may rather be induced by the bar than by supernova remnants since
the 6~cm radio emission morphology is more aligned with the 21~cm radio jet
\citep[see e.g.][]{1981ApJ...247..419U} and clear signs of
massive starformation in the central 100~pc remain undiscovered.\\




\noindent Thus, we conclude that excitation by X-rays and shock excitation
especially along the bar contribute to the H$_2$ and [FeII] fluxes observed.
According to figure \ref{h2_Mouri}, X-ray excitation in NGC~4151
may even be more dominant than in the Seyfert 1.5 galaxy NGC~1275 or the
ultra-luminous infrared galaxy NGC~6240.

\subsubsection{Total mass of molecular and ionized gas}

The luminosity in, e.g., 1-0S(1) is given by 
$$L_{1-0S(1)}=n_0g_Je^{-E_{1-0S(1)}/kT}Z(T)^{-1} A_{1-0S(1)} h\nu\textrm{     },$$
where $L_{1-0S(1)}$ is the luminosity in 1-0S(1), $n_0$ is the
number of H$_2$ molecules, $g_J$ the statistical weight for that
transition, $E_{1-0S(1)}$ the energy of the upper level, Z(T) the partition
function, T the temperature, $A_{1-0S(1)}$ the Einstein A coefficient
of the 1-0S(1) transition, and $\nu$ the frequency of the emitted photon.
The partition function is taken from \citet{1987A&A...182..348I}.
The total 1-0S(1) flux in our FoV is $F = 2.9\times 10^{-17} W/m^2$. 
Assuming a distance of d=13.25 Mpc and a 
temperature of 1700 K, we obtain a mass of hot molecular H$_2$ of
$M_{H_2}=100 M_{\odot}$, which also fully complies with the
measurements by \citet{2009MNRAS.394.1148S}. Since our FoV does not fully 
sample the whole inner region, the derived hot H$_2$ mass is a lower limit.
This amount is also rather small, but
as \citet{2005AJ....129.2197D} point out, the hot-to-cold mass ratio in centers
of galaxies ranges between 10$^{-7}$ and 10$^{-5}$, which matches the high gas concentrations of $\thicksim10^8$ $M_{\odot}$ as inferred from
mm-observations (NGC~4569: \citet{2007A&A...471..113B}; NGC~6951:
\citet{2007A&A...468L..63K} and 
others for the NUclei of GAlaxies (NUGA) group). Thus the
total amount of molecular gas in the central 100 pc of NGC~4151 may be
higher by several orders of magnitude.\\
Following \citet{2008MNRAS.385.1129R, 1982ApJ...253..136S}, 
the amount of ionized gas can be estimated:
$$M_{HII} = 2.88 \times 10^{22}
\left(\frac{F_{Br\gamma}}{W/m^2}\right)\left(\frac{d}{Mpc}\right)^2\left(\frac{N_e}{cm^{-3}}\right)^{-1}\textrm{     }.$$
Assuming an electron density 
of $N_e=100cm^{-3}$ and using the total narrow Br$\gamma$ flux of
$F_{Br\gamma}=4\times 10^{-17} W/m^2$, a total ionized hydrogen gas
mass of $M=2.1 \times 10^6 M_\odot$ can be derived, which fully
agrees with the measurements by \citet{2009MNRAS.394.1148S}.

\subsection{A lower limit to the HeI $2^1S$ column density}\label{subsec:helium}

As shown in figure \ref{he_radial}, the HeI emission is extended 
while the flux in the absorption component decreases exactly with the strength
of the continuum emission. Furthermore, the velocity offset of this absorption
complex remains constant throughout our FoV, and we assume that the absorption
component is related to a nuclear outflow that remains spatially unresolved.
The whole line complex appears P-Cygni-like with a symmetrical absorption
component and no broad emission component. The velocity offset of
the HeI absorption component is -280~km/s with a lorentzian HWHM of 400~km/s and an
equivalent width of 7.7$\AA$. 
These numbers differ from the velocities and FWHMs of absorption in UV
resonance lines, see table 1 of \citet{2001ApJ...551..671K}. These
authors used photoionization models to determine, e.g., the distance of
individual absorbers along the line-of-sight. As a result, 
absorption is usually broadened (FWHM higher than 430 km/s) for clouds
closer than 1 pc to the nucleus, but these clouds (D+E and D') also have
outflow velocities higher than about 500 km/s. Furthermore, Balmer and 
HeI$\lambda$3888$\AA$ absorption is not seen as a multi-component absorber 
as in the UV and the outflow velocities increase in nuclear high states 
\citep{2002AJ....124.2543H}, which has not been observed in the UV lines.
This may indicate that our HeI absorption originates from a different region 
than the observed UV resonance lines. Since the
opening cone of the ionizing radiation is close to the line-of-sight, it is 
reasonable to assume erosion of the edge of the obscuring torus.\\ 

\begin{figure}
\centering
\includegraphics[angle=90,width=8cm]{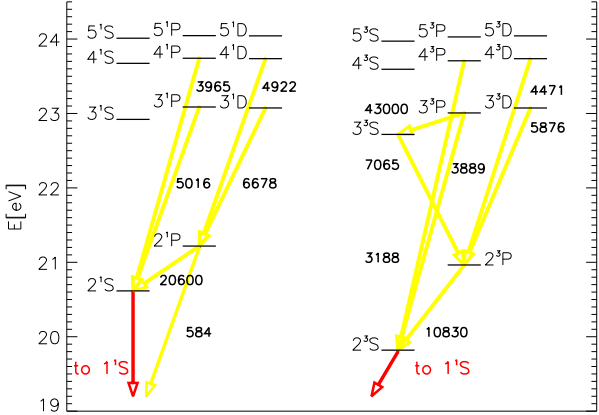}
\caption{Grotrian diagram of HeI. Prominent astronomical lines (plotted in yellow) 
are indicated with wavelengths in \AA. Transitions from metastable 
levels to the ground state are plotted in red.}
\label{he_grothrian}
\end{figure}

 
\noindent The blueshifted optically thin absorption component in the HeI
profiles of our nuclear spectra (see figure \ref{emission2}, top) can be used
to calculate the HeI $2^1S$ column density. Since only one of the many
prominent 
transitions from the lower HeI $2^1S$ level is covered by our observations,
this yields a lower limit to the column density of the HeI $2^1S$ level. 
Figure \ref{he_grothrian} shows a Grotrian diagram of HeI labeled with some
prominent astronomical transitions including the transitions from and to the
$2^1S$ level.  
The lowest singlet level $2^1S$ and especially the lowest triplet level $2^3S$
are both metastable with transition probabilities of $51.3 s^{-1}$ and $1.27
\times 10^{-4} s^{-1}$ to the ground state, $1^1S$. Depopulation of these
levels can take place by photoionization 
or by the highly forbidden emission of high-eV photons to the ground state.
Absorption of $\lambda$3889~\AA $ $ photons by $2^3S$ leads to enhanced
emission of $\lambda$10830~\AA $ $. Radiatively pumping the ground state
$1^1S$ to $2^1P$ by absorption of 
$\lambda$584~\AA $ $ photons may be essential to
produce emission of $\lambda$20581~\AA $ $ because
the decay of $2^1P$ by re-emitting a photon of the same wavelength is highly
favored (99.9$\%$) against the emission of a $\lambda$20581~\AA $ $ photon
\citep{1974ApJ...188..309R}. 
However, the $\lambda$20581~\AA $ $ line strength is limited by 
photoionization of hydrogen by the $\lambda$584~\AA $ $ photon
\citep{1980ApJ...235..889T}. Moreover, collisional excitation of the more
long-living level $2^3S$ to $2^1S$ may be of importance as well 
\citep[see e.g.][]{1989agna.book.....O}.\\

\noindent The total column density of $2^1S$ can be estimated using additional
absorption lines such as $\lambda$3965~\AA $ $ or $\lambda$5016~\AA. 
Unfortunately, these lines are blended with H$\epsilon$, [NIII], and [OIII]
and are also not covered by our observations. 
Here we confine ourselves to calculating a lower limit of the $2^1S$
column density following \citet{1997ApJ...478...80H} and
\citet{1999ApJ...516..750C}. 

\noindent The underlying continuum was fitted with a parabola using spectral
channels to the shorter and longer wavelength side of
the line complex, and we normalized each spectrum by dividing with the underlying
continuum emission. 
$I_r$ is then the residual flux at a particular radial velocity $v_r$. The
minimum covering factor $C_{los}$ in the line-of-sight, the fraction of
continuum that 
is absorbed by the occulter, is then $C_{los} >= 1-I_r$. 
Assuming screen extinction, the optical depth $\tau$ is then given by $\tau =
ln(1/I_r)$ if the covering
factor is one,  $\tau = ln(C_{los}/(I_r+C_{los}+1)$ otherwise.
The column density N is then obtained by integrating the optical depth across
the line profile,
$$N=\frac{m_e c}{\pi e^2 f_{12} \lambda}\int\tau(v_r) dv_r\textrm{     },$$
with $\lambda$ the laboratory wavelength, $m_e$
the electron mass, c the speed of light, e the electron charge and 
$f_{12}$ the (absorption) oscillator strength, where $f_{12}=0.375$
\citep{2009JPCRD..38..565W}. Assuming screen extinction and a
covering factor of $C_{los}=1$, we derive a column density of
$5 \times 10^{12}$~cm$^{-2}$. The minimum covering factor derived is 
$C_{los}$ = 0.09, which increases the column density to $1 \times
10^{14}$~cm$^{-2}$.
The f-value of the HeI 10830~\AA $ $ transition is about 0.18
and consequently comparable with the f-value of the HeI$\lambda$20581~\AA $ $ transition. Since
the transition probability for radiative decay of the $2^3S$ level is lower by a
factor of $10^{-5}$ than for the $2^1S$ level, and assuming that the 
population density of the $2^3S$ and $2^1S$ levels differ by a similar factor,
we expect a similarly deep HeI$\lambda$10830~\AA $ $ absorption complex, which
is indeed observed \citep{2009MNRAS.394.1148S}.\\

\section{Summary}

We have used near-infrared integral field data obtained with OSIRIS at the Keck
telescope to determine the morphology and dynamics of ionized and molecular
gas in the inner two arcseconds of NGC~4151. We compared our results with
previous measurements by \citet{2005AJ....130..945D}
and \citet{2010MNRAS.402..819S}. The main results can be summarized
as follows:\\

\noindent a) The stellar and non-stellar contributions to the total continuum
both peak at the same position and remain unresolved by our
observations. Br$\gamma$ and coronal lines peak a few parsec to the northwest/west of the continuum peak. Coronal lines also show a broad component
implying that these species may originate from an outflow.\\

\noindent b) The ionized gas extends toward the NLR while the molecular gas
strictly avoids the NLR. The NLR morphology is well compatible with
measurements by \citet{2010MNRAS.402..819S}. \\

\noindent c) We identified four dynamical components. First, emission at
systemic velocity, which is assumed to emerge from the galactic disk of
NGC~4151. Second, emission that is enhanced due to interaction with the jet.
Third, a nuclear inflow of gas (most likely within the galactic disk) 
that is observed in Br$\gamma$. And fourth, an outflow along the NLR bicone. \\

\noindent d) The outflow of the NLR gas as traced by [FeII] is well compatible
with the models proposed by \citet{2005AJ....130..945D} (the acceleration model)
and \citet{2010MNRAS.402..819S} (the constant-velocity model). We emphasize
that our FoV is smaller than the 
FoV provided in \citet{2010MNRAS.402..819S} so that we are unable to
confirm or reject their constant-velocity model for distances larger than
approximately one arcsecond from the nucleus. However, our data
indicate acceleration within the central arcsecond, which agrees with the model
proposed by \citet{2005AJ....130..945D}.\\

\noindent e) The 21 cm radio continuum jet originates from regions where H$_2$
gas is rotating into the path of the jet because of the galactic rotation.
The jet also interacts with gas at certain positions and locally enhances
emission of [FeII].\\ 

\noindent f) The hot H$_2$ gas appears to be thermal with a 
temperature of about 1700 K, and X-ray and shock excitation of H$_2$ are very likely
to occur. Derived column densities are about  
$10^{14}$ cm$^{-2}$. 
We detected a bridge of H$_2$ running through the nucleus. Although
the velocities along this bridge are systemic, we observed inflow in
Br$\gamma$ at speeds that match with speeds expected at the radius of the
sphere of influence.\\

\noindent g) We detected HeI emission that is accompanied by a blueshifted
optically thin absorption component in our nuclear spectra. The absorption
component has a velocity offset of -280 km/s and remains spatially
unresolved. We assume that this feature represents a nuclear 
outflow that occurs closer than a very few parsecs to the nucleus.
We calculated a covering factor of $C_{los}$ = 0.09 with a minimum column
density of $1 \times 10^{14}$ cm$^{-2}$ for the HeI $2^1S$ level.\\

\begin{acknowledgements}
The authors would like to sincerely thank the dedicated members of the Keck
Observatory staff (CARA), who greatly contributed to the success of the
commissioning of OSIRIS. We would like to acknowledge Sean
Adkins, Paola Amico, Randy Campbell, Al Conrad, Allan Honey, David Le Mignant,
Jim Lyke, Marcos van Dam, and Peter Wizinowich. Data presented here were
obtained at the W. M. Keck Observatory, which is operated as a scientific
partnership between the California Institute of Technology, the University of
California, and the National Aeronautics and Space Administration. The
Observatory was made possible by generous financial support of the W. M. Keck
Foundation. We also thank C. G. Mundell for providing us with the 21 cm radio
continuum data.

\end{acknowledgements}

\bibliographystyle{aa} 
\bibliography{mybibtex}

\begin{thebibliography}{68}
\expandafter\ifx\csname natexlab\endcsname\relax\def\natexlab#1{#1}\fi

\bibitem[{{Antonucci}(1993)}]{1993ARA&A..31..473A}
{Antonucci}, R. 1993, \araa, 31, 473

\bibitem[{{Bacon} {et~al.}(1995){Bacon}, {Adam}, {Baranne}, {Courtes}, {Dubet},
  {Dubois}, {Emsellem}, {Ferruit}, {Georgelin}, {Monnet}, {Pecontal},
  {Rousset}, \& {Say}}]{1995A&AS..113..347B}
{Bacon}, R., {Adam}, G., {Baranne}, A., {et~al.} 1995, \aaps, 113, 347

\bibitem[{{Beckwith} {et~al.}(1978){Beckwith}, {Persson}, {Neugebauer}, \&
  {Becklin}}]{1978ApJ...223..464B}
{Beckwith}, S., {Persson}, S.~E., {Neugebauer}, G., \& {Becklin}, E.~E. 1978,
  \apj, 223, 464

\bibitem[{{Black} \& {van Dishoeck}(1987)}]{1987ApJ...322..412B}
{Black}, J.~H. \& {van Dishoeck}, E.~F. 1987, \apj, 322, 412

\bibitem[{{Boone} {et~al.}(2007){Boone}, {Baker}, {Schinnerer}, {Combes},
  {Garc{\'{\i}}a-Burillo}, {Neri}, {Hunt}, {L{\'e}on}, {Krips}, {Tacconi}, \&
  {Eckart}}]{2007A&A...471..113B}
{Boone}, F., {Baker}, A.~J., {Schinnerer}, E., {et~al.} 2007, \aap, 471, 113

\bibitem[{{Brand} {et~al.}(1989){Brand}, {Toner}, {Geballe}, {Webster},
  {Williams}, \& {Burton}}]{1989MNRAS.236..929B}
{Brand}, P.~W.~J.~L., {Toner}, M.~P., {Geballe}, T.~R., {et~al.} 1989, \mnras,
  236, 929

\bibitem[{{Cappi} {et~al.}(2006){Cappi}, {Panessa}, {Bassani}, {Dadina}, {Di
  Cocco}, {Comastri}, {della Ceca}, {Filippenko}, {Gianotti}, {Ho}, {Malaguti},
  {Mulchaey}, {Palumbo}, {Piconcelli}, {Sargent}, {Stephen}, {Trifoglio}, \&
  {Weaver}}]{2006A&A...446..459C}
{Cappi}, M., {Panessa}, F., {Bassani}, L., {et~al.} 2006, \aap, 446, 459

\bibitem[{{Crenshaw} {et~al.}(1999){Crenshaw}, {Kraemer}, {Boggess}, {Maran},
  {Mushotzky}, \& {Wu}}]{1999ApJ...516..750C}
{Crenshaw}, D.~M., {Kraemer}, S.~B., {Boggess}, A., {et~al.} 1999, \apj, 516,
  750

\bibitem[{{Crenshaw} {et~al.}(2000{\natexlab{a}}){Crenshaw}, {Kraemer},
  {Hutchings}, {Bradley}, {Gull}, {Kaiser}, {Nelson}, {Ruiz}, \&
  {Weistrop}}]{2000AJ....120.1731C}
{Crenshaw}, D.~M., {Kraemer}, S.~B., {Hutchings}, J.~B., {et~al.}
  2000{\natexlab{a}}, \aj, 120, 1731

\bibitem[{{Crenshaw} {et~al.}(2000{\natexlab{b}}){Crenshaw}, {Kraemer},
  {Hutchings}, {Danks}, {Gull}, {Kaiser}, {Nelson}, \&
  {Weistrop}}]{2000ApJ...545L..27C}
{Crenshaw}, D.~M., {Kraemer}, S.~B., {Hutchings}, J.~B., {et~al.}
  2000{\natexlab{b}}, \apjl, 545, L27

\bibitem[{{Dale} {et~al.}(2005){Dale}, {Sheth}, {Helou}, {Regan}, \&
  {H{\"u}ttemeister}}]{2005AJ....129.2197D}
{Dale}, D.~A., {Sheth}, K., {Helou}, G., {Regan}, M.~W., \& {H{\"u}ttemeister},
  S. 2005, \aj, 129, 2197

\bibitem[{{Das} {et~al.}(2005){Das}, {Crenshaw}, {Hutchings}, {Deo}, {Kraemer},
  {Gull}, {Kaiser}, {Nelson}, \& {Weistrop}}]{2005AJ....130..945D}
{Das}, V., {Crenshaw}, D.~M., {Hutchings}, J.~B., {et~al.} 2005, \aj, 130, 945

\bibitem[{{Davies}(1973)}]{1973MNRAS.161P..25D}
{Davies}, R.~D. 1973, \mnras, 161, 25P

\bibitem[{{Draine} \& {Woods}(1990)}]{1990ApJ...363..464D}
{Draine}, B.~T. \& {Woods}, D.~T. 1990, \apj, 363, 464

\bibitem[{{Elvis}(2000)}]{2000ApJ...545...63E}
{Elvis}, M. 2000, \apj, 545, 63

\bibitem[{{Evans} {et~al.}(1993){Evans}, {Tsvetanov}, {Kriss}, {Ford},
  {Caganoff}, \& {Koratkar}}]{1993ApJ...417...82E}
{Evans}, I.~N., {Tsvetanov}, Z., {Kriss}, G.~A., {et~al.} 1993, \apj, 417, 82

\bibitem[{{Fernandez} {et~al.}(1999){Fernandez}, {Holloway}, {Meaburn},
  {Pedlar}, \& {Mundell}}]{1999MNRAS.305..319F}
{Fernandez}, B.~R., {Holloway}, A.~J., {Meaburn}, J., {Pedlar}, A., \&
  {Mundell}, C.~G. 1999, \mnras, 305, 319

\bibitem[{{Hamann} {et~al.}(1997){Hamann}, {Barlow}, {Junkkarinen}, \&
  {Burbidge}}]{1997ApJ...478...80H}
{Hamann}, F., {Barlow}, T.~A., {Junkkarinen}, V., \& {Burbidge}, E.~M. 1997,
  \apj, 478, 80

\bibitem[{{Hollenbach} \& {McKee}(1989)}]{1989ApJ...342..306H}
{Hollenbach}, D. \& {McKee}, C.~F. 1989, \apj, 342, 306

\bibitem[{{Hutchings} {et~al.}(1998){Hutchings}, {Crenshaw}, {Kaiser},
  {Kraemer}, {Weistrop}, {Baum}, {Bowers}, {Feinberg}, {Green}, {Gull},
  {Hartig}, {Hill}, \& {Lindler}}]{1998ApJ...492L.115H}
{Hutchings}, J.~B., {Crenshaw}, D.~M., {Kaiser}, M.~E., {et~al.} 1998, \apjl,
  492, L115

\bibitem[{{Hutchings} {et~al.}(2002){Hutchings}, {Crenshaw}, {Kraemer},
  {Gabel}, {Kaiser}, {Weistrop}, \& {Gull}}]{2002AJ....124.2543H}
{Hutchings}, J.~B., {Crenshaw}, D.~M., {Kraemer}, S.~B., {et~al.} 2002, \aj,
  124, 2543

\bibitem[{{Irwin}(1987)}]{1987A&A...182..348I}
{Irwin}, A.~W. 1987, \aap, 182, 348

\bibitem[{{Kaiser} {et~al.}(2000){Kaiser}, {Bradley}, {Hutchings}, {Crenshaw},
  {Gull}, {Kraemer}, {Nelson}, {Ruiz}, \& {Weistrop}}]{2000ApJ...528..260K}
{Kaiser}, M.~E., {Bradley}, II, L.~D., {Hutchings}, J.~B., {et~al.} 2000, \apj,
  528, 260

\bibitem[{{Knapen} {et~al.}(2003){Knapen}, {de Jong}, {Stedman}, \&
  {Bramich}}]{2003MNRAS.344..527K}
{Knapen}, J.~H., {de Jong}, R.~S., {Stedman}, S., \& {Bramich}, D.~M. 2003,
  \mnras, 344, 527

\bibitem[{{Knop} {et~al.}(1996){Knop}, {Armus}, {Larkin}, {Mathews}, {Shupe},
  \& {Soifer}}]{1996AJ....112...81K}
{Knop}, R.~A., {Armus}, L., {Larkin}, J.~E., {et~al.} 1996, \aj, 112, 81

\bibitem[{{Krabbe} {et~al.}(2004){Krabbe}, {Gasaway}, {Song}, {Iserlohe},
  {Weiss}, {Larkin}, {Barczys}, \& {Lafreniere}}]{2004SPIE.5492.1403K}
{Krabbe}, A., {Gasaway}, T., {Song}, I., {et~al.} 2004, in Society of
  Photo-Optical Instrumentation Engineers (SPIE) Conference Series, Vol. 5492,
  Society of Photo-Optical Instrumentation Engineers (SPIE) Conference Series,
  ed. A.~F.~M. {Moorwood} \& M.~{Iye}, 1403--1410

\bibitem[{{Krabbe} {et~al.}(2000){Krabbe}, {Sams}, {Genzel}, {Thatte}, \&
  {Prada}}]{kra00}
{Krabbe}, A., {Sams}, III, B.~J., {Genzel}, R., {Thatte}, N., \& {Prada}, F.
  2000, \aap, 354, 439

\bibitem[{{Kraemer} {et~al.}(2001){Kraemer}, {Crenshaw}, {Hutchings}, {George},
  {Danks}, {Gull}, {Kaiser}, {Nelson}, {Weistrop}, \&
  {Vieira}}]{2001ApJ...551..671K}
{Kraemer}, S.~B., {Crenshaw}, D.~M., {Hutchings}, J.~B., {et~al.} 2001, \apj,
  551, 671

\bibitem[{{Krips} {et~al.}(2007){Krips}, {Neri}, {Garc{\'{\i}}a-Burillo},
  {Combes}, {Schinnerer}, {Baker}, {Eckart}, {Boone}, {Hunt}, {Leon}, \&
  {Tacconi}}]{2007A&A...468L..63K}
{Krips}, M., {Neri}, R., {Garc{\'{\i}}a-Burillo}, S., {et~al.} 2007, \aap, 468,
  L63

\bibitem[{{Landini} {et~al.}(1984){Landini}, {Natta}, {Salinari}, {Oliva}, \&
  {Moorwood}}]{1984A&A...134..284L}
{Landini}, M., {Natta}, A., {Salinari}, P., {Oliva}, E., \& {Moorwood},
  A.~F.~M. 1984, \aap, 134, 284

\bibitem[{{Landsman}(1993)}]{1993ASPC...52..246L}
{Landsman}, W.~B. 1993, in Astronomical Society of the Pacific Conference
  Series, Vol.~52, Astronomical Data Analysis Software and Systems II, ed.
  R.~J. {Hanisch}, R.~J.~V. {Brissenden}, \& J.~{Barnes}, 246

\bibitem[{{Larkin} {et~al.}(2006){Larkin}, {Barczys}, {Krabbe}, {Adkins},
  {Aliado}, {Amico}, {Brims}, {Campbell}, {Canfield}, {Gasaway}, {Honey},
  {Iserlohe}, {Johnson}, {Kress}, {LaFreniere}, {Lyke}, {Magnone}, {Magnone},
  {McElwain}, {Moon}, {Quirrenbach}, {Skulason}, {Song}, {Spencer}, {Weiss}, \&
  {Wright}}]{2006SPIE.6269E..42L}
{Larkin}, J., {Barczys}, M., {Krabbe}, A., {et~al.} 2006, in Society of
  Photo-Optical Instrumentation Engineers (SPIE) Conference Series, Vol. 6269,
  Society of Photo-Optical Instrumentation Engineers (SPIE) Conference Series

\bibitem[{{Lepp} \& {McCray}(1983)}]{1983ApJ...269..560L}
{Lepp}, S. \& {McCray}, R. 1983, \apj, 269, 560

\bibitem[{{Maloney} {et~al.}(1996){Maloney}, {Hollenbach}, \&
  {Tielens}}]{1996ApJ...466..561M}
{Maloney}, P.~R., {Hollenbach}, D.~J., \& {Tielens}, A.~G.~G.~M. 1996, \apj,
  466, 561

\bibitem[{{Mouri}(1994)}]{1994ApJ...427..777M}
{Mouri}, H. 1994, \apj, 427, 777

\bibitem[{{Mundell} {et~al.}(1995){Mundell}, {Pedlar}, {Baum}, {O'Dea},
  {Gallimore}, \& {Brinks}}]{1995MNRAS.272..355M}
{Mundell}, C.~G., {Pedlar}, A., {Baum}, S.~A., {et~al.} 1995, \mnras, 272, 355

\bibitem[{{Mundell} {et~al.}(1999){Mundell}, {Pedlar}, {Shone}, \&
  {Robinson}}]{1999MNRAS.304..481M}
{Mundell}, C.~G., {Pedlar}, A., {Shone}, D.~L., \& {Robinson}, A. 1999, \mnras,
  304, 481

\bibitem[{{Mundell} \& {Shone}(1999)}]{1999MNRAS.304..475M}
{Mundell}, C.~G. \& {Shone}, D.~L. 1999, \mnras, 304, 475

\bibitem[{{Mundell} {et~al.}(2003){Mundell}, {Wrobel}, {Pedlar}, \&
  {Gallimore}}]{2003ApJ...583..192M}
{Mundell}, C.~G., {Wrobel}, J.~M., {Pedlar}, A., \& {Gallimore}, J.~F. 2003,
  \apj, 583, 192

\bibitem[{{Ogle} {et~al.}(2000){Ogle}, {Marshall}, {Lee}, \&
  {Canizares}}]{2000ApJ...545L..81O}
{Ogle}, P.~M., {Marshall}, H.~L., {Lee}, J.~C., \& {Canizares}, C.~R. 2000,
  \apjl, 545, L81

\bibitem[{{Onken} {et~al.}(2007){Onken}, {Valluri}, {Peterson}, {Pogge},
  {Bentz}, {Ferrarese}, {Vestergaard}, {Crenshaw}, {Sergeev}, {McHardy},
  {Merritt}, {Bower}, {Heckman}, \& {Wandel}}]{2007ApJ...670..105O}
{Onken}, C.~A., {Valluri}, M., {Peterson}, B.~M., {et~al.} 2007, \apj, 670, 105

\bibitem[{{Osterbrock}(1989)}]{1989agna.book.....O}
{Osterbrock}, D.~E. 1989, {Astrophysics of gaseous nebulae and active galactic
  nuclei}

\bibitem[{{Pedlar} {et~al.}(1992){Pedlar}, {Howley}, {Axon}, \&
  {Unger}}]{1992MNRAS.259..369P}
{Pedlar}, A., {Howley}, P., {Axon}, D.~J., \& {Unger}, S.~W. 1992, \mnras, 259,
  369

\bibitem[{{Peletier} {et~al.}(1999){Peletier}, {Knapen}, {Shlosman},
  {P{\'e}rez-Ram{\'{\i}}rez}, {Nadeau}, {Doyon}, {Rodriguez Espinosa}, \&
  {P{\'e}rez Garc{\'{\i}}a}}]{pel99}
{Peletier}, R.~F., {Knapen}, J.~H., {Shlosman}, I., {et~al.} 1999, \apjs, 125,
  363

\bibitem[{{Pickles}(1998)}]{1998PASP..110..863P}
{Pickles}, A.~J. 1998, \pasp, 110, 863

\bibitem[{{Pott} {et~al.}(2010){Pott}, {Malkan}, {Elitzur}, {Ghez}, {Herbst},
  {Sch{\"o}del}, \& {Woillez}}]{2010ApJ...715..736P}
{Pott}, J.-U., {Malkan}, M.~A., {Elitzur}, M., {et~al.} 2010, \apj, 715, 736

\bibitem[{{Prieto} {et~al.}(2005){Prieto}, {Marco}, \&
  {Gallimore}}]{2005MNRAS.364L..28P}
{Prieto}, M.~A., {Marco}, O., \& {Gallimore}, J. 2005, \mnras, 364, L28

\bibitem[{{Riffel} {et~al.}(2006){Riffel}, {Rodr{\'{\i}}guez-Ardila}, \&
  {Pastoriza}}]{2006A&A...457...61R}
{Riffel}, R., {Rodr{\'{\i}}guez-Ardila}, A., \& {Pastoriza}, M.~G. 2006, \aap,
  457, 61

\bibitem[{{Riffel} {et~al.}(2009){Riffel}, {Storchi-Bergmann}, \&
  {McGregor}}]{2009ApJ...698.1767R}
{Riffel}, R.~A., {Storchi-Bergmann}, T., \& {McGregor}, P.~J. 2009, \apj, 698,
  1767

\bibitem[{{Riffel} {et~al.}(2008){Riffel}, {Storchi-Bergmann}, {Winge},
  {McGregor}, {Beck}, \& {Schmitt}}]{2008MNRAS.385.1129R}
{Riffel}, R.~A., {Storchi-Bergmann}, T., {Winge}, C., {et~al.} 2008, \mnras,
  385, 1129

\bibitem[{{Robbins} \& {Bernat}(1974)}]{1974ApJ...188..309R}
{Robbins}, R.~R. \& {Bernat}, A.~P. 1974, \apj, 188, 309

\bibitem[{{Scoville} {et~al.}(1982){Scoville}, {Hall}, {Ridgway}, \&
  {Kleinmann}}]{1982ApJ...253..136S}
{Scoville}, N.~Z., {Hall}, D.~N.~B., {Ridgway}, S.~T., \& {Kleinmann}, S.~G.
  1982, \apj, 253, 136

\bibitem[{{S{\'e}rsic} \& {Pastoriza}(1965)}]{1965PASP...77..287S}
{S{\'e}rsic}, J.~L. \& {Pastoriza}, M. 1965, \pasp, 77, 287

\bibitem[{{Seyfert}(1943)}]{1943ApJ....97...28S}
{Seyfert}, C.~K. 1943, \apj, 97, 28

\bibitem[{{Simkin}(1975)}]{1975ApJ...200..567S}
{Simkin}, S.~M. 1975, \apj, 200, 567

\bibitem[{{Sternberg} \& {Dalgarno}(1989)}]{1989ApJ...338..197S}
{Sternberg}, A. \& {Dalgarno}, A. 1989, \apj, 338, 197

\bibitem[{{Storchi-Bergmann} {et~al.}(2010){Storchi-Bergmann}, {Lopes},
  {McGregor}, {Riffel}, {Beck}, \& {Martini}}]{2010MNRAS.402..819S}
{Storchi-Bergmann}, T., {Lopes}, R.~D.~S., {McGregor}, P.~J., {et~al.} 2010,
  \mnras, 402, 819

\bibitem[{{Storchi-Bergmann} {et~al.}(2009){Storchi-Bergmann}, {McGregor},
  {Riffel}, {Sim{\~o}es Lopes}, {Beck}, \& {Dopita}}]{2009MNRAS.394.1148S}
{Storchi-Bergmann}, T., {McGregor}, P.~J., {Riffel}, R.~A., {et~al.} 2009,
  \mnras, 394, 1148

\bibitem[{{Tadhunter} \& {Tsvetanov}(1989)}]{1989Natur.341..422T}
{Tadhunter}, C. \& {Tsvetanov}, Z. 1989, \nat, 341, 422

\bibitem[{{Thompson}(1995)}]{1995ApJ...445..700T}
{Thompson}, R.~I. 1995, \apj, 445, 700

\bibitem[{{Thompson} \& {Tokunaga}(1980)}]{1980ApJ...235..889T}
{Thompson}, R.~I. \& {Tokunaga}, A.~T. 1980, \apj, 235, 889

\bibitem[{{Turner} {et~al.}(1977){Turner}, {Kirby-Docken}, \&
  {Dalgarno}}]{1977ApJS...35..281T}
{Turner}, J., {Kirby-Docken}, K., \& {Dalgarno}, A. 1977, \apjs, 35, 281

\bibitem[{{Ulvestad} {et~al.}(1981){Ulvestad}, {Wilson}, \&
  {Sramek}}]{1981ApJ...247..419U}
{Ulvestad}, J.~S., {Wilson}, A.~S., \& {Sramek}, R.~A. 1981, \apj, 247, 419

\bibitem[{{V{\'e}ron-Cetty} \& {V{\'e}ron}(2006)}]{2006A&A...455..773V}
{V{\'e}ron-Cetty}, M.-P. \& {V{\'e}ron}, P. 2006, \aap, 455, 773

\bibitem[{{Wallace} \& {Hinkle}(1997)}]{1997ApJS..111..445W}
{Wallace}, L. \& {Hinkle}, K. 1997, \apjs, 111, 445

\bibitem[{{Wiese} \& {Fuhr}(2009)}]{2009JPCRD..38..565W}
{Wiese}, W.~L. \& {Fuhr}, J.~R. 2009, Journal of Physical and Chemical
  Reference Data, 38, 565

\bibitem[{{Wilson} \& {Ulvestad}(1982)}]{1982ApJ...263..576W}
{Wilson}, A.~S. \& {Ulvestad}, J.~S. 1982, \apj, 263, 576

\bibitem[{{Wizinowich} {et~al.}(2000){Wizinowich}, {Acton}, {Lai}, {Gathright},
  {Lupton}, \& {Stomski}}]{2000SPIE.4007....2W}
{Wizinowich}, P.~L., {Acton}, D.~S., {Lai}, O., {et~al.} 2000, in Society of
  Photo-Optical Instrumentation Engineers (SPIE) Conference Series, Vol. 4007,
  Society of Photo-Optical Instrumentation Engineers (SPIE) Conference Series,
  ed. P.~L. {Wizinowich}, 2--13

\end{thebibliography}

\begin{appendix}

\section{Data reduction software}

The data reduction software package consists of the data reduction pipeline
that performs all necessary basic steps to provide the
user with a standardized data cube and graphical user interfaces (GUI) to
plan the reduction strategy. The package is IDL-based (version
6.0 and higher, IDL is a product from ITT Visual Information Solutions) 
with some reduction steps programmed in C for reasons of
computation speed. The GUIs are Java-based.
The data reduction scheme itself is modular, providing individual reduction steps
for individual tasks such as sky subtraction, etc. The reduction scheme is 
coded as an XML-file and interpreted and executed by the backbone part of the 
software package. More detailed information about the data and command flow
can be found in \citet{2004SPIE.5492.1403K}. We point out that some of the
data reduction recipes below were specifically designed for reducing
commissioning data of the OSIRIS instrument. A comprehensive compendium of
data reduction strategies and a description of the current data reduction
pipeline can be found in the OSIRIS user manual.  

\subsection{Data reduction}

\subsubsection{Microlens arrays and the arrangement of spectra on the detector} 

\begin{figure}
\centering
\includegraphics[width=8cm]{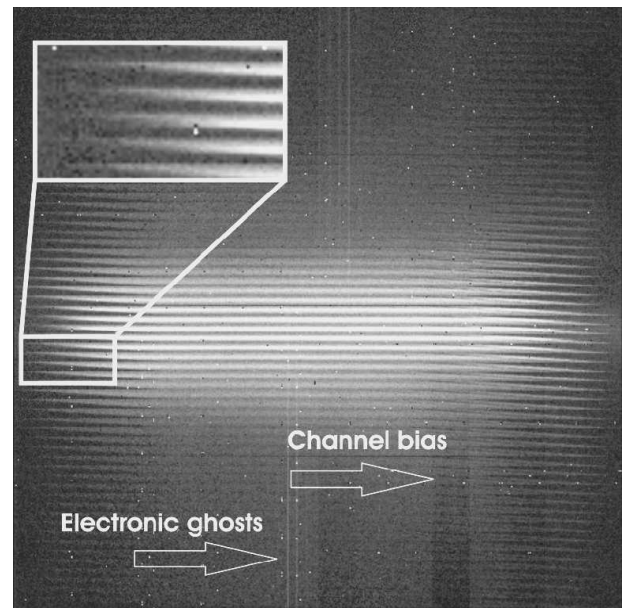}
\caption{Sky-subtracted detector image of NGC4151. The dispersion axis runs
  horizontally. Various electronic effects
  can be identified, such as bias variations in the multiplexer readout channels
and electronic ghosts. The staggered
  arrangement of the spectra can also be seen in the inset.}
\label{raw}
\end{figure}

\noindent A key device in integral field spectrographs is the image slicer that
samples and rearranges the field-of-view of the instrument in the focal plane
to create the entrance slit for the actual spectrograph. Various designs for
this device have been 
established. The most common ones for sampling the focal plane are optical
fibers, plane mirror slicers, and microlens arrays.\\
\noindent In OSIRIS, a microlens array samples the focal plane with an array of small
lenses. Each rectangular microlens has a width of 200 $\mu$m and focuses 
the light onto a pupil image that is
significantly smaller than the microlens' diameter. These images are then fed 
into the actual spectrograph. Overlapping of the individual spectra on the
detector is avoided by rotating the microlens array by 3.6$\degr$ relative to
the dispersion axis of the grating, which itself is aligned with the rows of
the detector. As a side effect, the spectra run diagonally across the detector
rows at the same angle. 
The arrangement of the spectra on the detector is staggered (see figure
\ref{raw} for a sky-subtracted detector image of NGC4151). 
In OSIRIS, spectra from adjacent microlenses of a row are shifted horizontally
(along the dispersion axis) by roughly 30 detector 
pixels, have a FWHM of two detector pixels 
perpendicular to it and are separated by two detector pixel.

\subsubsection{Reconstruction of the light distribution in the microlens array} 

\noindent The heart of each data reduction software for integral field
spectrographs is the reconstruction of the incident light on the slicer
from the recorded detector raw images via a mapping function. 
When using fibers or microlens 
arrays each point spread function (PSF) of every fiber/microlens and their 
location on the detector must be 
determined to reconstruct the distribution of incident light on the slicer. 
In OSIRIS, a movable scanning mask allows illumination of exactly one microlens
column. Illuminating the mask with a whitelight source results 
in continuum spectra that are well separated on the detector. 
From the measured PSFs the so-called rectification matrix is assembled, which 
contains spatially and spectrally 
the contribution of every microlens to the signal detected in any detector
pixel. This procedure is part of the instruments calibration and is performed
during daytime. Mathematically, for a given column or a given wavelength, the recorded signal in a detector  
pixel in row $i$ is $D_i$. We consider $I_k$ to be the incident light on the k-th 
microlens and $M_{ik}$ the
rectification matrix mapping how much light from the k-th microlens is focused
to the i-th detector pixel in the given row: $D_i = M_{ik} \times I_k$. Since
the detector has 2048 rows, $D_i$ is a vector with 2048 elements. The number of
microlenses in a given observing mode is $n$, so $I_k$ is a vector with n
elements. The rectification matrix itself therefore has 2048 $\times$ n
elements (for that given detector column). To solve for $I_k$ this matrix
equation is inverted with an iterative Gauss-Seidel algorithm in a process
hereafter called 'rectification'.\\ 
Since every microlens is illuminated with same intensity and spectrum, the
rectification matrix already accounts for the flatfield, making an additional
flatfielding step obsolete.

\subsubsection{Notes on selected reduction steps}

An updated and more complete list of reduction recipes including some
experimental ones can also be found in the 
OSIRIS user manual (Larkin et al. 2008). Here we only summarize special
reduction steps in the order applied to our data sets obtained during the
first commissioning run of OSIRIS.\\ 

\begin{center}
\textbf{Time-variable DC biases in the multiplexer readout channels\\}
\end{center}

\noindent The Hawaii II detector is equipped with eight readout channels (128
$\times$ 1024 detector pixels) per quadrant. The channels run horizontally in
the upper right and lower left quadrant and vertically in the other quadrants
(the dispersion axis runs horizontally on the detector).  
Occasionally, the readout electronics introduces time-variable DC biases in 
these channels that do not always cancel out in sky-subtraction, resulting in
residual biases of several \% of the maximum signal.
An automatic correction of these residual biases is only 
possible in dark areas of the detector (the very left and very right of the
detector) and horizontal channels (the lower left and upper right
quadrants). For the other quadrants we developed a manual correction method
that assumes that all measured continua with similar horizontal shift are
similar to each other, except for scaling, an assumption that fairly well
holds for our observation except for the bright AGN itself. 
In the lower/upper half of the detector the first/second half of a spectrum
crosses eight channels while the other half lies within one channel.  
Summing all spectra with similar horizontal shift of the lower and of the
upper half of the detector and comparing them to each other reveals well-defined steps in the two spectra, which are due to the mentioned channel biases.

\begin{center}
\textbf{Electronic ghosts\\}
\end{center}
\noindent Detector rows running along the fast-clock direction of the channel
that are highly discharged (having a high signal) are sometimes
reproduced at a lower signal level in other channels along their fast-clock
direction. This indicates that this crosstalk between channels is due to the
readout electronics. Since the alignment of the channels is vertical in the
upper left and lower right quadrant and horizontal in the others, this
crosstalk (electronic ghosts) may cross spectra on the detector, faking
emission lines. Our method seems to be the most suitable for our commissioning
data. Here, we median 50 rows at the bottom and at the top of the detector
vertically, shift the resulting vectors by a few detector pixels horizontally,  
and subtract them from each other. The difference should be
around zero with electronic ghosts showing up as positive peaks. Their value and
position can easily be determined and then subtracted.

\begin{center}
\textbf{Bad-pixel interpolation in raw data\\}
\end{center}
\noindent Owing to the staggered arrangement of the spectra on the detector, bad
pixels can only be interpolated in the raw image along the dispersion
axis. Depending on how many consecutive pixels need interpolation, the
interpolation is quadratic (isolated bad pixels) or linear (clusters of bad
pixels). Transient bad pixels in commissioning data require special attention.
If a pixel is bad in the rectification matrix but not in the detector image, 
rectification may in some cases diverges giving useless results for the whole 
microlens column at that wavelength, requiring interpolations in the matrix as
well.  
\footnote{In more recent versions of the rectification
algorithm for data taken after the first commissioning run we take full
advantage of the fact that the spectra are two pixels wide on 
the detector to reconstruct the value of bad pixels.}

\begin{center}
\textbf{Wavelength calibration\\}
\end{center}
\noindent Wavelength calibration is performed using neon, argon, krypton, and
xenon calibration lamps that illuminate the entrance pupil of the
instrument. Each spectrum is wavelength-calibrated individually. First, the
known emission lines are fitted with a Gaussian, then the fitted center
positions are interpolated with a cubic dispersion function. Interpolation
onto a regular wavelength grid is done linearly with a grid width close
to the intrinsic sampling of each filter band. The accuracy of the wavelength  
calibration was determined to be 2/10th of a detector pixel (~10 km/s) around
2.1 $\mu$m. \footnote{After commissioning of the instrument and several hardware
updates, the wavelength calibration is made in K band only and the solution is
resampled to the lower diffraction orders of the other wavelength bands.}

\begin{center}
\textbf{Bad-pixel identification and interpolation in data cubes\\}
\end{center}
\noindent Once a data cube is assembled, we can take the advantage of the fact
that neighboring angular resolution elements (spaxels)  
carry similar astronomical information. For any given spaxel all spectra of
neighboring spaxels are scaled in intensity to the spectrum of the central
spaxel. Any data element of a spectrum (spexel) that 
deviates by more than 3$\sigma$ from 
the median of all scaled spectra is considered as bad. Linear interpolation is
then performed along all three axes and the three interpolated values are
averaged. 

\begin{center}
\textbf{Correction of imbalanced sky subtraction\\}
\end{center}
\noindent Biases in sky-subtracted spectra, resulting from slightly imbalanced
sky subtraction or residual biases in the multiplexer readout channels,
can cause residuals when these spectra are divided by the 
telluric spectrum especially in the atmospheric transmission valleys between
2.00 and 2.08 $\mu$m. We
subsequently added a bias to each spectrum, divided by the telluric spectrum, and
fitted the result where the spectrum is free of emission and absorption
lines and where the transmission of the atmosphere is higher than 40\%, 
linearly. Assuming a linear continuum in this small
wavelength range, the appropriate bias was
found when the difference between the fit and the divided spectrum was the
smallest. This bias was then added to the whole spectrum.


\begin{center}
\textbf{Mosaicking\\}
\end{center}
\noindent Prior to mosaicking all data cubes are rotated to a common PA by
spatially oversampling each pixel and resampling to a new 
rotated grid. $\alpha$,$\delta$ offsets are determined from images created by 
collapsing the data cubes along the $\lambda$-axis and cross-correlating them
using the correl\_optimize routine from astrolib
\citep{1993ASPC...52..246L}. The shifted datasets are finally added.

\section{Influence of variable seeing on mosaics of data cubes}

The distribution of flux in a typical PSF of an adaptive optics system 
shows two components, a diffraction-limited spike on top of a more extended Gaussian 
seeing base.
Under different seeing conditions or different AO performances these two components
are pronounced and extended differently. However, the PSF should be radially symmetrical.
We obtained our five K-broadband data cubes under quite different seeing conditions. 
When combining the individual data cubes, circular radial profiles are only preserved
in a region where all data cubes overlap, the central arcsecond around the bright AGN.
Regions where only a fraction of the data cubes overlap are dominated 
by the combined radial flux distribution of those data cubes involved. Due to the 
rectangular shape of our FOV, the radial intensity profiles in the legs of the x-shaped 
mosaic are clearly dominated by the prevailing seeing conditions at the time of observation 
of the data cubes involved. Since the horizontal strip was obtained under poorer 
seeing conditions than the tilted strip, the combined radial profile has 
become of elongated shape, although intrinsically, the radial continuum profile is circular 
on this scale, see \citet{pel99}. \\

\noindent While mosaics of data cubes that contain extended emission suffer from variable seeing in the 
non-overlapping regions, localized emission of higher spatial frequency such as the nucleus 
are much less effected by seeing variations. Their shape is that of the combined PSF spikes 
and their locations within the FOV remain unchanged. As a result, under variable seeing 
conditions, localized emission and absorption features can reliably be traced throughout 
the combined data cube, while extended emission can only be reliably traced within the area 
shared by all contributing data cubes. We are therefore restricting our analysis of extended 
structures like the continuum to the central region in Figure \ref{bands}.

\section{Appendix: Spectral decomposition}

The decompositioning algorithm is based on the algorithm presented in
\citet{kra00}. In active and starburst galaxies, one can identify 
at least four sources of continuum emission: 
stellar ($I^{Star}_\lambda$), AGN power-law
($I^P_\lambda$), hot dust ($I^D_\lambda$), and free-free ($I^{FF}_\lambda$)
emission. The intrinsic continuum
emission $I^{Int}_\lambda$ at each wavelength is 
\begin{equation}
   I^{Int}_\lambda = w_SI^{Star}_\lambda + w_PI^P_\lambda + w_DI^D_\lambda +
   w_{FF}I^{FF}_\lambda ,
\end{equation} 

\noindent where $w_S$, $w_P$, $w_D$, and $w_{FF}$ represent the contributions
of the individual sources. 
A synthetic continuum spectrum $I^{Synth.}$ can be constructed by additionally
applying extinction. Here, we define the optical depth $\tau(\lambda,k)$ with
the reddening parameter $k$ as $\tau=k\lambda^{-1.85}$
\citep{1984A&A...134..284L} and apply a screen model.
We now construct a synthetic spectrum for every measured spectrum that fits
best in terms of minimal $\chi^2$ difference. 
Wavelengths with prominent emission/absorption lines due to non-continuum
sources (e.g. shocked ISM gas, photoionized gas clouds) are not considered in
the $\chi^2$ calculations.\\ 

\noindent The stellar component of the continuum emission was modeled 
using normalized stellar K-band spectra of \citet{1997ApJS..111..445W}. The
spectral 
resolution of our data was downsampled to the resolution of our stellar
template spectrum of HR8726, a K5Ib supergiant. The stellar template spectrum
was additionally redshifted and velocity broadened by convolving it with
a Gaussian line-of-sight velocity profile. The AGN power-law emission was
assumed to vary as $I^P_\lambda = \lambda^\alpha$ with the spectral index
$\alpha$. Emission from hot dust of temperature T was taken to be a black body
$B_\lambda(T)$ multiplied by a power-law emissivity of the form
$\epsilon_{Dust} \propto \lambda^{-1}$. Additionally, we assumed
that the amount of extinction and the amount of emission from hot dust are
correlated $I^D_\lambda \propto (1-e^{-\tau(\lambda,k)})$, giving
$I^D_\lambda = B_\lambda(T)\lambda^{-1} * (1.-e^{-\tau(\lambda,k)})$.\\
For reasons mentioned above, we ignored eventual free-free emission in our
calculations and limited the wavelength range when decomposing to 2.25 to 2.36
micron.
The complete set of fit parameters is $w_S$, $w_P$, $w_D$, $k$, $\alpha$, $T$,
the stellar redshift $z$, and the stellar velocity dispersion $\sigma$.
To increase the signal-to-noise ratio we convolved each slice of the datacube
with constant wavelength with a Gaussian seeing disk with a FWHM of 100
milli-arcseconds.

\end{appendix}

\end{document}